\documentclass[prd, twocolumn, lengthcheck, superscriptaddress, showpacs, letterpaper, nofootinbib]{revtex4}%{revtex4-1}

\usepackage{ifpdf}
\usepackage{color}
\usepackage{graphicx}  
\usepackage{float}
\usepackage{amsmath}
\usepackage{acronym}
\usepackage{multirow}
\usepackage{xspace}
\usepackage{units}
\usepackage{aas_macros}
\usepackage{hyperref}

\newcommand\optional[1]{}
\newcommand{\Msun}{\ensuremath{\mathrm{M}_\odot}}
\newcommand{\Mc}{\ensuremath{\mathcal{M}}}

\newacro{BNS}{binary neutron star}
\newacro{NSBH}{neutron star -- black hole binary}
\newacro{BBH}{binary black hole}
\newacro{SNR}{signal-to-noise ratio}
\newacro{PDF}{probability density function}
\newacro{PSD}{power spectral density}
\newacro{GW}{gravitational wave}
\newacro{CBC}{compact binary coalescence}

\begin{document}
\title{Parameter estimation for compact binary coalescence signals with the first generation gravitational-wave detector network}

\pacs{%
04.80.Nn, % gravitational-wave detectors and experiments
04.25.dg, % black-hole binaries
95.85.Sz, % Gravitational waves: astronomical observations
97.80.-d  % Stars: binary and multiple
}

\affiliation{LIGO - California Institute of Technology, Pasadena, CA  91125, USA}
\affiliation{California State University Fullerton, Fullerton CA 92831 USA}
\affiliation{SUPA, University of Glasgow, Glasgow, G12 8QQ, United Kingdom}
\affiliation{Laboratoire d'Annecy-le-Vieux de Physique des Particules (LAPP), Universit\'e de Savoie, CNRS/IN2P3, F-74941 Annecy-Le-Vieux, France}
\affiliation{INFN, Sezione di Napoli $^a$; Universit\`a di Napoli 'Federico II'$^b$, Complesso Universitario di Monte S.Angelo, I-80126 Napoli; Universit\`a di Salerno, Fisciano, I-84084 Salerno$^c$, Italy}
\affiliation{LIGO - Livingston Observatory, Livingston, LA  70754, USA}
\affiliation{Cardiff University, Cardiff, CF24 3AA, United Kingdom}
\affiliation{University of Sannio at Benevento, I-82100 Benevento, Italy and INFN (Sezione di Napoli), Italy}
\affiliation{Albert-Einstein-Institut, Max-Planck-Institut f\"ur Gravitationsphysik, D-30167 Hannover, Germany}
\affiliation{Leibniz Universit\"at Hannover, D-30167 Hannover, Germany}
\affiliation{Nikhef, Science Park, Amsterdam, the Netherlands$^a$; VU University Amsterdam, De Boelelaan 1081, 1081 HV Amsterdam, the Netherlands$^b$}
\affiliation{National Astronomical Observatory of Japan, Tokyo  181-8588, Japan}
\affiliation{University of Wisconsin--Milwaukee, Milwaukee, WI  53201, USA}
\affiliation{INFN, Sezione di Pisa$^a$; Universit\`a di Pisa$^b$; I-56127 Pisa; Universit\`a di Siena, I-53100 Siena$^c$, Italy}
\affiliation{University of Florida, Gainesville, FL  32611, USA}
\affiliation{INFN, Sezione di Roma$^a$; Universit\`a 'La Sapienza'$^b$, I-00185 Roma, Italy}
\affiliation{LIGO - Hanford Observatory, Richland, WA  99352, USA}
\affiliation{University of Birmingham, Birmingham, B15 2TT, United Kingdom}
\affiliation{Albert-Einstein-Institut, Max-Planck-Institut f\"ur Gravitationsphysik, D-14476 Golm, Germany}
\affiliation{Montana State University, Bozeman, MT 59717, USA}
\affiliation{European Gravitational Observatory (EGO), I-56021 Cascina (PI), Italy}
\affiliation{Syracuse University, Syracuse, NY  13244, USA}
\affiliation{LIGO - Massachusetts Institute of Technology, Cambridge, MA 02139, USA}
\affiliation{APC, AstroParticule et Cosmologie, Universit\'e Paris Diderot, CNRS/IN2P3, CEA/Irfu, Observatoire de Paris, Sorbonne Paris Cit\'e, 10, rue Alice Domon et L\'eonie Duquet, 75205 Paris Cedex 13, France}
\affiliation{Columbia University, New York, NY  10027, USA}
\affiliation{Stanford University, Stanford, CA  94305, USA}
\affiliation{IM-PAN 00-956 Warsaw$^a$; Astronomical Observatory Warsaw University 00-478 Warsaw$^b$; CAMK-PAN 00-716 Warsaw$^c$; Bia{\l}ystok University 15-424 Bia{\l}ystok$^d$; NCBJ 05-400 \'Swierk-Otwock$^e$; Institute of Astronomy 65-265 Zielona G\'ora$^f$,  Poland}
\affiliation{The University of Texas at Brownsville, Brownsville, TX 78520, USA}
\affiliation{San Jose State University, San Jose, CA 95192, USA}
\affiliation{Moscow State University, Moscow, 119992, Russia}
\affiliation{LAL, Universit\'e Paris-Sud, IN2P3/CNRS, F-91898 Orsay$^a$; ESPCI, CNRS,  F-75005 Paris$^b$, France}
\affiliation{NASA/Goddard Space Flight Center, Greenbelt, MD  20771, USA}
\affiliation{University of Western Australia, Crawley, WA 6009, Australia}
\affiliation{The Pennsylvania State University, University Park, PA  16802, USA}
\affiliation{Universit\'e Nice-Sophia-Antipolis, CNRS, Observatoire de la C\^ote d'Azur, F-06304 Nice$^a$; Institut de Physique de Rennes, CNRS, Universit\'e de Rennes 1, 35042 Rennes$^b$, France}
\affiliation{Laboratoire des Mat\'eriaux Avanc\'es (LMA), IN2P3/CNRS, Universit\'e de Lyon, F-69622 Villeurbanne, Lyon, France}
\affiliation{Washington State University, Pullman, WA 99164, USA}
\affiliation{INFN, Sezione di Perugia$^a$; Universit\`a di Perugia$^b$, I-06123 Perugia; Universit\`a di Camerino, Dipartimento di Fisica$^c$, I-62032 Camerino, Italy}
\affiliation{INFN, Sezione di Firenze, I-50019 Sesto Fiorentino$^a$; Universit\`a degli Studi di Urbino 'Carlo Bo', I-61029 Urbino$^b$, Italy}
\affiliation{University of Oregon, Eugene, OR  97403, USA}
\affiliation{Laboratoire Kastler Brossel, ENS, CNRS, UPMC, Universit\'e Pierre et Marie Curie, 4 Place Jussieu, F-75005 Paris, France}
\affiliation{University of Maryland, College Park, MD 20742 USA}
\affiliation{Universitat de les Illes Balears, E-07122 Palma de Mallorca, Spain}
\affiliation{University of Massachusetts - Amherst, Amherst, MA 01003, USA}
\affiliation{Canadian Institute for Theoretical Astrophysics, University of Toronto, Toronto, Ontario, M5S 3H8, Canada}
\affiliation{Tsinghua University, Beijing 100084 China}
\affiliation{University of Michigan, Ann Arbor, MI  48109, USA}
\affiliation{Louisiana State University, Baton Rouge, LA  70803, USA}
\affiliation{The University of Mississippi, University, MS 38677, USA}
\affiliation{Charles Sturt University, Wagga Wagga, NSW 2678, Australia}
\affiliation{Caltech-CaRT, Pasadena, CA  91125, USA}
\affiliation{INFN, Sezione di Genova;  I-16146  Genova, Italy}
\affiliation{Pusan National University, Busan 609-735, Korea}
\affiliation{Australian National University, Canberra, ACT 0200, Australia}
\affiliation{Carleton College, Northfield, MN  55057, USA}
\affiliation{The University of Melbourne, Parkville, VIC 3010, Australia}
\affiliation{INFN, Sezione di Roma Tor Vergata$^a$; Universit\`a di Roma Tor Vergata, I-00133 Roma$^b$; Universit\`a dell'Aquila, I-67100 L'Aquila$^c$, Italy}
\affiliation{University of Salerno, I-84084 Fisciano (Salerno)}
\affiliation{Instituto Nacional de Pesquisas Espaciais,  12227-010 - S\~{a}o Jos\'{e} dos Campos, SP, Brazil}
\affiliation{The University of Sheffield, Sheffield S10 2TN, United Kingdom}
\affiliation{Wigner RCP, RMKI, H-1121 Budapest, Konkoly Thege Mikl\'os \'ut 29-33, Hungary}
\affiliation{Inter-University Centre for Astronomy and Astrophysics, Pune - 411007, India}
\affiliation{University of Minnesota, Minneapolis, MN 55455, USA}
\affiliation{INFN, Gruppo Collegato di Trento$^a$ and Universit\`a di Trento$^b$,  I-38050 Povo, Trento, Italy;   INFN, Sezione di Padova$^c$ and Universit\`a di Padova$^d$, I-35131 Padova, Italy}
\affiliation{California Institute of Technology, Pasadena, CA  91125, USA}
\affiliation{Northwestern University, Evanston, IL  60208, USA}
\affiliation{Rochester Institute of Technology, Rochester, NY  14623, USA}
\affiliation{E\"otv\"os Lor\'and University, Budapest, 1117 Hungary}
\affiliation{University of Cambridge, Cambridge, CB2 1TN, United Kingdom}
\affiliation{University of Szeged, 6720 Szeged, D\'om t\'er 9, Hungary}
\affiliation{Rutherford Appleton Laboratory, HSIC, Chilton, Didcot, Oxon OX11 0QX United Kingdom}
\affiliation{Embry-Riddle Aeronautical University, Prescott, AZ   86301 USA}
\affiliation{Perimeter Institute for Theoretical Physics, Ontario, N2L 2Y5, Canada}
\affiliation{American University, Washington, DC 20016, USA}
\affiliation{University of New Hampshire, Durham, NH 03824, USA}
\affiliation{University of Southampton, Southampton, SO17 1BJ, United Kingdom}
\affiliation{Korea Institute of Science and Technology Information, Daejeon 305-806, Korea}
\affiliation{Hobart and William Smith Colleges, Geneva, NY  14456, USA}
\affiliation{Institute of Applied Physics, Nizhny Novgorod, 603950, Russia}
\affiliation{Lund Observatory, Box 43, SE-221 00, Lund, Sweden}
\affiliation{Hanyang University, Seoul 133-791, Korea}
\affiliation{Seoul National University, Seoul 151-742, Korea}
\affiliation{University of Strathclyde, Glasgow, G1 1XQ, United Kingdom}
\affiliation{The University of Texas at Austin, Austin, TX 78712, USA}
\affiliation{Southern University and A\&M College, Baton Rouge, LA  70813, USA}
\affiliation{University of Rochester, Rochester, NY  14627, USA}
\affiliation{University of Adelaide, Adelaide, SA 5005, Australia}
\affiliation{National Institute for Mathematical Sciences, Daejeon 305-390, Korea}
\affiliation{Louisiana Tech University, Ruston, LA  71272, USA}
\affiliation{McNeese State University, Lake Charles, LA 70609 USA}
\affiliation{Andrews University, Berrien Springs, MI 49104 USA}
\affiliation{Trinity University, San Antonio, TX  78212, USA}
\affiliation{University of Washington, Seattle, WA, 98195-4290, USA}
\affiliation{Southeastern Louisiana University, Hammond, LA  70402, USA}

\author{J.~Aasi$^\text{1}$}\noaffiliation
\author{J.~Abadie$^\text{1}$}\noaffiliation
\author{B.~P.~Abbott$^\text{1}$}\noaffiliation
\author{R.~Abbott$^\text{1}$}\noaffiliation
\author{T.~D.~Abbott$^\text{2}$}\noaffiliation
\author{M.~Abernathy$^\text{3}$}\noaffiliation
\author{T.~Accadia$^\text{4}$}\noaffiliation
\author{F.~Acernese$^\text{5a,5c}$}\noaffiliation
\author{C.~Adams$^\text{6}$}\noaffiliation
\author{T.~Adams$^\text{7}$}\noaffiliation
% ...

\begin{abstract}

Compact binary systems with neutron stars or black holes are one of the most promising sources for ground-based gravitational-wave detectors. 
Gravitational radiation encodes rich information about source physics; thus parameter estimation and model selection are crucial analysis steps for any detection candidate events. Detailed models of the anticipated waveforms enable inference on several parameters, such as component masses, spins, sky location and distance, that are essential for new astrophysical studies of these sources. However, accurate measurements of these parameters and discrimination of models describing the underlying physics are complicated by artifacts in the data, uncertainties in the waveform models and in the calibration of the detectors. Here we report such measurements on a selection of simulated signals added either in hardware or software to the data collected by the two LIGO instruments and the Virgo detector during their most recent joint science run, including a ``blind injection'' where the signal was not initially revealed to the collaboration. We exemplify the ability to extract information about the source physics on signals that cover the neutron-star and black-hole binary parameter space over the component mass range $1\,\Msun - 25\,\Msun$ and the full range of spin parameters. The cases reported in this study provide a snap-shot of the status of parameter estimation in preparation for the operation of advanced detectors.

\end{abstract}
\maketitle

\section{Introduction}
General relativity predicts that binary systems of compact objects lose energy through the emission of 
gravitational radiation, a prediction confirmed through binary-pulsar 
observations \cite{Einstein:1918,PhysRev.131.435,weisberg:2010}. During
this process they emit a characteristic `chirping' \ac{GW} signal of predominantly increasing
amplitude and frequency. For neutron stars and black holes the signal enters the observational 
band of the initial (at ~40\,Hz) and advanced (at ~10-20\,Hz) ground-based laser-interferometric detectors 
(e.g., ~\cite{Abbott:2007kv,Accadia:2012zz,Commissioning}), 
sweeping through the detection band for a few seconds (depending on the
masses of the objects) until coalescence.

The search for \ac{GW} signatures of compact binary coalescence in LIGO's most recent sixth science run and Virgo's science runs 2 \& 3 is described in \cite{Collaboration:S6CBClowmass,2013PhRvD..87b2002A}.  
No detection was reported. However, the operation of advanced instruments -- Advanced LIGO and Advanced Virgo -- from 2015+~\cite{Commissioning,Harry:2010zz,PSD:AV} suggests that the coalescence of compact binaries could be observed in the not too distant future~\cite{ratesdoc}. Once a detection candidate has been identified, the next step in the analysis is to measure the source parameters and test models that describe the underlying physics. This step is critical for enabling studies in astrophysics and fundamental physics in which \ac{GW} observations are expected to provide a new view of relativistic phenomena. One of the signals studied in this work represents an
end-to-end test of this process: a detection candidate was identified
with a false alarm rate of 1 in 7000 years ($1.4\times 10^{-4}$
yr$^{-1}$)~\cite{Collaboration:S6CBClowmass} and was fully characterized in terms of its physical
properties before it was revealed to be a ``blind'' hardware injection, i.e., a simulated signal added by coherently actuating the mirrors of the LIGO and Virgo detectors, 
without the knowledge of the data analysts~\cite{GW100916web}.

The ability to accurately estimate the parameters of coalescing binaries, including the masses of the components, 
their spins, the location of a binary on the sky and its distance, and the implications for new insights into the underlying physical processes have been at the centre of several studies. 
Accurate measurements of the masses provide information
about the mass distribution of (binary) black holes and neutron stars,
clues to determining the maximum mass of neutron stars \cite[e.g.,][]{Read:2009}, the underlying
neutron star equation of state, the minimum mass of stellar-mass black
holes and the presence or absence of the so-called ``mass gap''
\cite{Ozel:2010,Farr:2010,Kreidberg:2012}. Spin measurements are a direct window onto the critical stage of common envelope evolution, and the details of the supernova processes and kicks. Localizing the merger of compact binary systems on the sky may permit the identification of the host, and therefore an opportunity for in-depth studies of the environments in which compact binaries form \cite[e.g.,][]{Fairhurst2009,Nissanke:2011,2011arXiv1109.3498T,Virgo:2011aa,SwiftS6,2013ApJ...767..124N}. If the localization region in the sky is sufficiently small, a possible electro-magnetic counterpart could be found and multi-messenger studies of coalescing compact objects would become possible \cite[e.g.,][]{Bloom:2009vx,Mandel:2011,Metzger:2011bv}. For instance, the question of whether compact binaries are the progenitors of short gamma-ray bursts could be definitively answered. The direct determination of the luminosity distance to the source opens new possibilities for low redshift cosmography. In general, coalescing binaries are a laboratory for tests of the behavior of gravity in the strong-field regime.  Eventually, when multiple detections are available, studies of source populations will be possible \cite[e.g.,][]{BulikBelczynski:2003, MandelOshaughnessy:2010}, as well as more stringent tests of the underlying source dynamics~\cite{2012PhRvD..85h2003L,2012JPhCS.363a2028L}. 

Most of these studies use theoretical estimates of parameter uncertainty based on the Cramer-Rao bound~\cite{helmstrom-1968}, which should be valid in the limit of high \ac{SNR}.  Initial detections may be too weak for this bound to provide useful guidance.  Therefore, a complete Bayesian analysis like the one described below must be used to quantify parameter uncertainties.  
Other studies have relied on injections into synthetic data.  In this paper, we will use injections into real data, which introduces a new set of challenges,  including non-Gaussianity and non-stationarity.
 
The fact that gravitational waveforms used in the analysis are an approximation to the actual radiation produced by astrophysical sources and that the measured strain is affected by the uncertainties in the instrument calibration \cite{LIGOS5,VirgoS2,Vitale:2012} represent additional challenges for making robust inference on the underlying physics.
To study parameter estimation in this regime, we have analyzed several artificial \ac{CBC} signals added to real detector data, 
including the ``blind'' injection described above, added both in hardware and
software to the data collected by the two LIGO instruments (Hanford and Livingston) and the
Virgo detector during the most recent joint science run, S6/VSR2-3.
 The use of injections has been, and continues to be, an essential means to validate the detection process, and as we report here, has been naturally extended to the source characterization stage of the analysis.  Here we exemplify the ability to extract information about the source physics on a selected number of injections that cover the neutron star and black hole parameter space over the component mass range $1\,\Msun - 25\,\Msun$ and the full range of spin parameters. We consider a spectrum of realistic signal strengths, from candidates observed close to the detection threshold to high \ac{SNR} events, and various relative strengths across the instruments of the network. We analyzed the signals using a range of waveform models that demonstrate the interplay between (some) systematic bias and statistical uncertainty. To help validate our results, we carry out the analysis with several independent techniques; these are implemented within a specially-developed software package part of the LSC Algorithm Library, \emph{LALInference}~ \cite{LALPE}.

The paper is organized as follows. In Section \ref{sec:Analysis} we give a brief overview of the analysis method. While no detections were claimed in \cite{Collaboration:S6CBClowmass}, simulated signals (`injections') were added to the data, both at a hardware level as the data was being taken and in software afterwards. The hardware injections were performed to validate the end-to-end analysis, including parameter estimation on detection candidates, whereas the injections in software serve as a useful comparison, free of any calibration error in the detectors.
Here we report on the analysis using six hardware and software injections, including waveform models for \ac{BNS}, \ac{NSBH} and \ac{BBH} simulations, described in section \ref{sec:Simulations}.
One of these hardware injections was performed without the knowledge of the data analysis teams as part of the `blind injection challenge'; it was successfully detected as reported in \cite{Collaboration:S6CBClowmass}.
We use these injections to illustrate the possible implications for \ac{GW} astronomy in section \ref{sec:implications} and we conclude in section \ref{sec:astro}.

\section{Analysis}
\label{sec:Analysis}

\subsection{Bayesian Inference}
\label{sec:bayes}

Each data segment containing an injected signal was analyzed using a Bayesian parameter estimation pipeline to calculate the \ac{PDF} of the unknown parameters of the waveform model.
We will call $\vec{\theta}$ the vector containing those parameters. The actual content and dimension of $\vec{\theta}$, i.e. the dimensionality of the parameter space, depend on the waveform model used for the analysis (see section~\ref{sec:waveform}).

The posterior distribution of $\vec\theta$ given a model $H$ is given by Bayes' theorem,
\begin{equation}\label{eq.PDF}
p(\vec{\theta}|\{d\},H)=\frac{p(\vec{\theta}|H)p(\{d\}|\vec{\theta},H)}{P(\{d\}
|H)}\ ,
\end{equation}
where  $p(\vec{\theta}|H)$ is the \emph{prior} distribution of $\vec\theta$, describing 
knowledge about the parameters within a model $H$ before the data is analyzed, and $p(\{d\}|\vec{\theta},H)$ is the \emph{likelihood function}, denoting the probability under model $H$ of obtaining the dataset $\{d\}$  for a given parameter set $\vec{\theta}$. The likelihood is a function of the noise-weighted residuals after subtracting the model from the data, and is thus a direct measure of the goodness of fit of the model to the data.

The \emph{Optimal Network \ac{SNR}} is defined as
\begin{equation}\label{eq.SNR}
\mathrm{SNR}=\sqrt{\sum_{det} \int_{f_{Low}}^{f_{High}}\frac{|s_{det}(f,\vec\theta)|^2}{S_{det}(f)}df},
\end{equation}
where the sum is taken over each detector $det$, with $s_{det}$ the signal in that detector and $S_{det}(f)$ its noise \ac{PSD}.

Our model for the likelihood function is based on the assumption that the noise is stationary and Gaussian, and uncorrelated in different frequency bins.  Although we do not expect this assumption to be precisely true for real detector noise, limited investigations suggest that this is an acceptable approximation when the data is of good quality \cite{Raymond:2010}.

The denominator of eq. \ref{eq.PDF}, $P(\{d\}|H) \equiv Z_{H}$, is the \emph{evidence} for the model $H$. 
As it is a normalization constant, the evidence does not affect the estimation of the parameters for a particular model $H$, but it does allow us to compare the ability of different models to describe the data. 
The Bayes factor between two models, which quantifies the relative support given to each model by the data, is given by
\begin{equation}
\frac{Z_1}{Z_2} \equiv \frac{P(\{d\}|H_1)}{P(\{d\}|H_2)}=\frac{\int_{\Theta}p(\vec{\theta}|H_1)p(\{d\}
|\vec{\theta},H_1)d\vec{\theta}}{\int_{\Lambda}p(\vec{\lambda}|H_2)p(\{d\}|\vec{
\lambda},H_2)d\vec{\lambda}}\ .
\end{equation}
where $H_1$ and $H_2$ represent the two different models and $\vec{\theta}$,
$\vec{\lambda}$ their respective model parameter vectors. The parameter spaces can be completely unrelated, even of different dimensionality, which allows us to compare models which range from 9 parameters in the case of non-spinning
systems, up to 15 in the case of a system with two spinning components.
Throughout we will quote evidence relative to the Gaussian noise model, $\ln(Z) = \ln(Z_{H})-\ln(Z_{\mathrm{Gaussian}})$.

The high dimensionality of the parameter space and the complicated structure of the likelihood function make it impractical to exhaustively calculate posterior quantities. Instead we rely
on stochastic sampling of the posterior \ac{PDF}, which provides us with an approximation of the underlying true distributions, \emph{e.g.} by binning the samples to produce histograms.
The analysis was performed using \emph{LALInference}~\cite{LALPE}, which allows calculation of the prior, likelihood and templates using standardised functions.
Because accurate sampling of the posterior \ac{PDF} is a difficult task, we used two independent sampling algorithms on all but a few cases, cross-comparing results to confirm convergence.
These were based on Markov Chain Monte Carlo (MCMC) \cite{Sluys:2008a,Sluys:2008b} and Nested Sampling \cite{Veitch:2010} techniques, and were included in the \emph{LALInference} package.
We found good agreement between the two techniques giving us confidence in the results.
For a sample of the cases we used a third algorithm, MultiNest \cite{Feroz:2009}, as an additional check.
The Nested Sampling and MultiNest algorithms directly produce an estimate of the evidence \cite{Skilling:2006}; we also computed the evidence from MCMC results using direct integration~\cite{Weinberg:2009}.

\subsection{Waveforms and source parameters}
\label{sec:waveform}

Unless otherwise specified we shall use a system of units in which $c = G = 1$. The waveform as measured at a generic detector can be written in the frequency domain as \cite{lrr-2009-2}:

\begin{eqnarray}\label{eq.signalIFO}
    s(f,\vec\theta)&=& [F_+(\alpha,\delta,\psi) h_+(f,\vec\theta') \\ \nonumber
    &&+F_{\times}(\alpha,\delta,\psi)\; h_{\times}(f,\vec\theta')] e^{2 \pi i f \Delta t}
\end{eqnarray}

where:

\begin{itemize}
    \item $F_+(\alpha,\delta,\psi)$ and $F_{\times}(\alpha,\delta,\psi)$ are the known antenna beam pattern of the detector, which gives the amplitude response of the antenna to the $+$ and $\times$ polarizations.
    \item $\alpha$ and $\delta$ are the right-ascension and declination of the source;
    \item $\psi$ is the polarization angle (see, e.g., \cite{thorne.k:1987});
    \item $\Delta t$ is the time delay between the arrival of the signal at the detector and at a common reference frame (e.g., the center of the Earth), a known function of $\alpha$ and $\delta$;
    \item $h_+(f,\vec\theta')$ and $h_\times(f,\vec\theta')$ are the two independent polarizations of the signal, with $\vec\theta'=\vec\theta\setminus\left\{\alpha,\delta,\psi\right\}$ (i.e. $\vec\theta'$ includes all the waveform parameters except for right-ascension, declination and polarization angle).
\end{itemize}

The actual form of $h_+(f,\vec\theta')$ and $h_\times(f,\vec\theta')$ depends on which model one is considering. 

The \ac{GW} signal can be written using the post-Newtonian (pN) expansion~\cite{Blanchet:2006av}.
We used a total of nine different waveform models (or \emph{approximants}) in our analysis, including the TaylorF2 
(which gives a pN-based, analytic expression for the frequency-domain waveform)~\cite{BuonannoIyerOchsnerYiSathya2009,2012arXiv1210.6666A}, 
SpinTaylorT4 (which is a time-domain pN model supplemented with spin precession equations)~\cite{BuonannoChenVallisneri:2003b} 
and IMRPhenomB (which is a TaylorF2-like waveform model supplemented with a merger-ringdown and calibrated to agree with numerical relativity simulations)~\cite{Ajith:2009bn} approximants with binary spin parameters either disabled, set aligned to the orbital angular momentum, or enabled and allowed full freedom of orientation. Table~\ref{tab:waveforms} lists the approximants used.
For example, the TaylorF2 waveform at lowest amplitude order can be written in the frequency domain combined with the stationary phase approximation \cite{lrr-2009-2}:

\begin{eqnarray}
    h_+(f)&=& \frac{1+\cos^2\iota}{2 d_L} \;h(f)\label{eq.tf2a}\\
    h_\times(f)&=& i \frac{\cos\iota}{d_L} \;h(f)\label{eq.tf2b}
\end{eqnarray}

where:
\begin{eqnarray}
h(f)&=&\frac{\Mc^\frac{5}{6}}{\pi^\frac{2}{3}}\sqrt{\frac{5}{24}} f^{-7/6} e^{-i \psi(f)} \\
\label{eq.psioff}
\psi(f)&=& 2\pi f t_c - \phi_c - \frac{\pi}{4} + \frac{3}{128 \eta v^5}  \sum_{k=0}^N \alpha_k v^k; \\
&&\mbox{ with  }v\equiv (\pi (m_1+m_2) f)^\frac{1}{3}. \nonumber
\end{eqnarray}

We see that the two polarizations depend on six parameters (the $\alpha_k$ are known functions of $\Mc$ and $\eta$,~\cite{BuonannoIyerOchsnerYiSathya2009}, up to $N=7$) :

\begin{itemize}
    \item $\Mc$. The chirp mass of the system: $\Mc=(m_1 m_2)^{3/5}(m_1+m_2)^{-1/5}$, where $m_1$ and $m_2$ are the component masses;
    \item $\eta$. The symmetric mass ratio, defined as: $\eta=(m_1 m_2)/(m_1+m_2)^2$;
    \item $\iota$. The inclination angle, i.e. the angle between the orbital angular momentum vector and the line of sight;
    \item $d_L$. The luminosity distance to the source;
    \item $t_c$. An arbitrary reference time, usually chosen to be the time of coalescence of the binary;
    \item $\phi_c$. The phase of the waveform at the reference time $t_c$. 
\end{itemize}

Together with $\alpha$, $\delta$ and $\psi$, these six parameters form the 9-dimensional vector $\vec\theta$ for the TaylorF2 model.
The upper value of the index $k$ in eq.~(\ref{eq.psioff}) determines the phase pN order, with $N=0$ being the lowest order (0pN) and $N=7$ the 3.5pN order.
We assume that the eccentricity of the binary system is negligible, as radiation reaction circularises orbits efficiently before the signal frequency enters the detector's observational band~\cite{2011A&A...527A..70K}.
We can also see from equations~\ref{eq.tf2a} and \ref{eq.tf2b} that the inclination $\iota$ and distance $d_L$ are strongly correlated. 
This correlation is illustrated in section~\ref{sec:Simulations}.

When spins are present and we constrain the spin vectors to be aligned with the orbital angular momentum of the binary, we need two additional parameters bringing $\vec\theta$ to 11 parameters. We use the spin magnitudes $a_1$ and $a_2$, defined as $a_i\equiv (\vec{s_i}\cdot\hat{L})/m_i^2$ where $\vec{s_i}$ and $m_i$ are the spin and mass of the object $i$, and $\hat{L}$ is the unit vector along the orbital angular momentum. To account for both aligned and anti-aligned cases, we allow $a_i$ to be in range $[-1,1]$. When generic spins are considered, we need six additional parameters, and $\vec\theta$ becomes a 15-dimensional vector. In this case the spin magnitudes $a_i$, defined as $a_i\equiv |\vec{s_i}|/m_i^2$ and in the range $[0,1]$, and four angles specify the orientations.

The presence of arbitrary spins will alter both the amplitude and phase of the signal. The pN expansion of the phase evolution is changed to include additional terms dependent on the components of the spin vectors. If the spins are not aligned with the orbital angular momentum, then the spin-orbit and spin-spin coupling will cause the orbital angular momentum vector to precess around the total angular momentum vector. As this implies that $\iota$ (and $\psi$) become time dependent, the values of these quantities at the \ac{GW} frequency 40\,Hz are used for parameter estimation. We refer the reader to~\cite{BuonannoChenVallisneri:2003b} for more details about SpinTaylorT4.

The pN approximation is valid for the \emph{inspiral} phase of the binary, when the two objects are still far apart~\cite{Blanchet:2002xy,2008PhRvD..77l4006Y}. 
For this reason, the TaylorF2 and SpinTaylorT4 waveforms must terminate at an approximate point 
where they begin to break down, with the TaylorF2 ending at the \emph{innermost stable circular orbit}~\cite{BuonannoIyerOchsnerYiSathya2009,2012arXiv1210.6666A} 
frequency of a test particle orbiting a Schwarzschild black hole, 
and the SpinTaylorT4 terminating at the \emph{minimum energy circular orbit}~\cite{BuonannoChenVallisneri:2003b}.
IMRPhenomB waveforms extend the signal into the merger and ring-down stages of the binary coalescence, making use of a phenomenological model tuned by comparison with numerical relativity simulations. These later stages take place at higher frequencies than the inspiral, hence are more important for higher-mass systems where the frequency scale is lower, and where the merger and ring-down contribute a larger proportion of \ac{SNR} by being in a more sensitive band of the detectors.

For the analysis of non-blind hardware and software injections (the first seven models of Table~\ref{tab:waveforms}), we considered the pN expansion either up to the 2pN order in phase or up to the 3.5pN order in phase (2.5pN for spin effects when included), and 0pN in amplitude. For the blind hardware injection we used approximants at 2.5pN order in phase, as the blind injection only included terms up to this order~\cite{Collaboration:S6CBClowmass,GW100916web} (see Sect.~\ref{bigdog} below).
The different waveforms use different post-Newtonian expansions of the binary phase, and do not give identical results in general. When analyzing a real \ac{GW} signal, the ignorance of the higher order pN terms will cause a systematic bias in the inference of the binary parameters, and using a range of different approximants here will give us an indication of the size of this error compared to the statistical uncertainty in the results.

\subsection{Priors}

As shown in~eq. \ref{eq.PDF}, the posterior distribution of $\vec\theta$ depends both on the likelihood and prior distributions of $\vec\theta$. We used the same prior density functions (and range) for all the analyses, uniform in the
component masses with the range $1\,\Msun\le m_{1,2}\le 30\,\Msun$, and with the total mass constrained by $m_1 + m_2 \le 35\,\Msun$. This range encompasses the low-mass search range used in \cite{Collaboration:S6CBClowmass}, where $1\,\Msun\le m_{1,2}\le 24\,\Msun$ and $m_1 + m_2 \le 25\,\Msun$.
The prior density function on the location of the source was taken to be
uniform in volume, constrained between luminosity distances $d_L \in [1,100]$~Mpc.  We used an
isotropic prior on the orientation of the orbital plane of the binary. 
For analyses using waveform models that account for possible spins, the prior on
the spin magnitudes, $a_1,a_2$, was taken to be uniform in the range $[0,1]$ (range $[-1,1]$ in the spin-aligned cases), and the spin
angular momentum vectors were taken to be isotropic.

The computational cost of the parameter estimation pipeline precludes us from running it on all times; therefore, the parameter estimation analysis relies on an estimate of the coalescence time as provided by the detection pipeline \cite{Collaboration:S6CBClowmass}.  In practice, a 200\,ms window centered on this trigger time is sufficient to guard against the uncertainty in the coalescence time estimates from the detection pipeline, see for instance \cite{Brown:2004vh,ihopePaper:2012}.
Our results are not significantly affected by other astrophysically sensible choices of priors. For the \ac{SNR}s used in this paper, our posteriors are much narrower than our priors for most parameters.

\begin{table}

\begin{tabular}{|c|l|c|c|c|}
\hline
Model & Name  & Spin & Merger and & Ref. \\
 & & effects & ring-down & \\
\hline
TF2 & TaylorF2 & no & no & \cite{BuonannoIyerOchsnerYiSathya2009} \\
TF2\_2 & TaylorF2 @ 2pN & no & no & \cite{BuonannoIyerOchsnerYiSathya2009} \\
TF2\_RS & TaylorF2 RedSpin & aligned & no & \cite{2012arXiv1210.6666A} \\
ST\_NS & Non-spinning STPN & no & no & \cite{BuonannoChenVallisneri:2003b} \\
ST\_SA & Aligned spin STPN & aligned & no & \cite{BuonannoChenVallisneri:2003b} \\
ST & Full STPN & yes & no & \cite{BuonannoChenVallisneri:2003b} \\
IMRPB & IMRPhenomB & aligned & yes & \cite{Ajith:2009bn}\\
TF2\_25~\footnotemark[1] & TaylorF2 @ 2.5pN & no & no & \cite{BuonannoIyerOchsnerYiSathya2009} \\
ST\_25~\footnotemark[1] & Full STPN @ 2.5pN & yes & no & \cite{BuonannoChenVallisneri:2003b} \\
\hline
\end{tabular}
\footnotetext[1]{Used only for the blind hardware injection}
\caption{\label{tab:waveforms}List of the waveform models used for the analysis. `Aligned' refers to both spin vectors being aligned to the orbital angular momentum of the binary. `STPN' refers to the SpinTaylor post-Newtonian waveform for precessing binaries in LAL \cite{LALPE}. In all models, the inspiral phase of the binary evolution is described by the post-Newtonian (pN) expansion to 3.5pN order (2.5pN for spin effects when included) in phase, and 0pN in amplitude, unless otherwise specified.}
\end{table}

\subsection{Data description}

Data from the multi-detector network consisting of two LIGO instruments (H1 and L1) and
Virgo (V1) were used coherently in the analysis.
From each detector a total of 32 seconds of data [-30,\,+2]\,s around the GPS time of the injection were analyzed.
The initial frequency of the analysis was 40\,Hz, which is low enough so that the detectors are not significantly sensitive to the start of the signals and template. 
The sampling rate was 2048\,Hz (4096\,Hz for the \ac{BNS}, section \ref{hwinj:bns}), corresponding to a 1024\,Hz (2048\,Hz) Nyquist frequency, high enough to include the entire waveform for all models except IMRPhenomB, which has a negligible contribution to the signal-to-noise ratio at higher frequencies.
The 32-second segments of time-domain data were Tukey windowed, with a 0.4\,s roll-off on either side of the segment.
The \ac{PSD} of the instrumental noise was estimated using 1024\,s
of data after the end of the analyzed segment.  We verified that
varying the methods of \ac{PSD} estimation (using $\pm512$ seconds spanning the
signal trigger time, and using median and mean estimation methods), had a negligible
effect on the parameter estimation results, smaller than the systematic uncertainties of parameter estimation, 
although varying the time during which the \ac{PSD} is estimated can have a significant effect (see section \ref{hwsw}).

Calibration errors, which can influence the reconstructed amplitude, phase, and
timing of the data \cite{LIGOS5}, have the potential to affect parameter estimation results.  
An analysis of the effect of calibration errors, which considered mock errors similar to those expected during the S6 and VSR2/3 runs, concluded that such errors are unlikely to cause a significant deterioration in parameter estimation accuracy \cite{Vitale:2012} at the moderate \ac{SNR}s considered in this paper.

\section{Simulations}
\label{sec:Simulations}

\begin{table}

\begin{tabular}{|c|l|c|c|c|c|c|c|c|c|}
\hline
\textsection & HW/ &  $\Mc$ & $m_1$ & $m_2$ & $d_L$ & $\iota $ & $|a_1|$ & $|a_2|$ & SNR\\
 & SW &  $(\Msun)$ & $(\Msun)$ & $(\Msun)$ & $(Mpc)$ & $(^\circ)$ &  &  & \\
\hline
\ref{hwinj:bbh} & HW & 3.865 & 4.91 & 4.02 & 36.2 & 26 & 0 & 0 & 13 \\
\ref{hwinj:bns} & HW & 1.502 & 1.808 & 1.647 & 5.94 & 138 & 0 & 0 & 36 \\
\hline
\ref{swinj:bbh} & SW &  4.76  & 6  & 5 & 30.0  & 1.1 & 0.6  & 0.8 & 19 \\
\ref{swinj:nsbh} & SW &  2.99 & 10.0 & 1.4  & 16.0 & 0.5 & 0.7 & 0 & 13 \\
\hline
\ref{bigdog} & HW &  4.96 & 24.81 & 1.74 & 24.37 & 109 & 0.57 & 0.16 & 16 \\
\hline
\end{tabular}
\caption{\label{tab:injections}Parameters of hardware (HW) and software (SW) injected signals discussed in section~\ref{sec:Simulations}. Non-spinning injections were generated using the EOBNR~\cite{Buonanno:2007pf} waveform model, whereas spinning injections used the SpinTaylor model. The \ac{SNR} column shows optimal network \ac{SNR}, eq.~\ref{eq.SNR}.}
\end{table}

\begin{table}

\begin{tabular}{|c|l|c|c|c|c|c|c|c|c|}
\hline
 \textsection & TF2 & TF2 & TF2 & TF2 & ST & ST & ST & ST & IMR\\
 &  & \_2 & \_25 & \_RS & \_NS & \_SA &  & \_25 & PB\\
\hline
\ref{hwinj:bbh} &  69 & 70 & - & 70 & 69 & 71 & 71 & - & 73\\
\ref{hwinj:bns} & 686 & 699 & - & 694 & 685 & 697 & 694 & - & 668\\
\hline
\ref{swinj:bbh} &  154  & 146 & - & 153 & 158  & 158 & 157 & - & 155\\ 
\ref{swinj:nsbh} &  48 & 48 & - & 50 & 50 & 52 & 64 & - & 52\\
\hline
\ref{bigdog} & - & - & 136 & - & - & - & - & 213 & -\\
\hline
\end{tabular}
\caption{\label{tab:evidences}Logarithm of the evidence $\ln(Z)$ relative to the Gaussian noise model for each injection and each waveform family obtained via direct integration. The numbers come with statistical error bars of $\pm 5$. 
}
\end{table}

Over the course of the LIGO S6 and Virgo VSR2/3 science runs a series of hardware injections have been carried out where the arm lengths of the three detectors were physically changed to simulate the passing of a \ac{GW}.  

As an end-to-end test of the search pipeline during the science runs, a signal was added to the data via a hardware injection, without the data analysts' knowledge (`blind').  The parameters and template family were revealed only after the search was complete, and we include a retrospective analysis of this blind injection in section \ref{bigdog} (see \cite{GW100916web} for parameter estimation carried out before the injection was revealed).

We have also added several simulated software injections into real detector noise from those runs. Below, we describe the results of parameter estimation analysis on several of these injections, whose parameters are listed in Table \ref{tab:injections}. 
We present the posterior probability density functions on the source parameters
using a coherent, multi-detector data model with the seven waveform families described above.
In Appendix ~\ref{sec:credible-intervals} we present 90\% credible intervals
obtained using the waveform models in Table~\ref{tab:waveforms} for
each injection in Table~\ref{tab:injections}.

\subsection{Non-spinning hardware injections}

\subsubsection{Binary black hole}\label{hwinj:bbh}

%\begin{figure*}[p]
%%\vspace{-10pt}
%\begin{minipage}[b]{0.49\textwidth}
%\centering
%\includegraphics[width=\textwidth]{Images/pdf/EOBNR_mchirp_90.pdf}
%\end{minipage}
%\begin{minipage}[b]{0.49\textwidth}
%\centering
%\includegraphics[width=\textwidth]{Images/pdf/EOBNR_m2-m1.pdf}
%\end{minipage}
%%\vspace{-10pt}
%\caption{\label{fig:bbh_mass}(left) Posterior probability distributions for the chirp mass $\Mc$ of the non-spinning \ac{BBH} hardware injection (section \ref{hwinj:bbh}) for the seven signal models considered. The injected value is marked with a vertical red line. (right) Overlay of 90\% probability regions for the joint posterior distribution on the component masses $m_1$, $m_2$ of the binary. The true value is marked by the blue star. Models which allow for non-zero spins find wider \ac{PDF}s for the coupled mass parameters. 
%}
%\end{figure*}

\begin{figure*}[p]

\includegraphics[width=\textwidth]{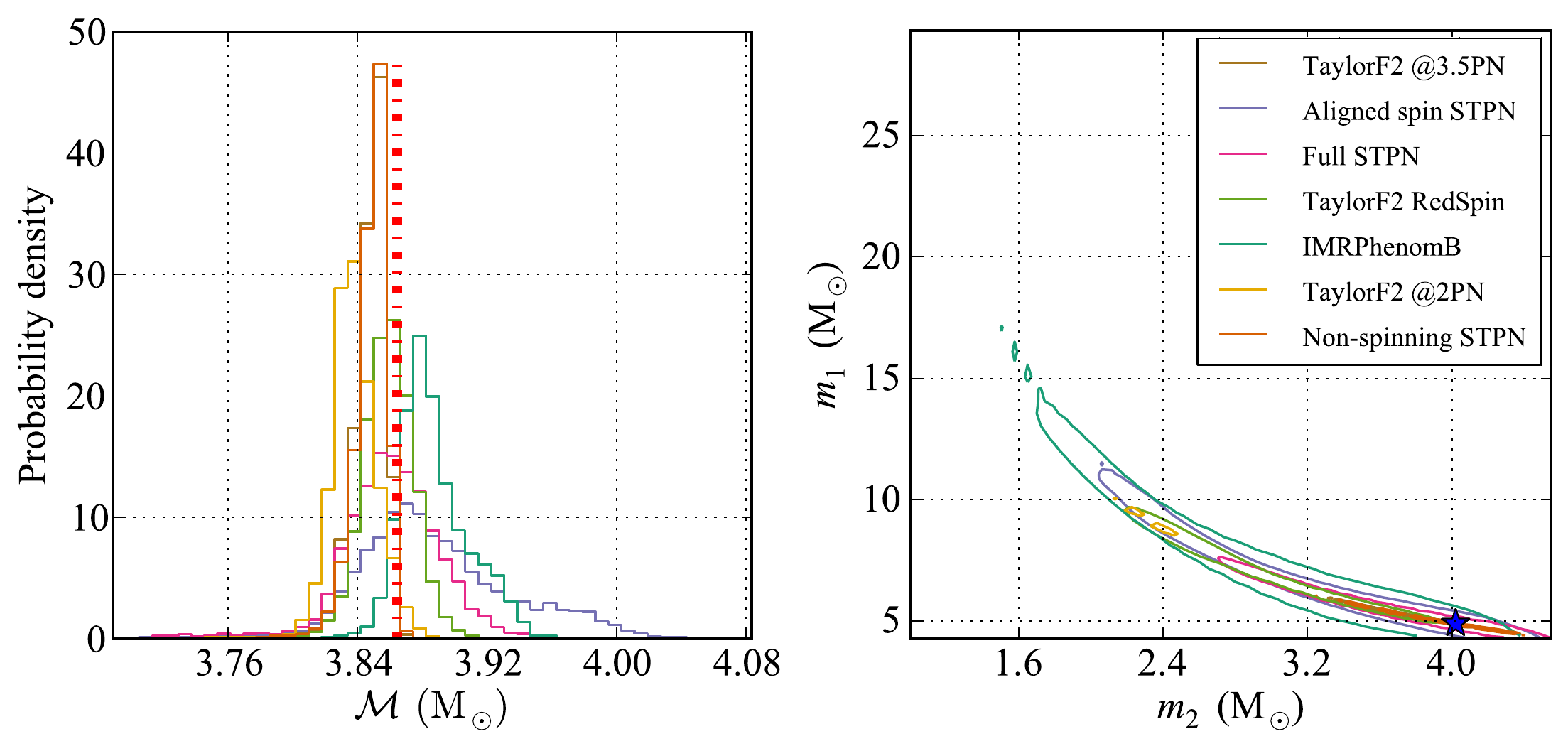}

\caption{\label{fig:bbh_mass}(left) Posterior probability distributions for the chirp mass $\Mc$ of the non-spinning \ac{BBH} hardware injection (section \ref{hwinj:bbh}) for the seven signal models considered. The injected value is marked with a vertical red line. (right) Overlay of 90\% probability regions for the joint posterior distribution on the component masses $m_1$, $m_2$ of the binary. The true value is marked by the blue star. Models which allow for non-zero spins find wider \ac{PDF}s for the coupled mass parameters. 
}
\end{figure*}

%\begin{figure*}
%\vspace{-10pt}
%\begin{minipage}[b]{0.48\textwidth}
%\centering
%\includegraphics[width=\textwidth]{Images/pdf/EOBNR_dec_ra.pdf}
%\end{minipage}
%\begin{minipage}[b]{0.48\textwidth}
%\centering
%\includegraphics[width=\textwidth]{Images/pdf/EOBNR_dist_iota.pdf}
%\end{minipage}
%\vspace{-10pt}
%\caption{\label{fig:bbh_location}Joint posterior probability regions for the location and inclination angle of the non-spinning \ac{BBH} hardware injection (section \ref{hwinj:bbh}) for the seven signal models considered. (left) The binary is constrained to two neighboring regions of the sky. (right) The distance and inclination, like the sky location, are estimated with a similar accuracy in models that include or exclude spins.
%}
%\end{figure*}
%
%\begin{figure*}
%\vspace{-10pt}
%\begin{minipage}[b]{0.48\textwidth}
%\centering
%\includegraphics[width=\textwidth]{Images/pdf/EOBNR_a1.pdf}
%\end{minipage}
%\begin{minipage}[b]{0.48\textwidth}
%\centering
%\includegraphics[width=\textwidth]{Images/pdf/EOBNR_a2.pdf}
%\end{minipage}
%\vspace{-10pt}
%\caption{\label{fig:bbh_spin}Posterior probability distributions for the dimensionless spin magnitude of the heavier (left) and lighter (right) components of the binary from the non-spinning \ac{BBH} hardware injection (section \ref{hwinj:bbh}), as inferred in the model ST (table~\ref{tab:waveforms}), full-spin STPN. The injection was made with $a_1=a_2=0$.
%}
%\end{figure*}

\begin{figure*}
\includegraphics[width=\textwidth]{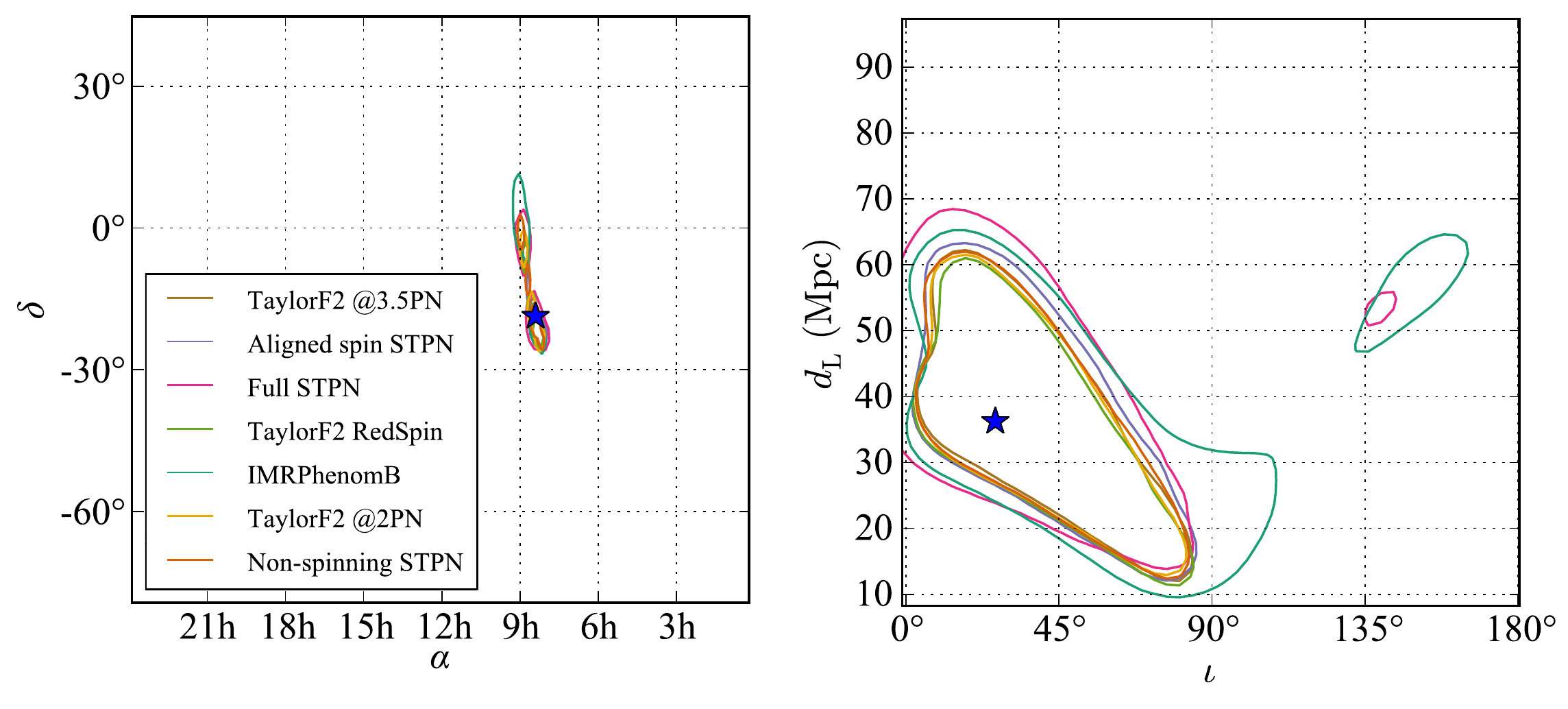}
\caption{\label{fig:bbh_location}Joint posterior probability regions for the location and inclination angle of the non-spinning \ac{BBH} hardware injection (section \ref{hwinj:bbh}) for the seven signal models considered. (left) The binary is constrained to two neighboring regions of the sky. (right) The distance and inclination, like the sky location, are estimated with a similar accuracy in models that include or exclude spins.
}
\end{figure*}

\begin{figure*}
\includegraphics[width=\textwidth]{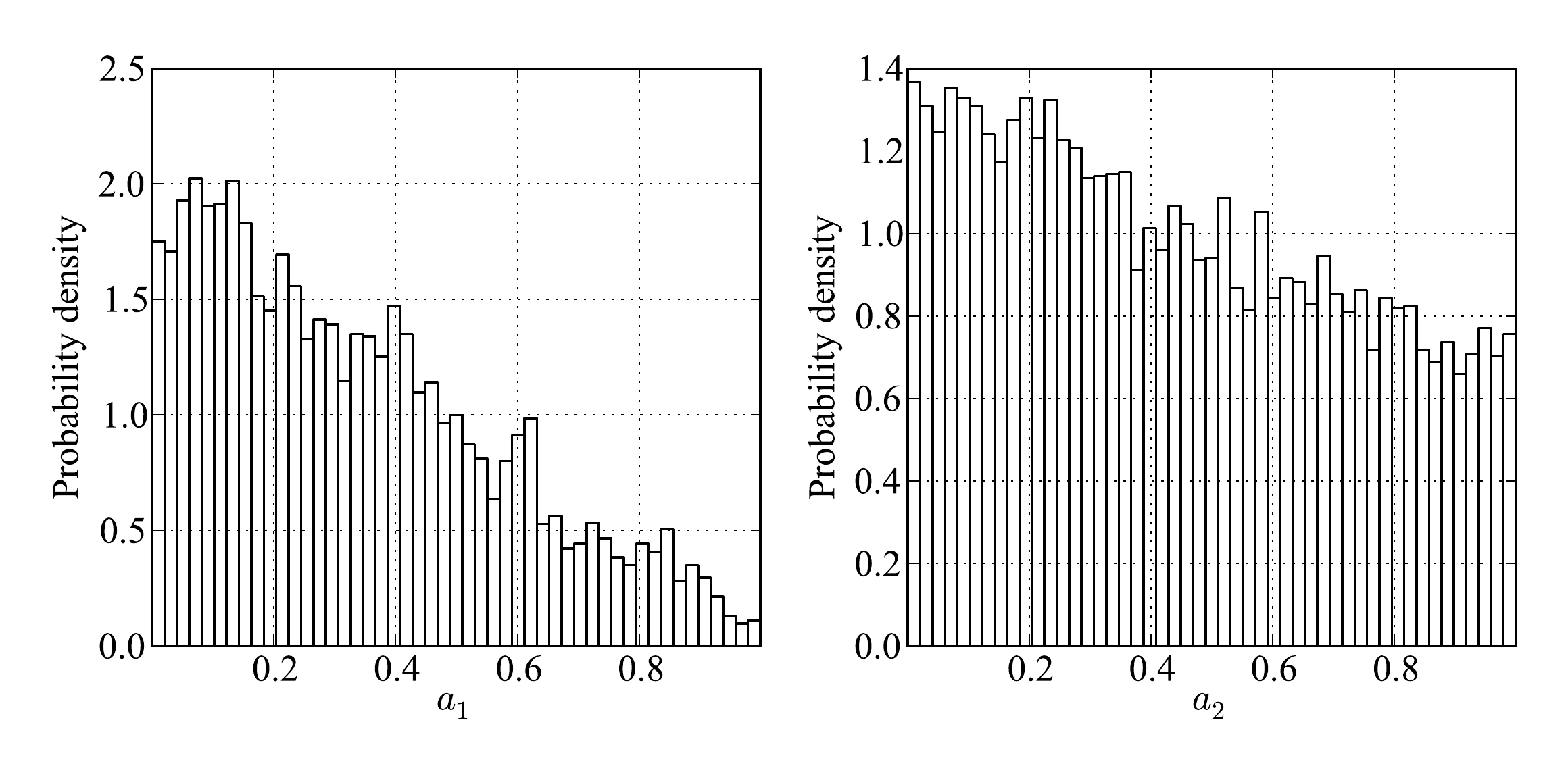}
\caption{\label{fig:bbh_spin}Posterior probability distributions for the dimensionless spin magnitude of the heavier (left) and lighter (right) components of the binary from the non-spinning \ac{BBH} hardware injection (section \ref{hwinj:bbh}), as inferred in the model ST (table~\ref{tab:waveforms}), full-spin STPN. The injection was made with $a_1=a_2=0$.
}
\end{figure*}

We first describe the analysis of a hardware injection corresponding to a binary black hole with non-spinning $4.91\,\Msun$ and $4.02\,\Msun$ components, injected at a network \ac{SNR} of 13 (\ac{SNR}s of $8.7$ in the Hanford and Livingston detectors and $4.4$ in the Virgo detector).  This injection was made with effective-one-body -- numerical-relativity (EOBNR) waveforms, using the \emph{EOBNRv1} version as described in~\cite{Buonanno:2007pf}.

Figure \ref{fig:bbh_mass} shows a comparison of the posterior \ac{PDF} of the mass parameters for different models. The chirp mass has a low statistical uncertainty of $\sim 1\,\%$, with the greatest statistical uncertainty claimed by models that allow for non-zero spin magnitudes due to inter-parameter degeneracies.  The $90\%$ credible intervals on the chirp mass obtained with TaylorF2 templates just exclude the true value, an indication of the systematic bias due to the waveform differences between the injected EOBNR waveform and these templates.  These differences are also responsible for the particularly strong bias in the mass ratio for 2.0pN TaylorF2 templates, exceeding the statistical measurement uncertainty in this parameter.  In contrast, we find that by using the 3.5pN TaylorF2 templates the systematic bias is reduced to less than the statistical uncertainty at the 90\,\% credible interval.  The uncertainty in the mass ratio is typically larger than that in the chirp mass, leading to the characteristic thin, 
correlated joint distribution for $m_1$ and $m_2$ evident in figure~\ref{fig:bbh_mass} (right).

Figure \ref{fig:bbh_location} shows a selection of \ac{PDF}s for the extrinsic parameters of the \ac{BBH} source: the recovered sky position, distance and inclination angle.  
The binary can be constrained to two neighboring regions of the sky representing the reflection of the source location through the plane of the detectors, which cuts this contiguous region in two.
The correlations between intrinsic parameters (masses and spins, see~\cite{Cutler:1994,Poisson:1995ef,2013PhRvD..87b4035B}) and extrinsic parameters are generally weaker than correlations between different parameters within the same group.  Thus both models that include and models that exclude non-zero spins have similar statistical measurement uncertainties for extrinsic parameters.

Finally, figure \ref{fig:bbh_spin} shows individual posteriors for the two dimensionless spin magnitudes, obtained using the model ST (table~\ref{tab:waveforms}), full-spin STPN waveforms. Although spin is not very strongly constrained, particularly for the lower-mass secondary, both spin measurements are consistent with the true value of $0$ spin. 
The absence of strong constraint is due to both the small difference in masses and nearly face-on inclination.
This inability to constrain the spin is reflected in the evidence for each template family, shown in table \ref{tab:evidences}, where all models have the same evidence within the error bars.

\subsubsection{Binary neutron star}\label{hwinj:bns}

\begin{figure*}
\includegraphics[width=\textwidth]{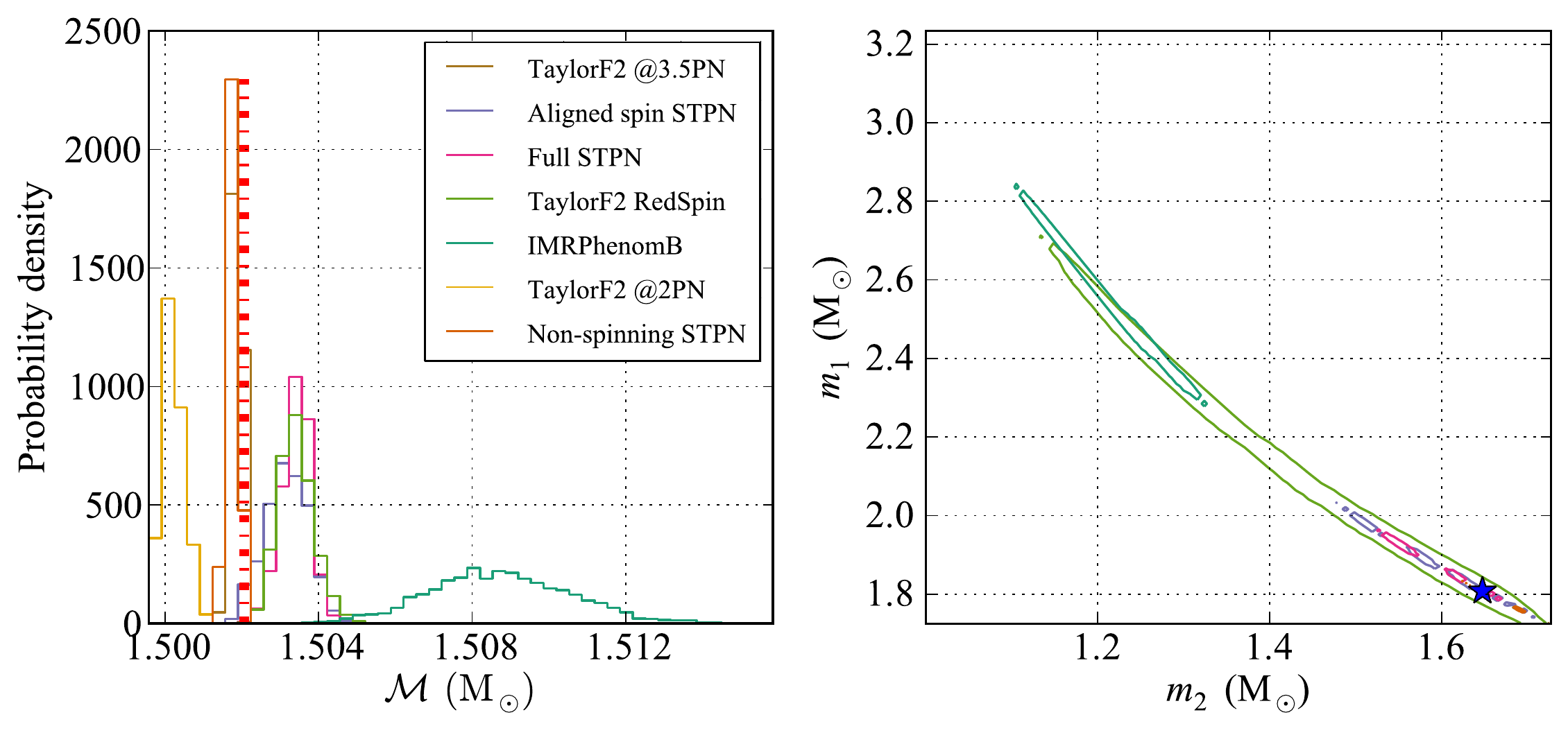}
\caption{\label{fig:bns_mass}(left) Posterior probability distributions for the chirp mass $\Mc$ of the non-spinning \ac{BNS} hardware injection (section \ref{hwinj:bns}) for the seven signal models considered. The injected value is marked with a vertical red line. (right) Overlay of 90\% probability regions for the joint posterior distribution on the component masses $m_1$, $m_2$ of the binary. The true value is marked by the blue star. 
}
\end{figure*}

\begin{figure*}
\includegraphics[width=\textwidth]{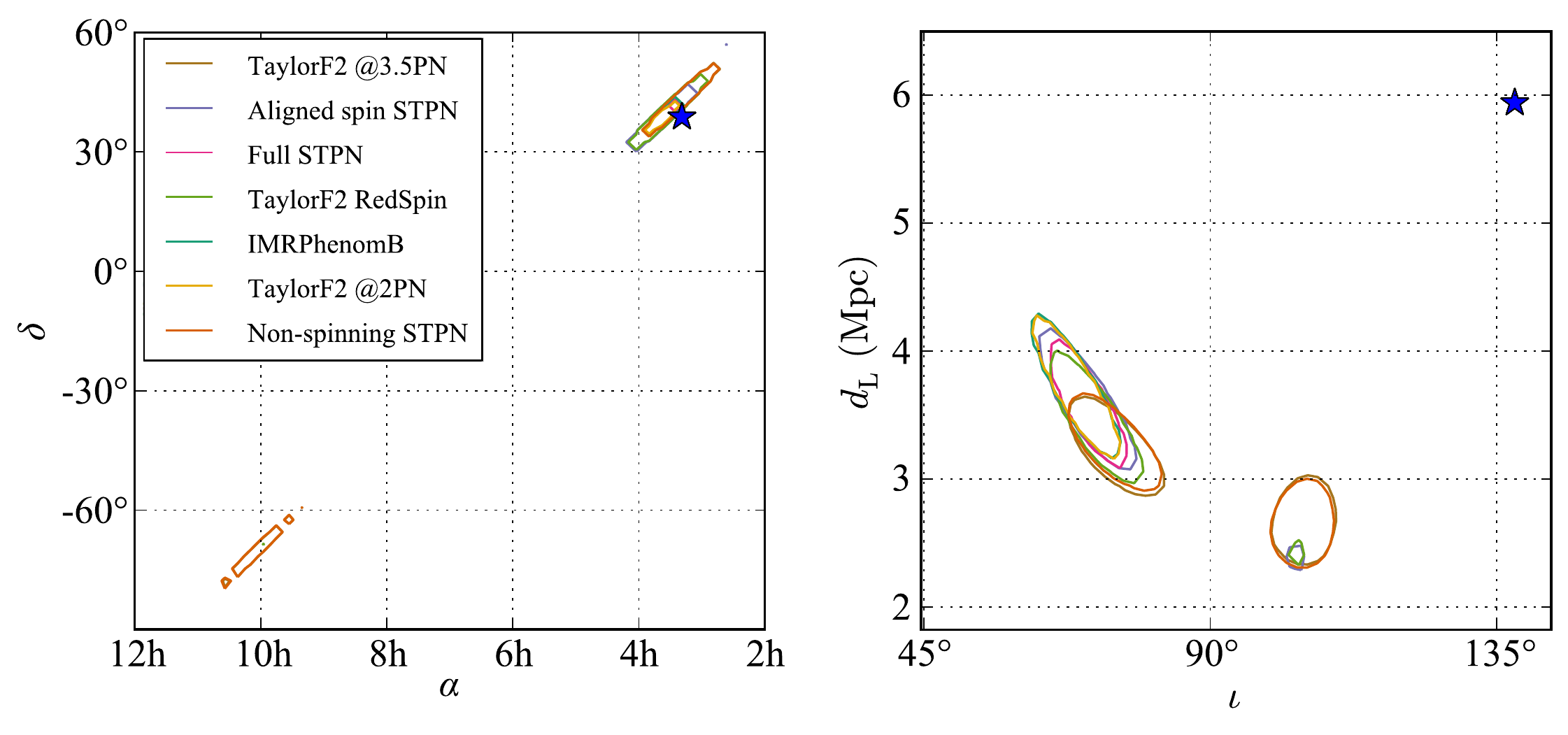}
\caption{\label{fig:bns_location}Joint posterior probability regions for the location and inclination angle of the non-spinning \ac{BNS} hardware injection (section \ref{hwinj:bns}) for the seven signal models considered. (left) The binary can be constrained to two regions of the sky representing the reflection of the source location through the plane of the detectors. (right) The distance and inclination, like the sky location, are estimated with a similar accuracy in models that include or exclude spins. The data were such that the injected distance and inclination angle are far outside the 90\,\% credible intervals.
}
\end{figure*}

We also analyzed a hardware injection simulating a non-spinning neutron star binary system with $1.81\,\Msun$ and $1.65\,\Msun$ components, injected at a network \ac{SNR} of 36 (\ac{SNR} of 26 in the Hanford detector, 17 in the Livingston detector and 20 in the Virgo detector).  This injection was made with the same EOBNR waveform family as in section~\ref{hwinj:bbh} above.

Figure \ref{fig:bns_mass} shows a comparison of the posterior \ac{PDF} of the mass parameters for the different models. The chirp mass has a low statistical uncertainty of $\lesssim 0.1\%$ thanks to the high network \ac{SNR}, with the greatest statistical uncertainty claimed by models that allow for non-zero spin magnitudes due to inter-parameter degeneracies, similar to section~\ref{hwinj:bbh}. The greater statistical uncertainty of the IMRPhenomB waveform model stems from this model not describing such a low-mass system ($m_1 + m_2 < 10\,\Msun$) very well \cite{Ajith:2009bn}. Correspondingly it has the lowest Bayes factor in table~\ref{tab:evidences}.  The $90\%$ credible intervals on the chirp mass obtained with TaylorF2 templates include the true value. Unlike in section~\ref{hwinj:bbh}, the merger present in the injected EOBNR waveform happens at a frequency (~1.2kHz) where the detectors are not very sensitive.
Figure \ref{fig:bns_location} shows the extrinsic parameters of the \ac{BNS} source: sky position, distance and inclination angle.
In this case, the data were such that the injected distance and inclination angle lay far outside the 90\,\% credible intervals.  This could be due to a particularly unlikely noise realization, some non-stationarity or non-Gaussianity in the noise (i.e., a "glitch"), or some issue with the way the hardware injection was carried out.

\subsection{Spinning software injections}

\subsubsection{Binary black hole}\label{swinj:bbh}

\begin{figure*}
\includegraphics[width=\textwidth]{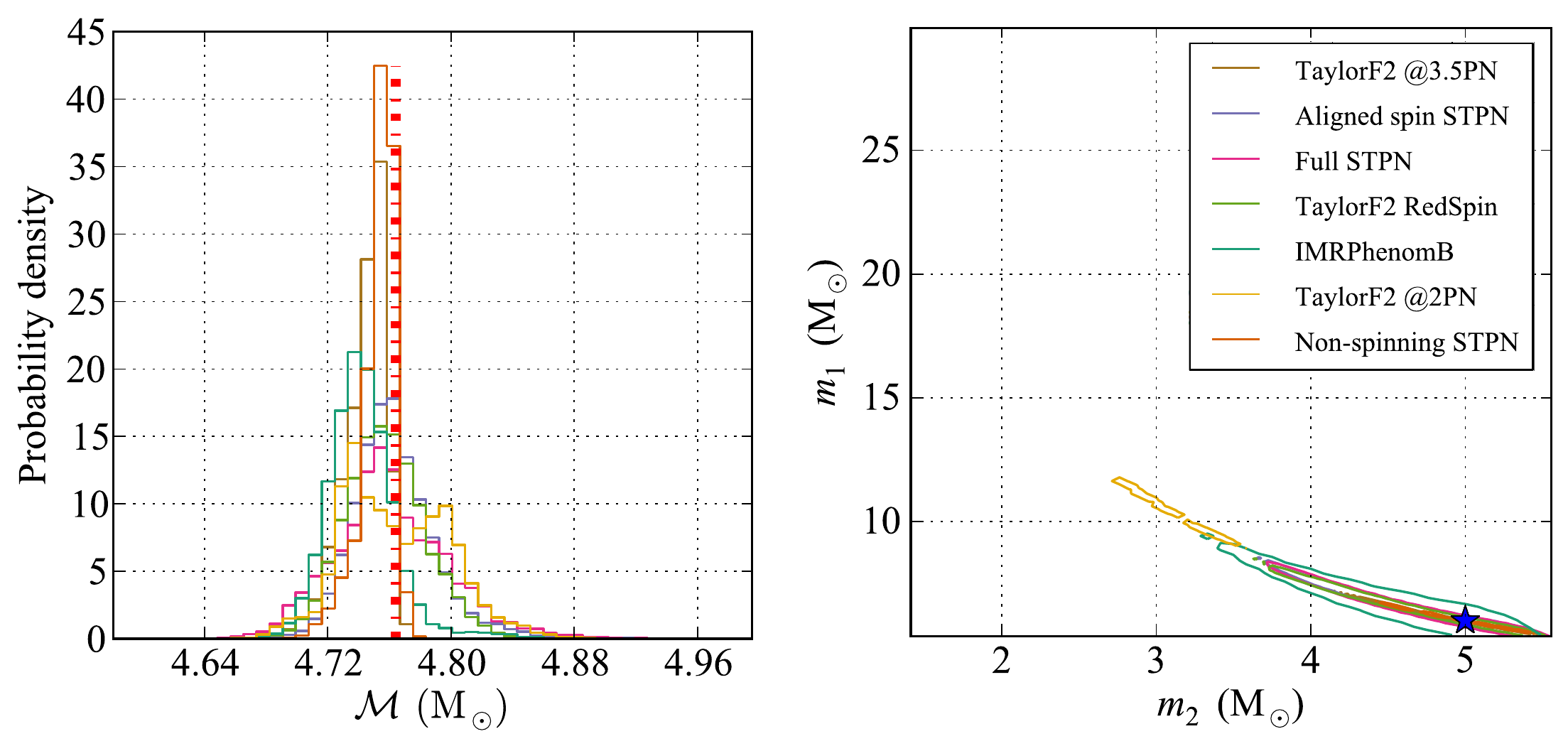}
\caption{\label{fig:swbbh_mass}(left) Posterior probability distributions for the chirp mass $\Mc$ of the spinning \ac{BBH} software injection (section \ref{swinj:bbh}) for the seven signal models considered. The injected value is marked with a vertical red line. (right) Overlay of 90\% probability regions for the joint posterior distribution on the component masses $m_1$, $m_2$ of the binary.
}
\end{figure*}

\begin{figure*}
\includegraphics[width=\textwidth]{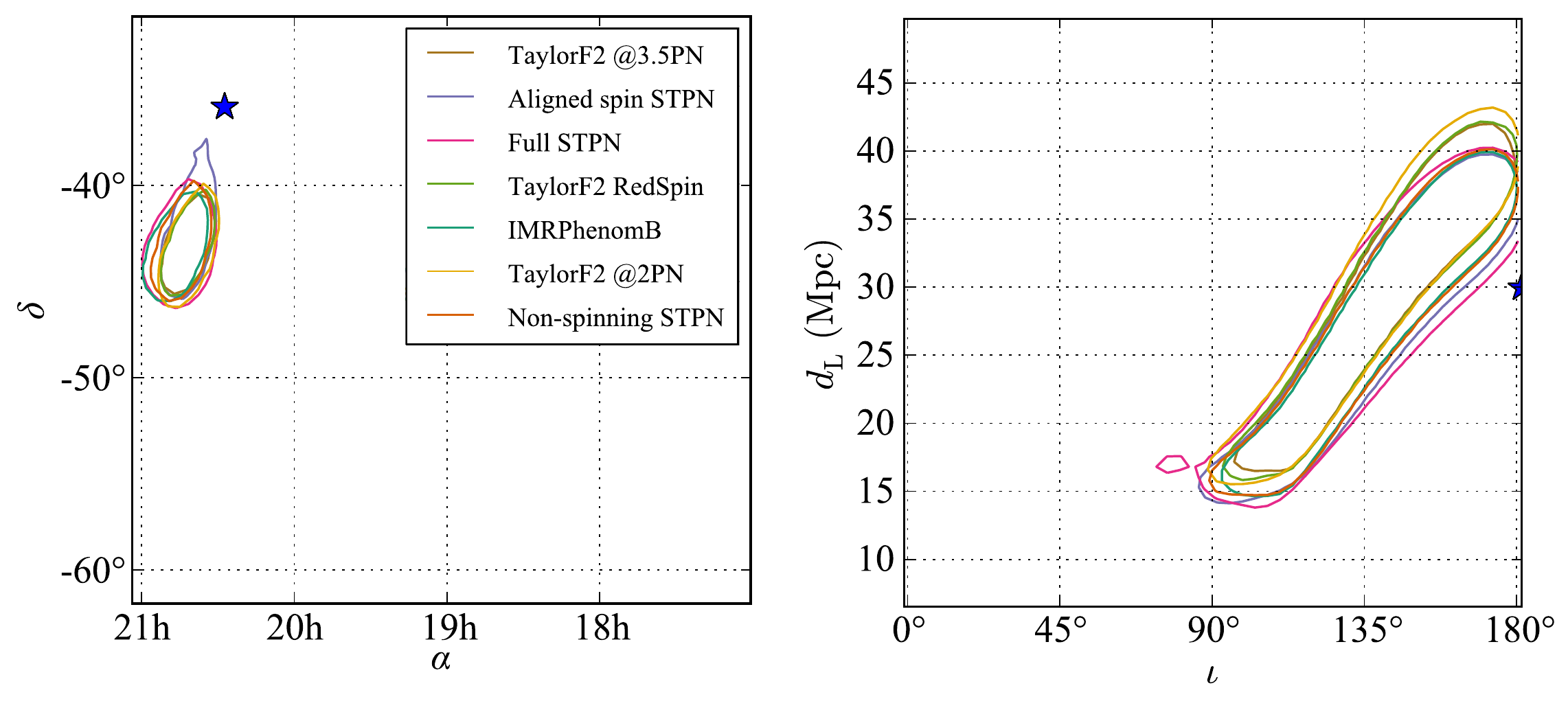}
\caption{\label{fig:swbbh_location}Joint posterior probability regions for the location and inclination angle of the spinning \ac{BBH} software injection (section \ref{swinj:bbh}). (left) The binary's true location lies just outside of the 90\% credible interval.  (right) The degeneracy in distance and inclination prevents either parameter from being accurately constrained individually.}
\end{figure*}

\begin{figure*}
\includegraphics[width=\textwidth]{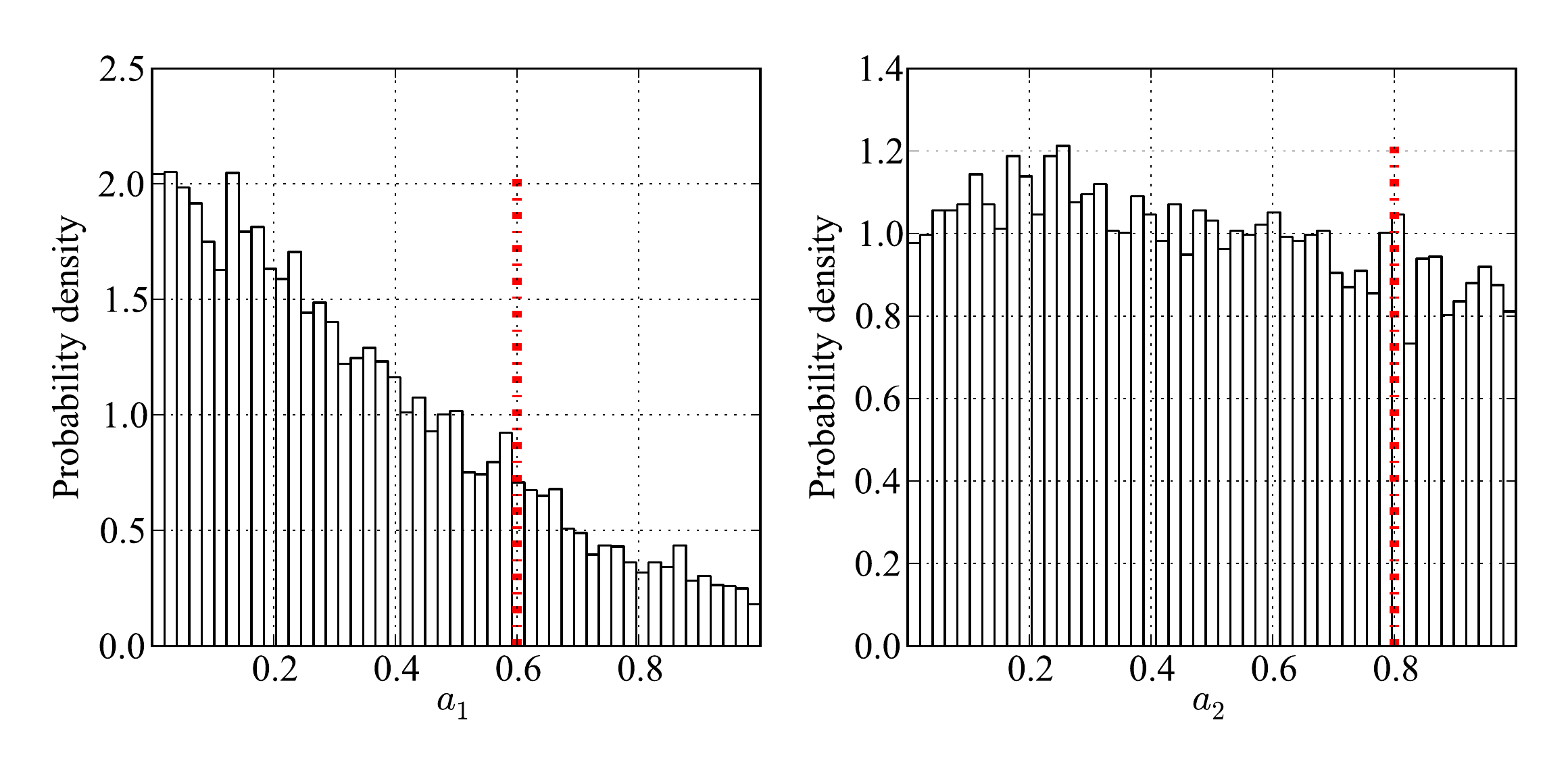}
\caption{\label{fig:swbbh_spin}Posterior probability distributions for the dimensionless spin magnitude of the heavier (left) and lighter (right) components of the binary from the spinning \ac{BBH} software injection (section \ref{swinj:bbh}), as inferred in the model ST (table~\ref{tab:waveforms}), full-spin STPN; the true values are shown with vertical red lines.
}
\end{figure*}

We simulated the signal from a binary black hole with misaligned spinning components with a SpinTaylor software injection.  This binary consists of $6\,\Msun$ and $5\,\Msun$ black holes, with dimensionless spin magnitudes of $0.6$ and $0.8$, misaligned with the orbital angular momentum by angles of $40^\circ$ and $150^\circ$, respectively, and an optimal network \ac{SNR} of 19 (\ac{SNR} of 17 in the Hanford detector, 6.5 in the Livingston detector and 4.9 in the Virgo detector). The misalignment between the spins and the orbital angular momentum causes the plane of the binary to precess, producing both amplitude and phase modulations in the received \ac{GW} signal (see section~\ref{sec:waveform}).

Figure \ref{fig:swbbh_mass} shows a comparison of the posterior \ac{PDF}s of the mass parameters inferred with different template models.  The mass ratio is again severely biased for the 2pN TaylorF2 model, which is not surprising for a 3.5 pN injection with spinning components as both the pN order and the spin alter the phase evolution of the signal.

Figure \ref{fig:swbbh_location} shows the recovered sky location, distance and inclination angle of the source. For all those parameters the injection value lies just outside of the 90\,\% credible regions.

The spin magnitudes are again poorly constrained, with non-zero support across the entire allowed range, as shown in figure \ref{fig:swbbh_spin}. Even though the injected signal was simulated from a system with high spin magnitudes (0.6 and 0.8), the low difference in masses, the near anti-alignment of the spins with the orbital angular momentum and especially the face-on inclination conspire to give the poor spin estimates. 
In addition, with this weak precession effect, the spin tilts (angles between the spin vector and the orbital angular momentum) are also poorly constrained. A comparison of the evidences in table \ref{tab:evidences} indicates that all models have the same evidence within the error bars. 

\subsubsection{Neutron star - black hole}\label{swinj:nsbh}

\begin{figure*}
\includegraphics[width=\textwidth]{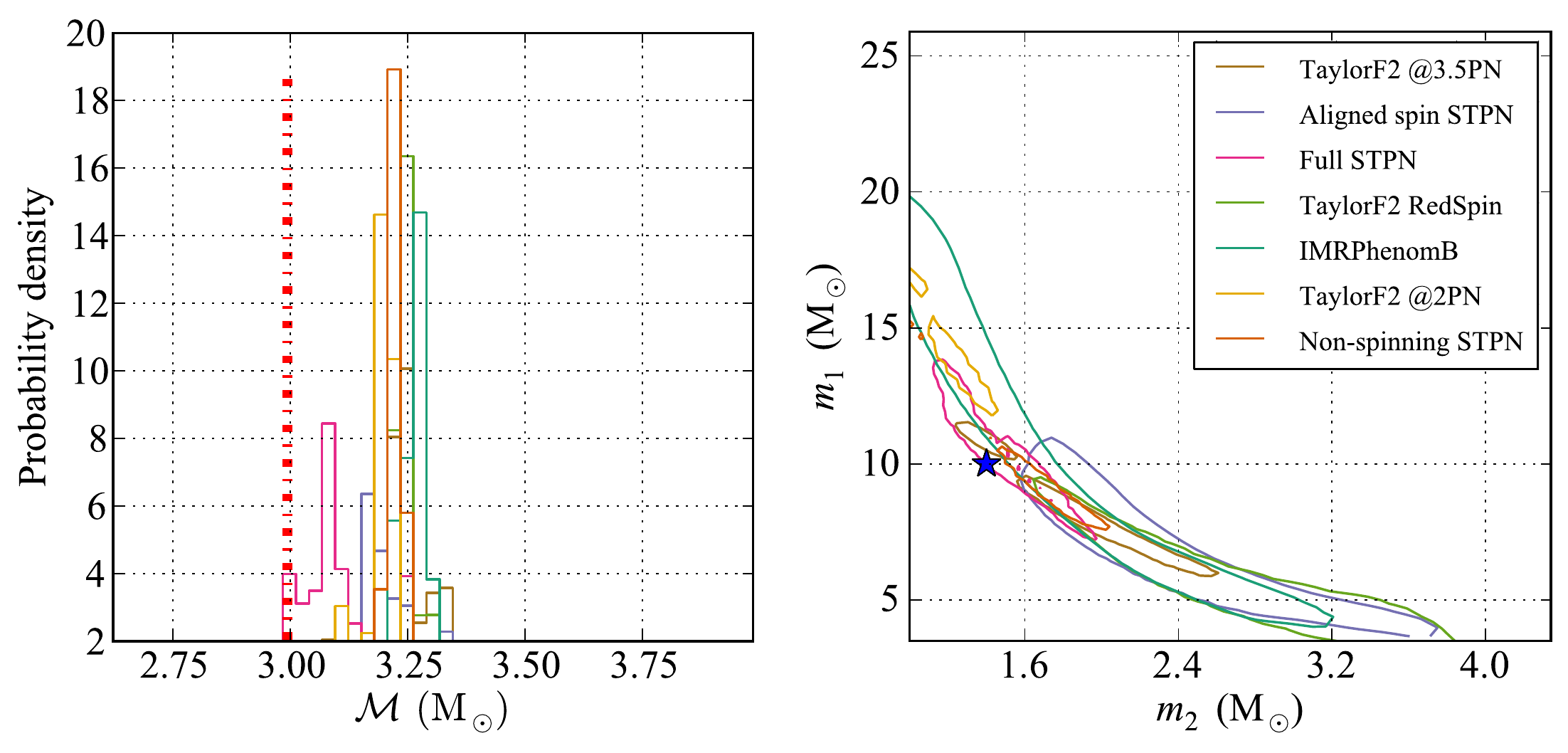}
\caption{\label{fig:swnsbh_mass}(left) Posterior probability distributions for the chirp mass $\Mc$ of the spinning \ac{NSBH} software injection (section \ref{swinj:nsbh}) for the seven signal models considered. The injected value is marked with a vertical red line. (right) Overlay of 90\% probability regions for the joint posterior distribution on the component masses $m_1$, $m_2$ of the binary.
}
\end{figure*}

\begin{figure*}
\includegraphics[width=\textwidth]{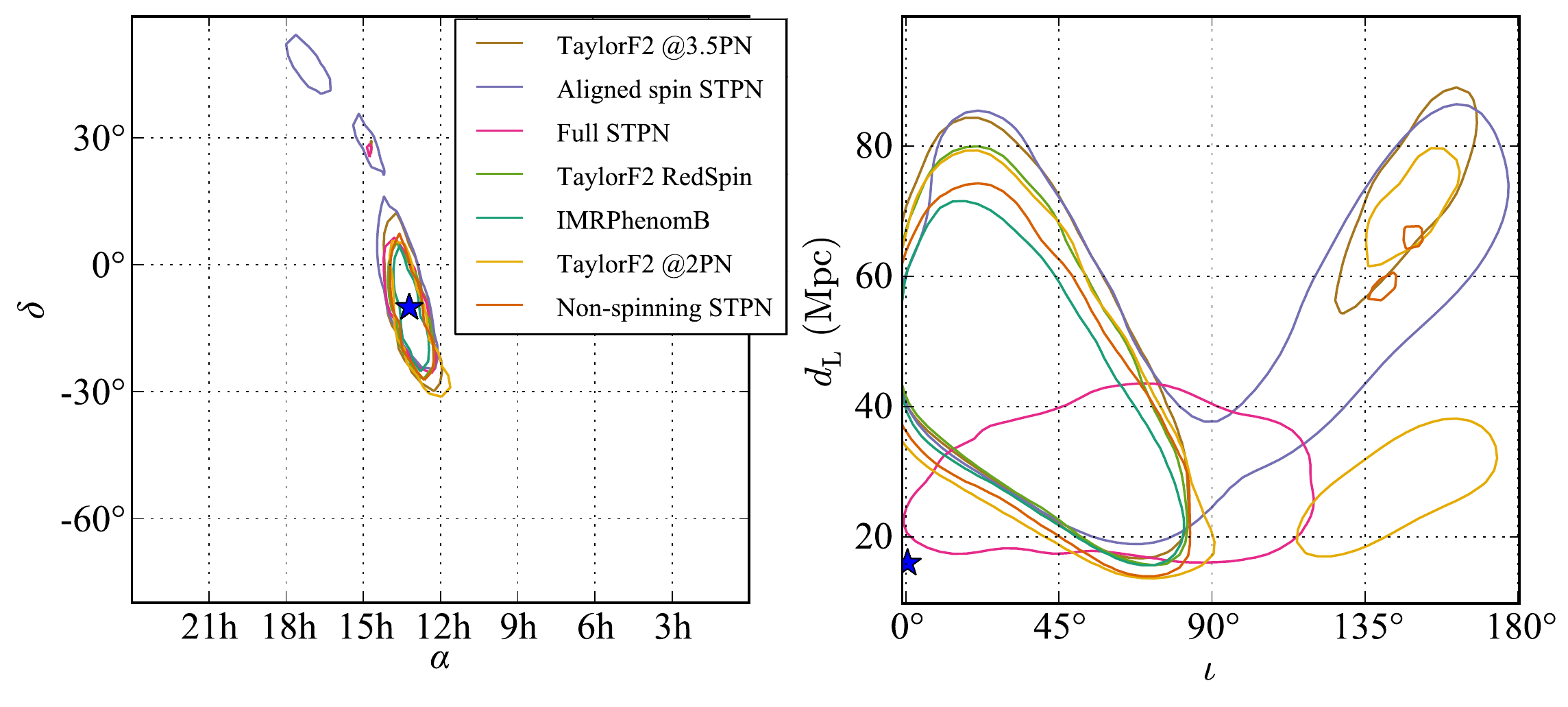}
\caption{\label{fig:swnsbh_location}Joint posterior probability regions for the location and inclination angle of the spinning \ac{NSBH} software injection (section \ref{swinj:nsbh}). (left) The binary is localized well on the sky.  (right) In this case, the true value lies outside of the 90\% credible interval of the joint distance-inclination marginalized probability density function.}
\end{figure*}

\begin{figure*}
\includegraphics[width=\textwidth]{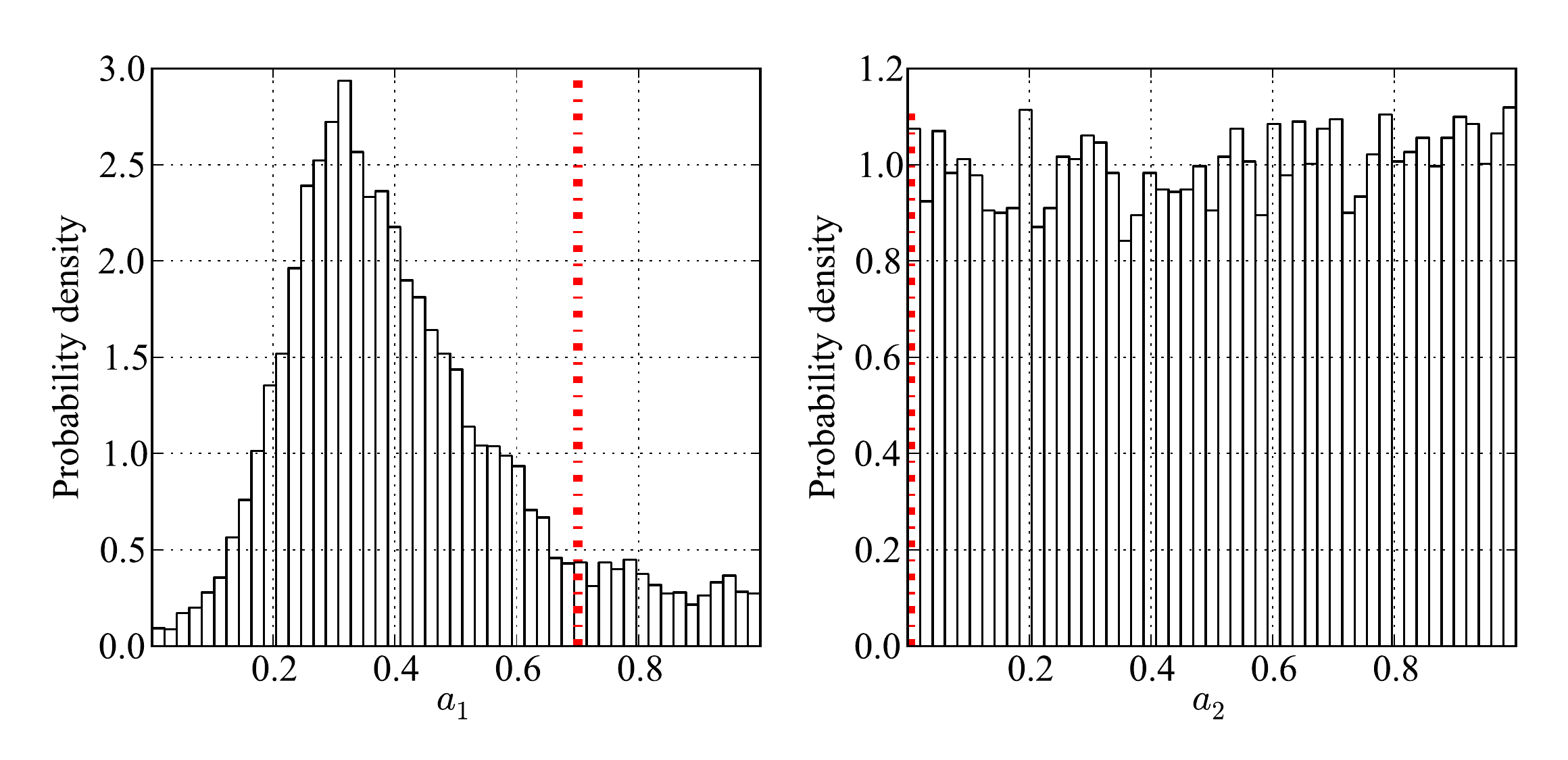}
\caption{\label{fig:swnsbh_spin}Posterior probability distributions for the dimensionless spin magnitude of the heavier (left) and lighter (right) components of the binary from the spinning \ac{NSBH} software injection (section \ref{swinj:nsbh}), as inferred in the model ST (table~\ref{tab:waveforms}), full-spin STPN; the true values are shown with vertical red lines.
}
\end{figure*}

We simulated the signal from a neutron star - black hole system with misaligned black hole spin components with a SpinTaylor software injection.  This binary consists of a $10\,\Msun$ black hole with dimensionless spin magnitude of $0.7$, misaligned with the orbital angular momentum by an angle of $130^\circ$, and a $1.4\,\Msun$ non-spinning neutron star, at an optimal network \ac{SNR} of 13 (\ac{SNR} of 7.9 in the Hanford detector, 9.2 in the Livingston detector and 3.6 in the Virgo detector).

Figure \ref{fig:swnsbh_mass} shows a comparison of the posterior \ac{PDF}s of the mass parameters inferred with different template models.  
Figure \ref{fig:swnsbh_location} shows the recovered sky location, distance and inclination angle of the source. 
The joint distance-inclination posterior degeneracy prevents either parameter from being measured precisely. The model describing most accurately the data, with the highest Bayes factor in table~\ref{tab:evidences}, Full STPN, delivers the most accurate parameter estimates, as expected since this model was used for the injection.

Figure \ref{fig:swnsbh_spin} shows the posterior \ac{PDF}s for the spin parameters. The spin magnitude of the neutron star is unconstrained due to the large difference in masses. The spin of the more massive black hole encodes more information in the waveform, but is also poorly constrained due to the almost face-on inclination.

\subsection{Blind hardware injection}\label{bigdog}

\begin{figure*}
\includegraphics[width=\textwidth]{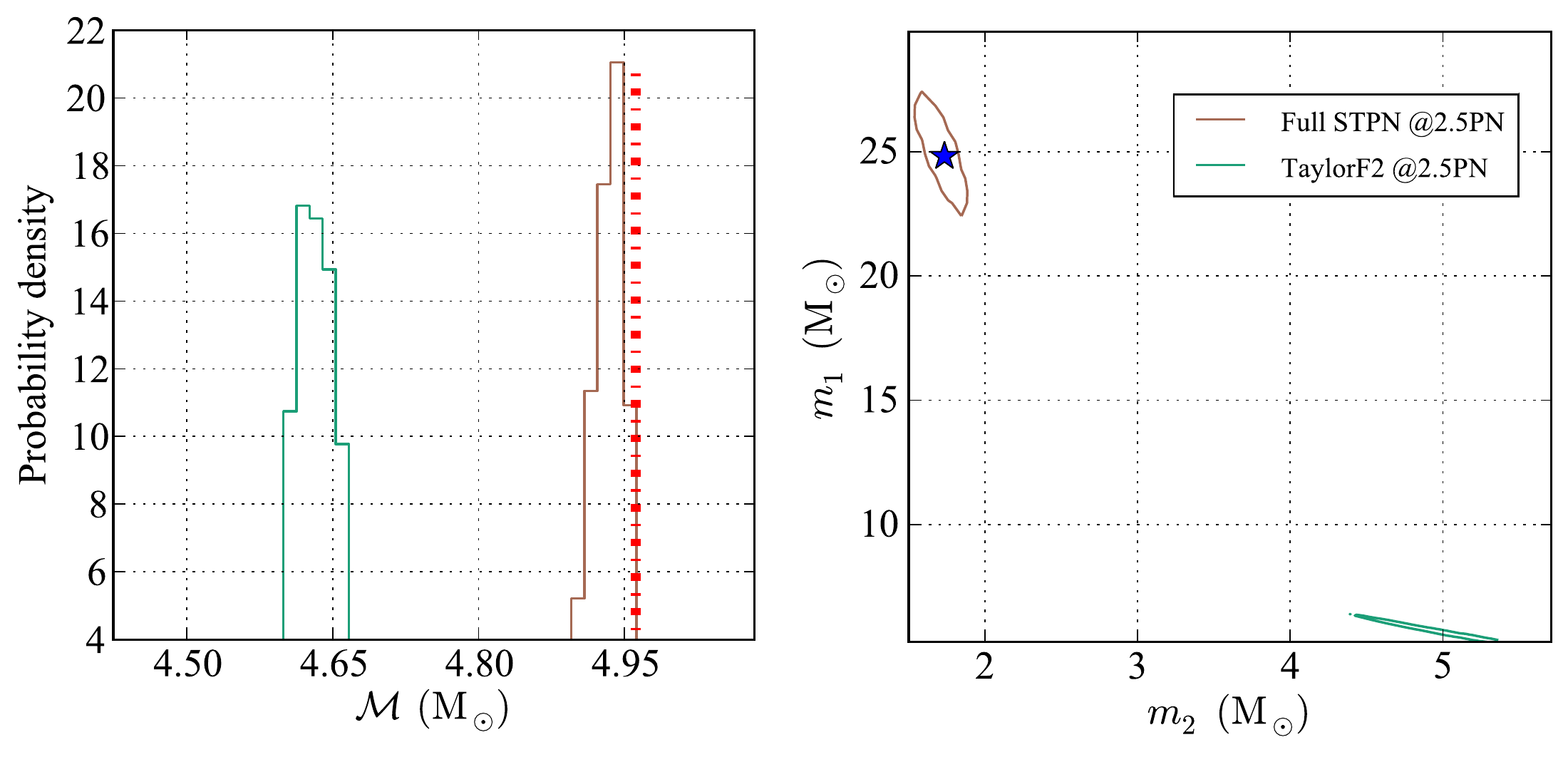}
\caption{\label{fig:bd_mass}(left) Posterior probability distributions for the chirp mass $\Mc$ of the blind injection (section \ref{bigdog}) for signal models at 2.5pN. The injected value is marked with a vertical line. (right) Overlay of 90\% probability regions for the joint posterior distribution on the component masses $m_1$, $m_2$ of the binary. The bias introduced by an analysis with a model which disallows spin is clear.
}
\end{figure*}

\begin{figure*}
\includegraphics[width=\textwidth]{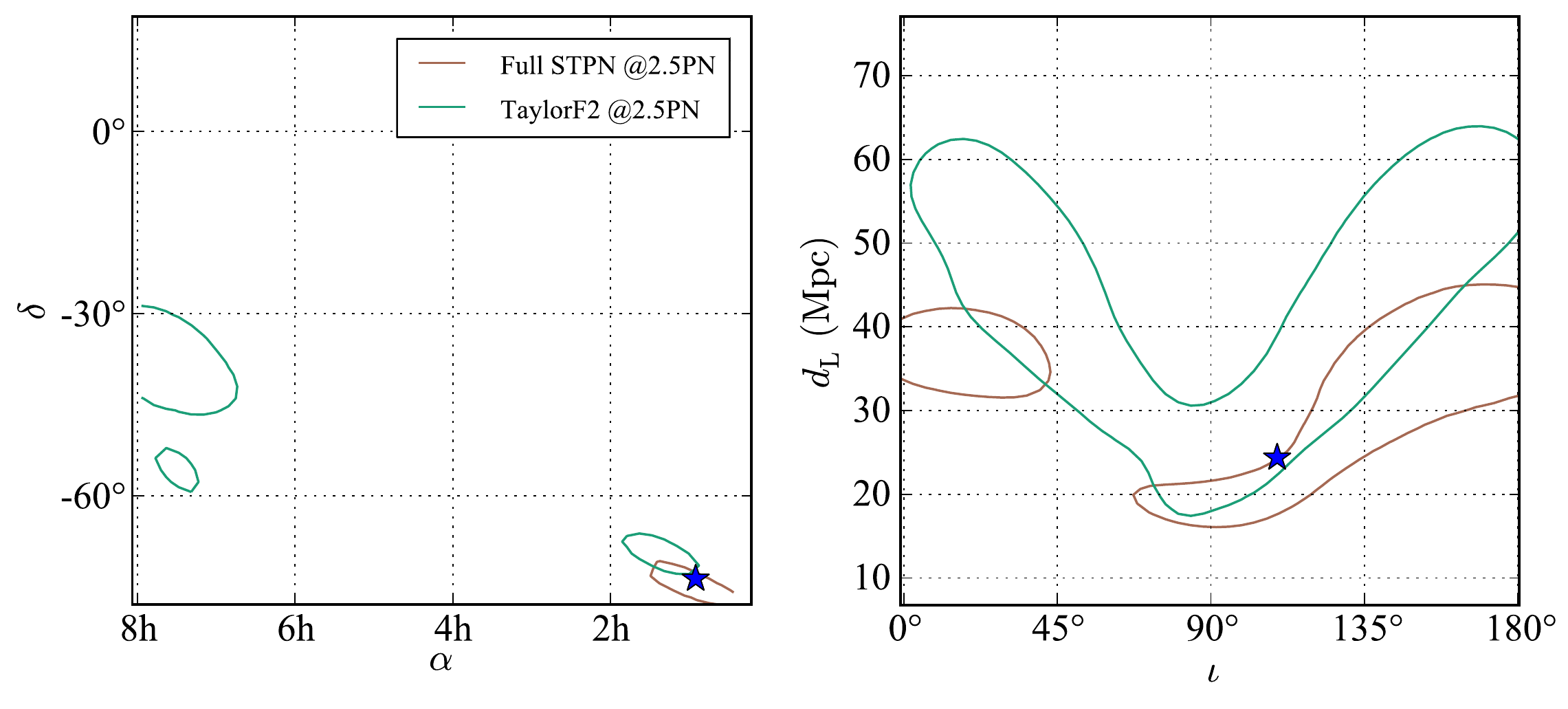}
\caption{\label{fig:bd_location}Joint posterior probability regions for the location and inclination angle of the blind injection (section \ref{bigdog}). (left) The sky location is constrained to several distinct regions lying along a half circle on the sky. 
(right) Again, the characteristic V-shape degeneracy in distance and inclination is evident.
}
\end{figure*}

\begin{figure*}
\includegraphics[width=\textwidth]{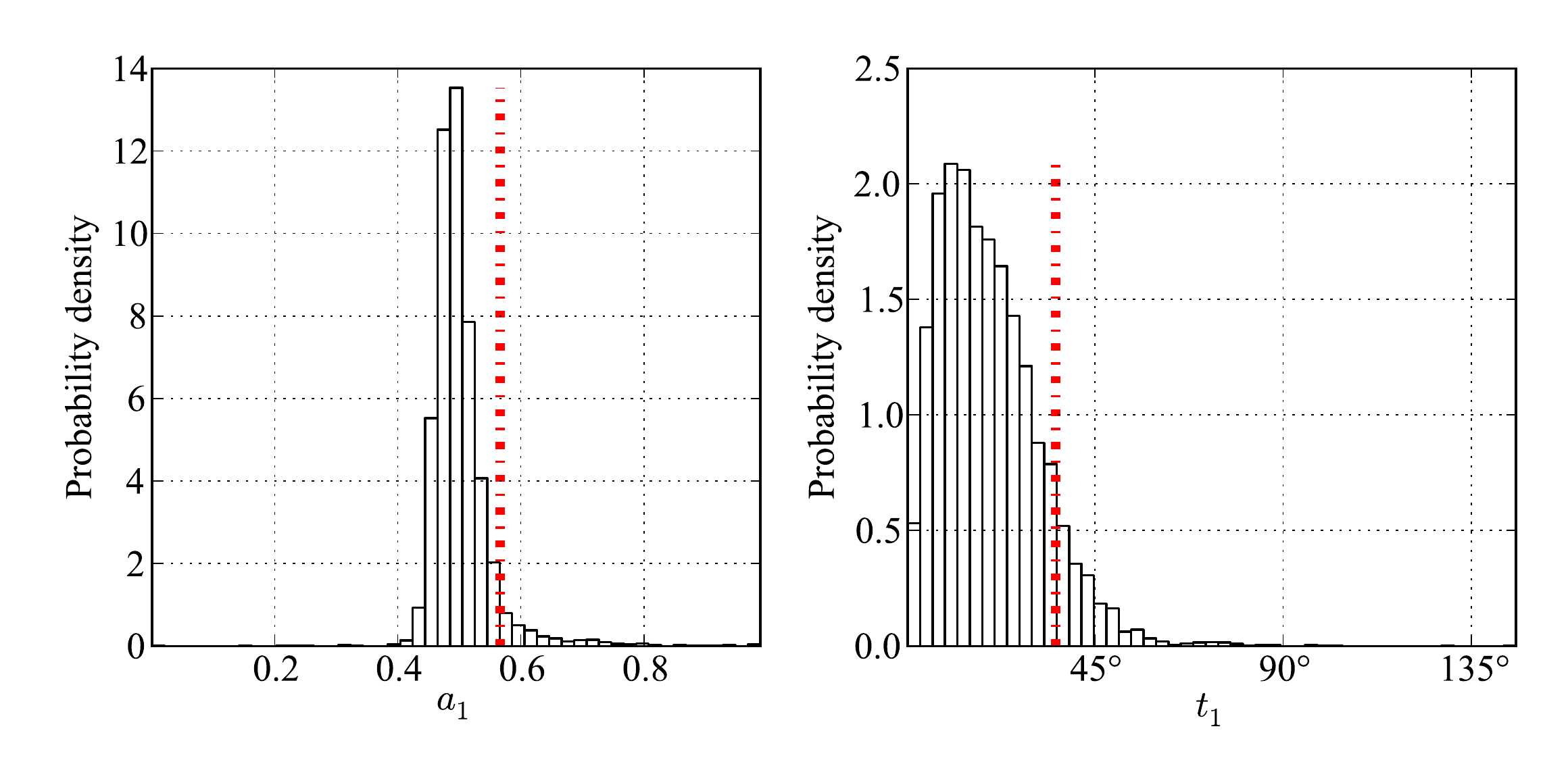}
\caption{\label{fig:bd_spin}Posterior probability distributions for the dimensionless spin magnitude (left) and tilt angle (angle between the spin vector and the orbital angular momentum, measured at 40Hz, right) of the heavier component of the binary from the blind injection (section \ref{bigdog}), as inferred with model full-spin STPN at 2.5 pN order in phase (table~\ref{tab:waveforms}); the true values are shown with vertical red lines.  The large difference in masses allows for the spin of the massive component to be measured. On the other hand, the spin of the light component is unconstrained.
}
\end{figure*}

The search pipeline described in \cite{Collaboration:S6CBClowmass} 
identified a \ac{GW} candidate occurring on 16 September
2010 at 06:42:23 UTC. 
A Bayesian analysis was performed using the algorithms and implementations 
described above, where parameter estimates varied significantly depending on 
the exact model used for the gravitational waveform. 

Following the completion of the analysis, the event was revealed to be a blind injection. 
Further investigation revealed several problems with the pre-un-blinding parameter estimation:
\begin{itemize}
\item{}The template signal included phase corrections only up to 2.5pN order, which is an outlier in the post-Newtonian expansion, see \cite{BuonannoChenVallisneri:2003a}.  At that time (before the blind injection was revealed as such) parameter estimation was not carried out with templates at this order, leading to a significant bias in the mass ratio, and hence the component masses.
\item{}The signal to be injected in the Hanford and Livingston sites had the wrong sign, making the signal incoherent between the LIGO and Virgo detectors. This caused strong biases in the estimated values of the extrinsic parameters.
\item{}Lastly, an injection software bug artificially set to zero one of the phase terms in the injected waveform.
\end{itemize}

We present here an analysis using templates at 2.5pN order in phase, performed after the un-blinding.  We artificially introduce a sign flip in the templates for Hanford and Livingston detectors in order to match the injected waveform. Figure \ref{fig:bd_mass} shows a comparison of the posterior \ac{PDF}s of the mass parameters for TaylorF2 and SpinTaylor waveform models.  The fully spinning, precessing analysis with SpinTaylor templates at 2.5 pN order in phase (i.e., the same template as used in the injection, up to the injection software bug) centers on the correct masses for this neutron star -- black hole system, $m_1=24.81\,\Msun$ and $m_2=1.74\,\Msun$.  However, the systematic bias due to fixing spin to zero in the analysis of this spinning injection is very significant, and leads to the wrong conclusion that a \ac{BBH} system is observed if the TaylorF2 model is used.  

Figure \ref{fig:bd_location} shows the recovered sky position, distance and inclination angle of the source. The sky location recovered is contained in several distinct regions, spread in an arc on the sky. This behavior is consistent with two detectors contributing the majority of the \ac{SNR} $15$ for this source (\ac{SNR} of 11 in the Hanford detector, 9.8 in the Livingston detector and 4.1 in the Virgo detector). On top of this first-order constraint, spin projection effects can break the degeneracy, as discussed in \cite{Raymond:2009}.

The large difference in masses, $m_1=24.81\,\Msun$, $m_2=1.74\,\Msun$, in combination with the high inclination allows for the magnitude of the spin of the massive component to be measured with an accuracy $\sim 10\%$, as shown in figure~\ref{fig:bd_spin}.  Meanwhile, the spin of the light component is unconstrained, and the posterior on its magnitude is consistent with the prior. Correspondingly, the evidences in table~\ref{tab:evidences}
favor the spinning model. Signals from face-on binaries are louder than signals from high-inclination sources, making a face-on binary more likely to be detected. As such the poor spin constraints in the previous sections are more representative of the expected observations. However, the study presented here does not include a statistically sufficient number of signals to cover the range of possible measurements. A wider study is ongoing to predict more accurately the constraints the LIGO/Virgo network will be able to put on spins.

\begin{figure*}
\includegraphics[width=0.48\textwidth]{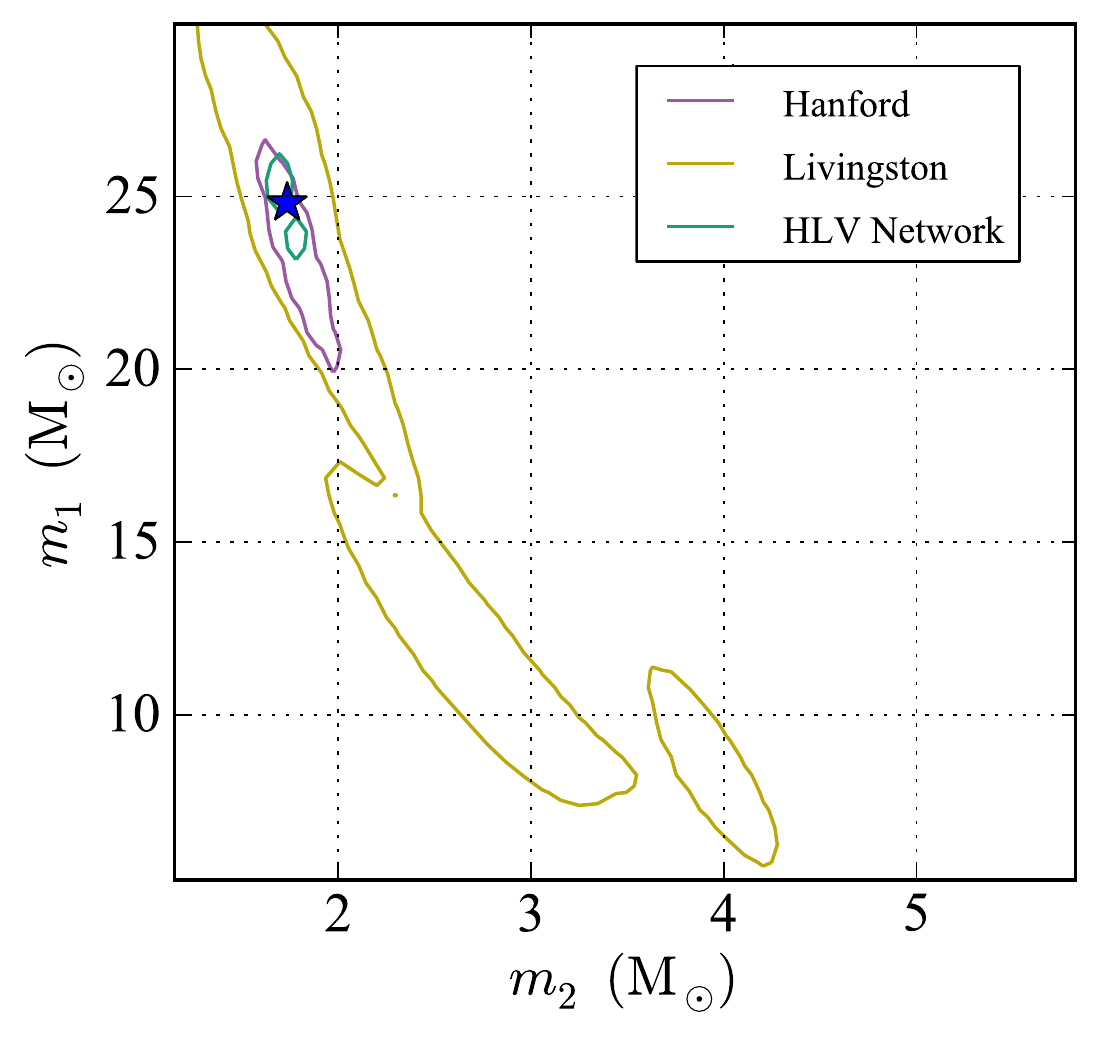}
\caption{\label{fig:bd_hlv}Overlay of 90\% probability regions for the joint posterior distribution on the component masses $m_1$, $m_2$ of the binary for the blind injection (section \ref{bigdog}), as inferred with model ST\_25 (table~\ref{tab:waveforms}), full-spin STPN (at 2.5 pN order in phase), using data from the Hanford detector only, the Livingston detector only or the whole Hanford-Livingston-Virgo network (HLV Network).
}
\end{figure*}

A deeper analysis of the data around the blind hardware injection revealed some non-Gaussian behavior of the Livingston detector before and during the injected signal. 
From figure \ref{fig:bd_hlv} it can be seen that the analysis of the Hanford detector data or Livingston detector data alone 
produces mass estimates consistent with the full network analysis using data 
from all three detectors, but the Livingston detector data cannot constrain the parameters as 
precisely as Hanford detector data, showing multiple modes. This leads us to conclude that 
detector glitches have the potential to reduce parameter estimation accuracy. 
However, the coherent analysis of the data from the multi-detector network 
increases robustness against the effect of a glitch in a single instrument by 
requiring that the recovered parameters are consistent with all datasets. 
This effect could be mitigated by including such glitches in the model of the data, as suggested in ~\cite{PhysRevD.82.103007}.

%\begin{figure}
%\vspace{-10pt}
%\centering
%\includegraphics[width=0.48\textwidth]{Images/pdf/BD_HLV.pdf}
%\vspace{-10pt}
%\caption{\label{fig:bd_hlv}Overlay of 90\% probability regions for the joint posterior distribution on the component masses $m_1$, $m_2$ of the binary for the blind injection (section \ref{bigdog}), as inferred with model ST\_25 (table~\ref{tab:waveforms}), full-spin STPN (at 2.5 pN order in phase), using data from the Hanford detector only, the Livingston detector only or the whole Hanford-Livingston-Virgo network (HLV Network).
%}
%\end{figure}

\subsection{The effects of different noise realizations}\label{hwsw}

%\begin{figure*}
%\vspace{-10pt}
%\begin{minipage}[b]{0.48\textwidth}
%\centering
%\includegraphics[width=\textwidth]{Images/pdf/compare_m2-m1.pdf}
%\end{minipage}
%\begin{minipage}[b]{0.48\textwidth}
%\centering
%\includegraphics[width=\textwidth]{Images/pdf/compare_iota-dist.pdf}
%\end{minipage}
%\vspace{-10pt}
%\caption{\label{fig:bbh_hwsw_psd}Overlay of 90\,\% probability regions for the joint posterior probability distribution of the component masses $m_1$, $m_2$ of the binary (left) and the distance and inclination (right) for a series of signal injection times. Both sets of distributions show a spread comparable to the spread observed between different waveform models, as shown in figures \ref{fig:bbh_mass} and \ref{fig:bbh_location}.
%}
%\end{figure*}

\begin{figure*}
\includegraphics[width=\textwidth]{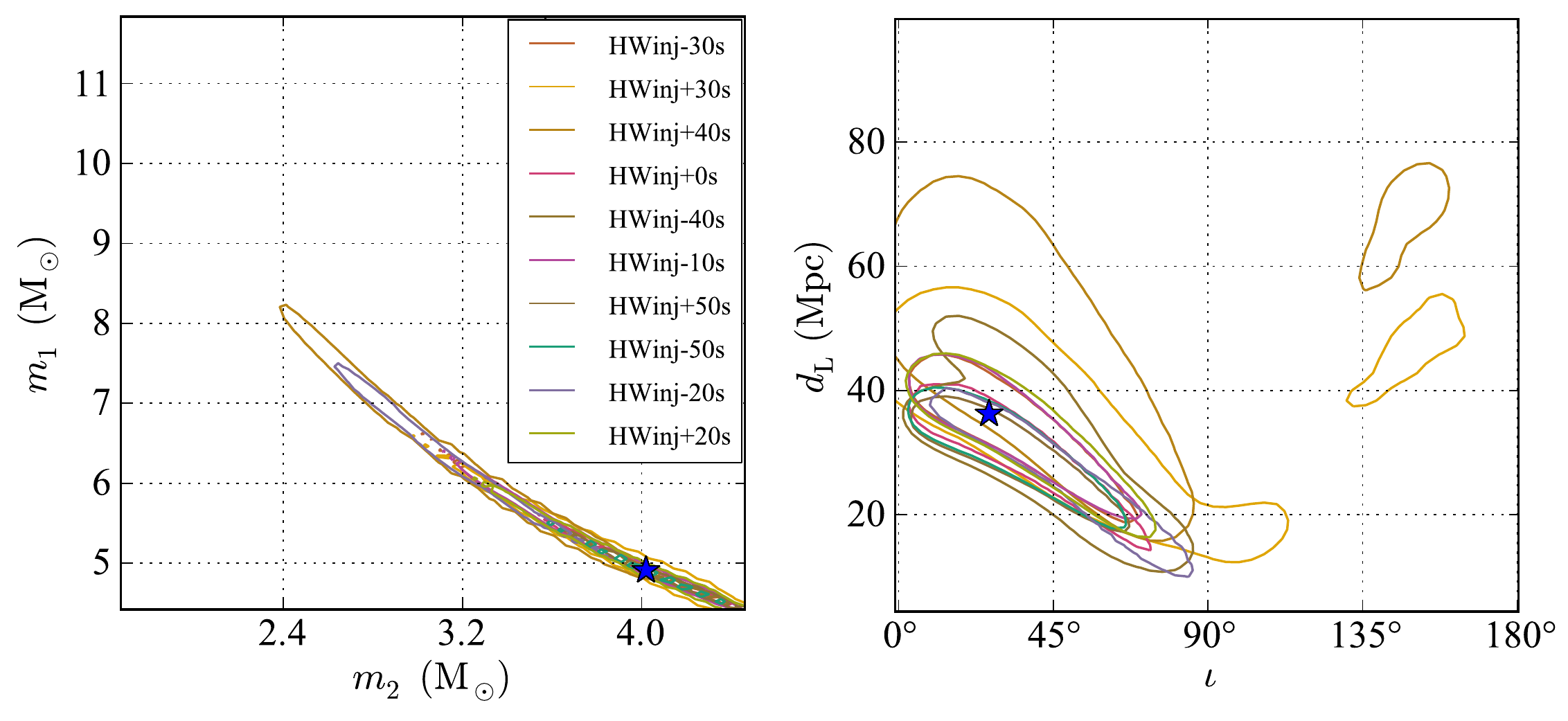}
\caption{\label{fig:bbh_hwsw_psd}Overlay of 90\,\% probability regions for the joint posterior probability distribution of the component masses $m_1$, $m_2$ of the binary (left) and the distance and inclination (right) for a series of signal injection times. Both sets of distributions show a spread comparable to the spread observed between different waveform models, as shown in figures \ref{fig:bbh_mass} and \ref{fig:bbh_location}.
}
\end{figure*}

%\begin{figure*}
%\begin{minipage}[b]{0.31\textwidth}
%\centering
%\includegraphics[width=\textwidth]{Images/pdf/H1_PSD_COMPARE_GRID_RED_PDF.pdf}
%\end{minipage}
%\begin{minipage}[b]{0.31\textwidth}
%\centering
%\includegraphics[width=\textwidth]{Images/pdf/L1_PSD_COMPARE_GRID_RED_PDF.pdf}
%\end{minipage}
%\begin{minipage}[b]{0.31\textwidth}
%\centering
%\includegraphics[width=\textwidth]{Images/pdf/V1_PSD_COMPARE_GRID_RED_PDF.pdf}
%\end{minipage}
%\caption{\label{fig:compare_psd} The plots show in red 10 overlaid noise noise power spectral densities (from left to right, LIGO-Hanford, LIGO-Livingston, and Virgo, respectively), each computed over 1024 sec of data, and separated by 10 sec. In black is the PSD used in Figs.~\ref{fig:bbh_mass} and~\ref{fig:bbh_location}.  The PSDs were used in the analyses whose results are shown in Figs.~\ref{fig:bbh_hwsw_psd} and~\ref{fig:bbh_hwsw_noise} and provide an indication of the fluctuation of the noise in the instruments over a time-span of minutes.
%}
%\end{figure*}

\begin{figure*}
\begin{minipage}[b]{0.48\textwidth}
\centering
\includegraphics[width=\textwidth]{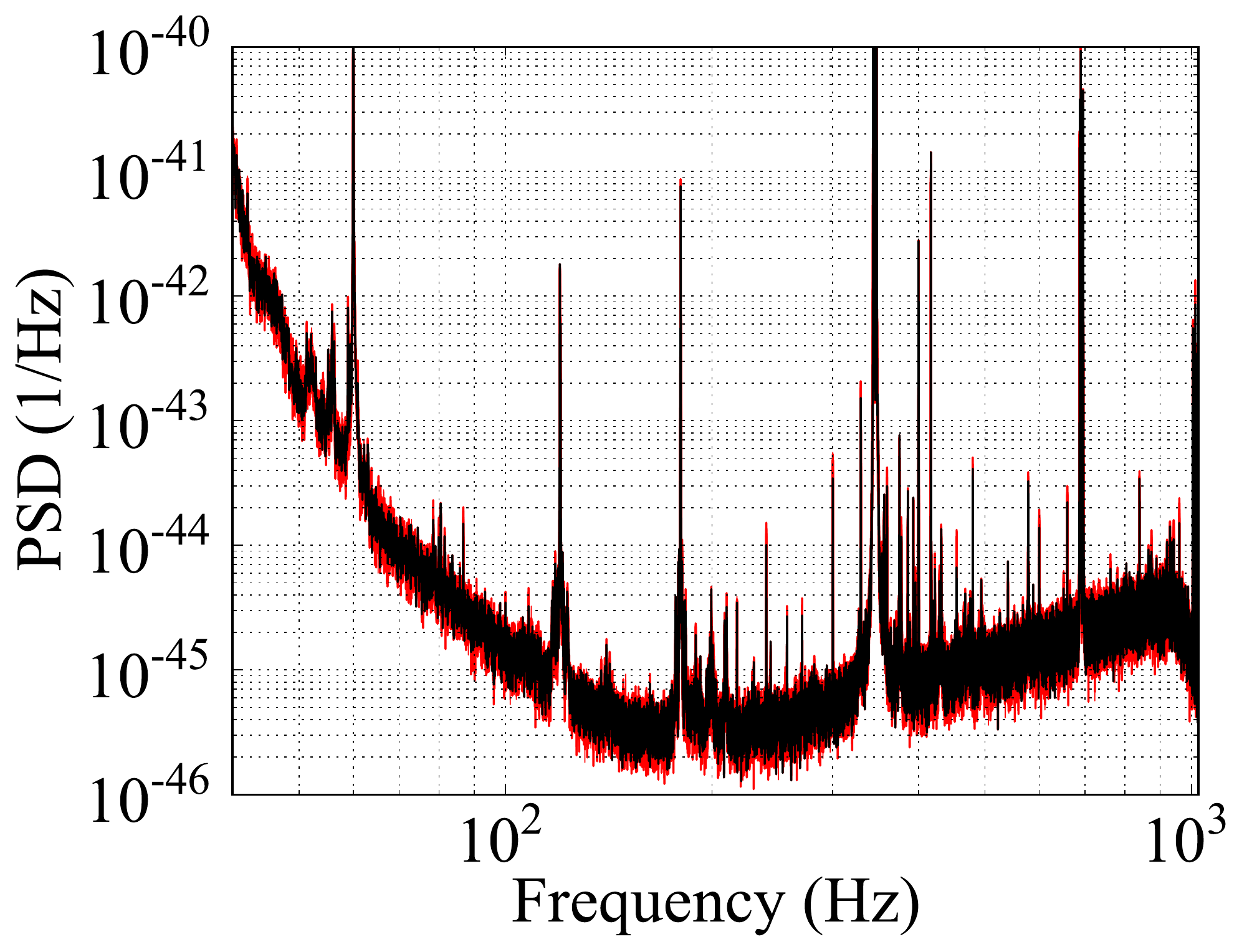}
\end{minipage}
\begin{minipage}[b]{0.48\textwidth}
\centering
\includegraphics[width=\textwidth]{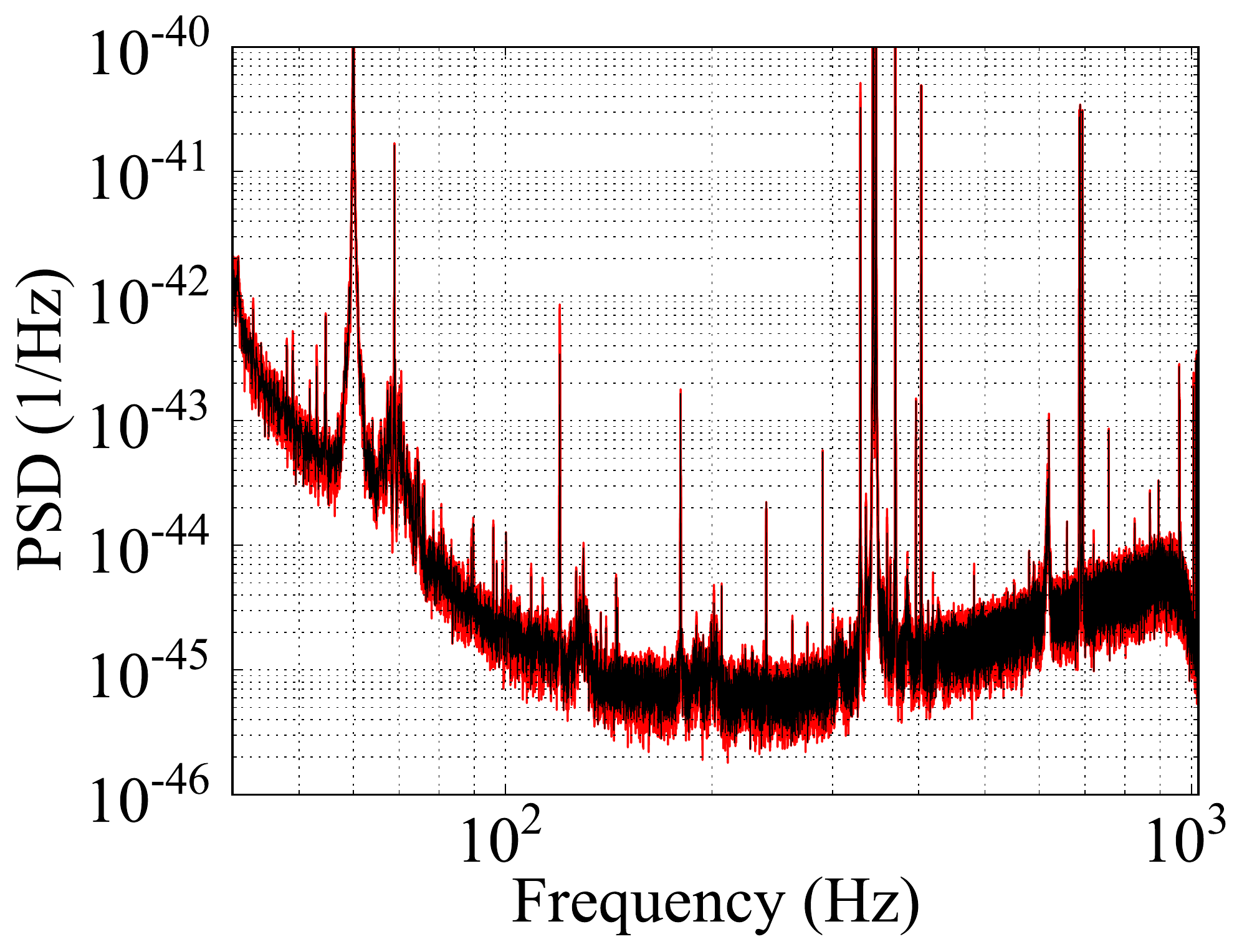}
\end{minipage}
\begin{minipage}[b]{0.48\textwidth}
\centering
\includegraphics[width=\textwidth]{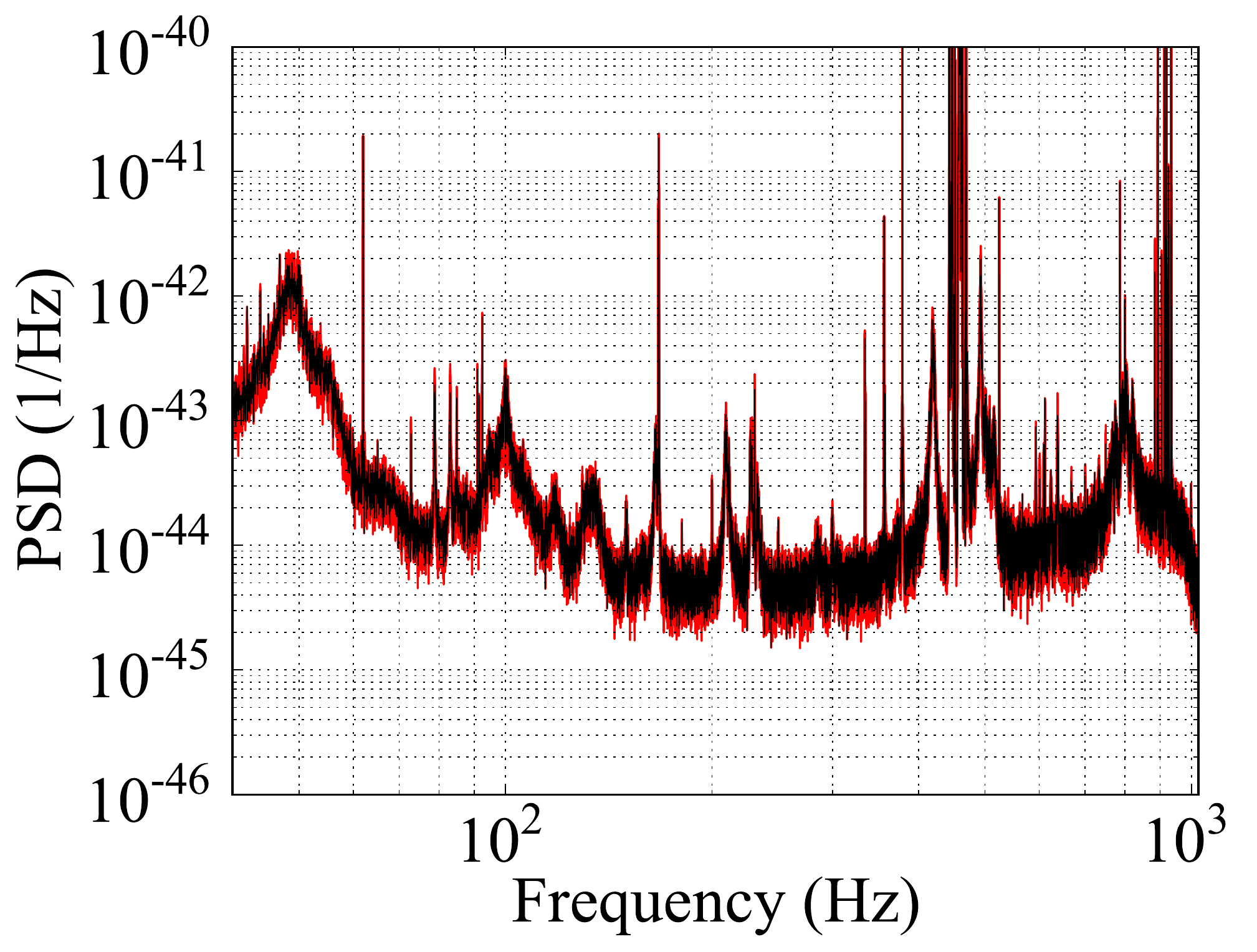}
\end{minipage}
\caption{\label{fig:compare_psd} The plots show in red 10 overlaid noise noise power spectral densities (top left LIGO-Hanford, top right LIGO-Livingston, and bottom Virgo), each computed over 1024 sec of data, and separated by 10 sec. In black is the PSD used in Figs.~\ref{fig:bbh_mass} and~\ref{fig:bbh_location}.  The PSDs were used in the analyses whose results are shown in Figs.~\ref{fig:bbh_hwsw_psd} and~\ref{fig:bbh_hwsw_noise} and provide an indication of the fluctuation of the noise in the instruments over a time-span of minutes.
}
\end{figure*}

%\begin{figure*}
%\vspace{-10pt}
%\begin{minipage}[b]{0.48\textwidth}
%\centering
%\includegraphics[width=\textwidth]{Images/pdf/EOBNR_SW-HW_mchirp_90.pdf}
%\end{minipage}
%\begin{minipage}[b]{0.48\textwidth}
%\centering
%\includegraphics[width=\textwidth]{Images/pdf/EOBNR_SW-HW_dist_iota.pdf}
%\end{minipage}
%\vspace{-10pt}
%\caption{\label{fig:bbh_hwsw_noise}(left) Posterior probability distributions for the chirp mass $\Mc$ of the non-spinning \ac{BBH} injection for signal model TF2 (table~\ref{tab:waveforms}), TaylorF2 at 3.5 pN order, for the hardware injection (section \ref{hwinj:bbh}) and 3 software injections of the same parameters injected 100, 500 and 1000 seconds before the hardware injection (see section \ref{hwsw}). The injected value is marked with a vertical red line. (right) Overlay of 90\% probability regions for the joint posterior distribution on the inclination and distance of the binary. Different realizations of the noise, rather than differences between hardware and software injections, are the likely reason for the variations in recovered \ac{PDF}s.
%}
%\end{figure*}

\begin{figure*}
\includegraphics[width=\textwidth]{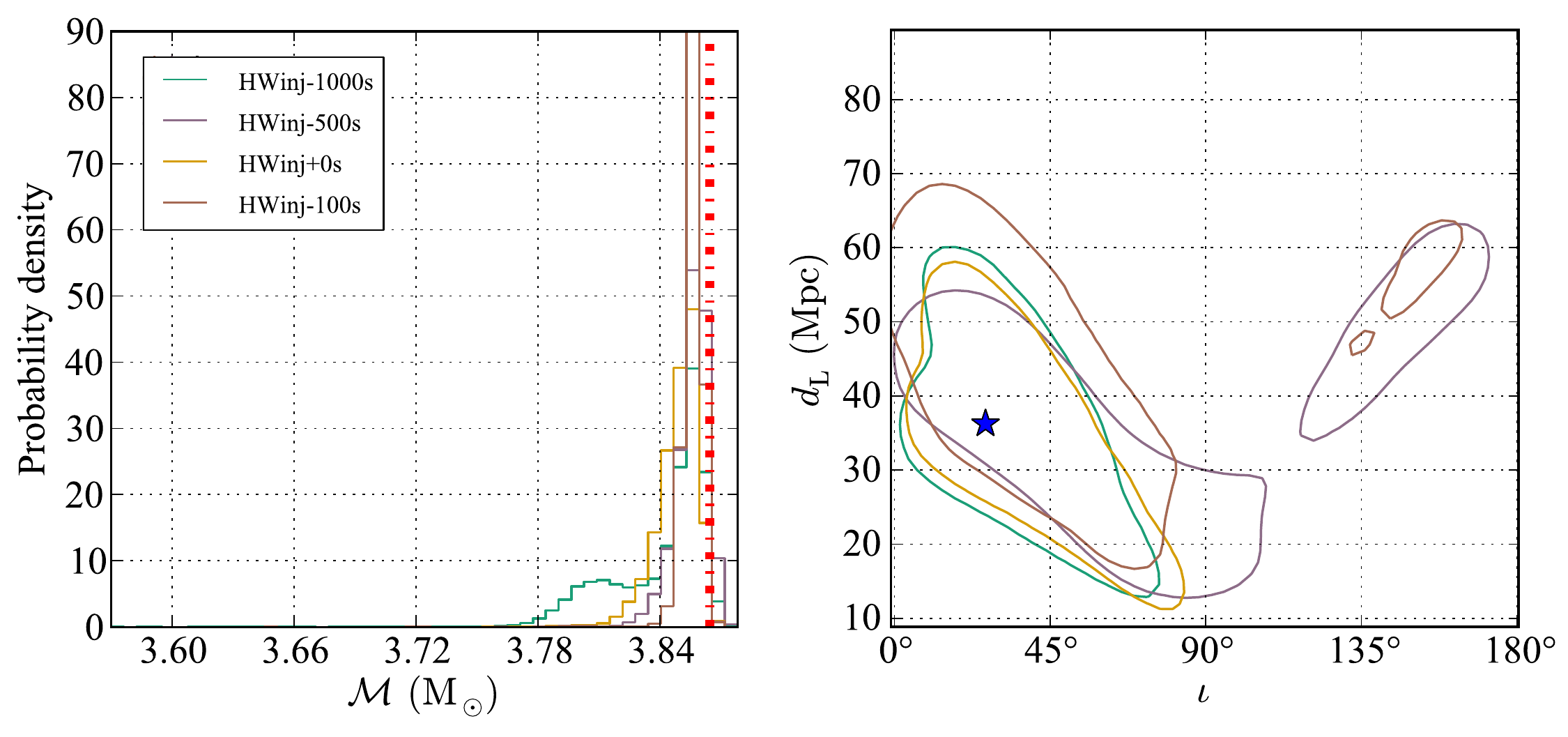}
\caption{\label{fig:bbh_hwsw_noise}(left) Posterior probability distributions for the chirp mass $\Mc$ of the non-spinning \ac{BBH} injection for signal model TF2 (table~\ref{tab:waveforms}), TaylorF2 at 3.5 pN order, for the hardware injection (section \ref{hwinj:bbh}) and 3 software injections of the same parameters injected 100, 500 and 1000 seconds before the hardware injection (see section \ref{hwsw}). The injected value is marked with a vertical red line. (right) Overlay of 90\% probability regions for the joint posterior distribution on the inclination and distance of the binary. Different realizations of the noise, rather than differences between hardware and software injections, are the likely reason for the variations in recovered \ac{PDF}s.
}
\end{figure*}

To exemplify the effect of noise on the recovered \ac{PDF}s, ten software injections (see figure \ref{fig:bbh_hwsw_psd}) with parameters identical to the non-spinning \ac{BBH} hardware injection (section \ref{hwinj:bbh}) were created with injection times in succession at 10-seconds intervals. With each injection at a different time, the estimated \ac{PSD}s were also (slightly) different (see figure~\ref{fig:compare_psd}), reflecting the slow change of the \ac{PSD} as a function of time. All the signals were both injected and recovered using the TaylorF2 signal model to eliminate systematic effects from using the wrong waveform family. Figure \ref{fig:bbh_hwsw_psd} shows the joint distributions of the recovered component masses as well as the distance and inclination. The spread of the 90\,\% credible intervals for different injection times is comparable to the spread observed using different waveform models, as shown in figures \ref{fig:bbh_mass} and \ref{fig:bbh_location}, illustrating the dependence on the specific model 
and realization of the noise used. Fixing the injection time for 
a series of different 
\ac{PSD}s gives narrower \ac{PDF}s than in figure \ref{fig:bbh_hwsw_psd}. For the particular data segment considered here, the same is also true for a fixed \ac{PSD} and a series of injection times but to a lesser extent. We plan to address this in the future by performing the analysis with a model in which the \ac{PSD} is not fixed, but parametrized as a function of a suitable set of unknown parameters to be marginalised over. For the ten software injections the true injected parameter is found within the given 90\,\% credible interval in 90\,\% of the injections, as expected. 
Figure~\ref{fig:bbh_hwsw_noise} shows a direct comparison between a non-spinning \ac{BBH} hardware injection and three software injections replicated with the same parameters. All are evaluated using the \ac{PSD} of the hardware injection with the software injections at 100, 500 and 1000 seconds before the hardware injection. Again, there are variations in the recovered \ac{PDF}s but the hardware injection is not an outlier. The variations are caused by the inherent dependency of the analysis on the particular noise realization used. Statistical fluctuations appear to be able to dominate any systematic bias introduced by the method of performing the hardware injection, such as the actuation function used in modelling the impulse applied to the mirrors to produce the desired signal.

\subsection{Computational cost}
The choice of approximant has strong implications for the computational cost of the analysis, which varied from a few hours to three weeks of real time, using up to three 8-core CPUs. Frequency domain templates such as TaylorF2 and IMRPhenomB generally require much less computational time than the time domain SpinTaylorT4, as they do not require numerical solution of the orbital frequency and spin evolution equations or a Fourier transform before computing the likelihood. The time domain SpinTaylorT4 can include the full six spin parameters, allowing precession of the orbital plane, whereas the frequency domain templates can only include spin aligned with the orbit.
The computational cost is also strongly influenced by the length of the templates, which ranged from $2.16$\,s for the blind injection (section~\ref{bigdog}) to $18$\,s for the \ac{BNS} hardware injection (section~\ref{hwinj:bns}), using a $40$\,Hz starting frequency. An additional effect is the higher frequency at which a \ac{BNS} will merge, forcing us to analyze the data at a higher sampling rate. A higher signal-to-noise ratio also increases the run-time with more iterations required to reach the more contrasted posterior distribution. This effect can be somewhat managed by increasing the number of parallel tempering chains in the MCMC algorithm \cite{Sluys:2008a,Sluys:2008b}, or by using parallel nested sampling runs, reducing the overall run-time at the expense of using more compute units in parallel.  Other promising cost-saving techniques include waveform interpolation to ease the burden of computing millions of templates \cite[e.g.,][]{SmithSVD:2012}.  

In a future triggered analysis, where the parameter estimation is run as a follow-up to a search algorithm, some guidance on the chirp mass from the search pipeline could be used to choose the length of the data segment to be analyzed so as to minimize the cost while allowing enough time for the full waveform to be generated. In the analysis presented here, the run-time varied from hours for the low \ac{SNR}, short segment (high mass) injections using frequency-domain templates, up to several weeks for the high \ac{SNR}, long segments (low mass) using time-domain fully precessing templates.

\section{Implications for gravitational-wave astronomy}\label{sec:implications}

Although the 5 injections in Table \ref{tab:injections} are not statistically sufficient to cover the range of source possibilities, we do have samples representing the three expected classes, \ac{BNS}, \ac{NSBH} and \ac{BBH}, at a range of \ac{SNR} and some including spin. This allows us to highlight some of the general features of parameter estimates that are likely to be encountered with future analyses of signal candidates in advanced gravitational-wave detectors.

We have found, in accordance with expectations, that the chirp mass parameter can be resolved with a very small statistical uncertainty of only a few percent or even lower, even when the \ac{SNR} is relatively low, as in the case of the \ac{BBH} hardware injection \ref{hwinj:bbh} and the \ac{NSBH} software injection~\ref{swinj:nsbh}, which had network \ac{SNR}s of $13$ (see figures \ref{fig:bbh_mass} and \ref{fig:swnsbh_mass}). With the lower mass and higher \ac{SNR} \ac{BNS} hardware injection, the systematic differences between waveform models dominate the statistical uncertainty on the chirp mass, as the signal accumulates more cycles in the sensitive band of the detectors. 

Previous parameter estimation studies have mainly focused on estimating the typical size of the errors, primarily using the Fisher information matrix~\cite{rao2009linear}.
The inverse of the Fisher matrix provides lower bounds for the covariance matrix %errors
of maximum likelihood estimators (Cramer-Rao lower bound)~\cite{helmstrom-1968}.
In~\cite{Arun:2006yw} a single detector analysis of non-spinning signals suggested that the chirp mass $\Mc$ could be estimated by LIGO with a fraction of percent accuracy and the mass ratio $\eta$ with an accuracy of a few percent. We find in our analysis (section~\ref{sec:Simulations}) that for the 5 injections considered this accuracy is reached only for the high \ac{SNR} example in section~\ref{hwinj:bns}. In general, the full analysis recovers chirp mass $\Mc$ and the mass ratio $\eta$ at the percent and tens of percent levels, respectively.
In addition, \cite{Arun:2006yw} underlined the dependence of the accuracy on the phase order of the post-Newtonian expansion, which we illustrate in the figures of section~\ref{sec:Simulations}.
Multiple detector studies, based on both Fisher matrix estimates and numerical simulations~\cite{2011PhRvD..84j4020V,2012PhRvD..85j4045V} showed the difference of accuracy in parameter estimation, between intrinsic parameters (e.g., masses) and extrinsic parameters (e.g., sky position, the strongly correlated inclination and distance). The estimation of the extrinsic parameters mainly depends on the number and positions of the GW detectors, while the accuracy for intrinsic
parameters depends mainly on the \ac{SNR}, and they are nearly equally well estimated with a single interferometer, for the same \ac{SNR}. 
The Fisher matrix calculation has known limitations~\cite{2008PhRvD..77d2001V}, and tends to underestimate the errors at low \ac{SNR}~\cite{2010PhRvD..81l4048Z}, which our results confirm. 

In all examples, the mass ratio parameter is strongly affected by systematic differences between the waveform models. For those models which included spin, the mass ratio parameter had a larger uncertainty due to correlations between the spins and mass ratio (see~\cite{Cutler:1994,Poisson:1995ef,2013PhRvD..87b4035B}), but these broader posteriors did encompass the true values. This indicates that even for sources which are expected to have an insignificant spin, i.e. \ac{BNS} systems, a conservative analysis with fully spinning templates is desirable. For systems involving black holes, performing the full spinning parameter estimation is essential, and further improvements in waveform modelling, informed by numerical relativity simulations, are crucial for reducing systematic biases and therefore providing more precise mass and spin estimates. Particularly necessary are fast to compute and accurate waveform models which include for arbitrary spins, and inspiral, merger and ring-down phases.

For the extrinsic parameters, of which the most interesting are the source location and distance, the accuracy of inference is limited by the statistical uncertainty coming from the finite \ac{SNR} of the source. The overall signal amplitude in each detector may be measured with a fractional accuracy $\sim 1/\mathrm{SNR}$, but the amplitude in each detector is a complicated function of the location, distance and polarization angle (see eqn. \ref{eq.signalIFO}), which produces the complicated credible interval shapes seen above. Most notably, the precision to which the distance can be measured is limited by the presence of correlation with the inclination angle, which is true for any detector network. On the other hand, the degeneracy in sky position can be greatly reduced by observing the signal strongly in multiple detectors; in the case of the three-detector network used here the sky position is restricted to two locations, symmetric when reflected across the plane containing the detectors, and with 
opposite inclinations (see fig. \ref{fig:bns_location}). This degeneracy is broken in the case of the spinning \ac{BBH} signal (\ref{swinj:bbh}) where the inclination angle can be determined from the precession of the orbital plane, eliminating one of the two sky locations (see figure \ref{fig:swbbh_location}). As the size of the two sky locations is primarily determined by the location of the detectors and the \ac{SNR}, these results are expected to be qualitatively similar for the advanced detector network. When the signal amplitude is relatively low in one detector (see figure \ref{fig:swnsbh_location}), more of the ring-like structure is observable, and the location cannot be well-determined.

The evidence values listed in Table~\ref{tab:evidences} can in principle allow us to discriminate between different waveform models, \textit{e.g.} whether the data support the presence of spins in the compact objects that generate the detected signal. However, in the analysis reported here the noise model that was used (Gaussian and stationary with a \ac{PSD} estimated using nearby data) is likely not realistic enough to draw sufficiently robust conclusions, with the exception of the most clear cases (for example the dominance of the spinning model for the injection of section~\ref{bigdog}). Work in is progress to include in the Bayesian analysis our uncertainty about the \ac{PSD}, and a more flexible noise model,~\cite{RoverMeyerChristensen:2011,PhysRevD.82.103007,Littenberg:2013gja}.

\section{Conclusions}\label{sec:astro}

In this paper, we have applied a suite of Bayesian parameter estimation tools to hardware and software injections from the last initial LIGO and Virgo science runs. 
The primary challenge of parameter estimation for \ac{CBC} signals lies in efficiently locating and sampling the modes of the posterior \ac{PDF} in a multi-dimensional parameter space.
We have shown that we are able to explore and compare these complicated, correlated and degenerate \ac{PDF}s for a variety of waveform models which, taken together, include the full range of physical effects expected to be of importance in a real analysis: inclusion of spins, precession of the orbital plane, and the merger-ringdown signal. We have verified the consistency of our results by cross-comparison among different sampling algorithms.

Although there has been rapid progress in the field of \ac{GW} parameter estimation, a number of key questions remain.  Some of these are already being addressed, or will be addressed over the next few years before advanced detectors come online. Some of the most pressing questions are:
\begin{itemize}
\item{}How precisely will we be able to measure spin magnitudes and spin tilt angles in precessing \ac{NSBH} and \ac{BBH} systems? 
\item{}How important are systematic waveform biases due to imperfect waveform knowledge for various system classes?  What are the accuracy requirements on waveform families for parameter estimation?
\item{}What is the best way to handle the long-duration signals in advanced detectors? Their extended bandwidth can cause current implementations to increase in runtime by factors of twenty-five or more.
\item{}How accurately can we measure finite-size and tidal-dissipation effects for systems involving neutron stars?
\item{}How is parameter estimation affected when the background noise is not stationary and Gaussian?
\end{itemize}

We anticipate that the analysis methods used in this paper will be directly applicable to the first detections from the advanced detector era, allowing us to carry out astrophysical inference and to measure masses, spins, and sky locations of coalescing compact binaries.  This, in turn, will improve searches for electromagnetic counterparts, and ultimately allow us to solve a key inverse problem of \ac{GW} astrophysics: to reconstruct binary evolution and dynamical binary formation from the observed distributions of masses and spins of merging compact binaries.

\acknowledgments

The authors gratefully acknowledge the support of the United States
National Science Foundation for the construction and operation of the
LIGO Laboratory, the Science and Technology Facilities Council of the
United Kingdom, the Max-Planck-Society, and the State of
Niedersachsen/Germany for support of the construction and operation of
the GEO600 detector, and the Italian Istituto Nazionale di Fisica
Nucleare and the French Centre National de la Recherche Scientifique
for the construction and operation of the Virgo detector. The authors
also gratefully acknowledge the support of the research by these
agencies and by the Australian Research Council, 
the International Science Linkages program of the Commonwealth of Australia,
the Council of Scientific and Industrial Research of India, 
the Istituto Nazionale di Fisica Nucleare of Italy, 
%%---- modified Feb2012:
% the Spanish Ministerio de Educaci\'on y Ciencia, 
the Spanish Ministerio de Econom\'ia y Competitividad,
%%----------------------
the Conselleria d'Economia Hisenda i Innovaci\'o of the
Govern de les Illes Balears, the Foundation for Fundamental Research
on Matter supported by the Netherlands Organisation for Scientific Research, 
the Polish Ministry of Science and Higher Education, the FOCUS
Programme of Foundation for Polish Science,
the Royal Society, the Scottish Funding Council, the
Scottish Universities Physics Alliance, the National Aeronautics and
Space Administration, 
%% added May 2012
the National Research Foundation of Korea,
Industry Canada and the Province of Ontario through the Ministry of Economic Development and Innovation, 
the National Science and Engineering Research Council Canada,
%%%
the Carnegie Trust, the Leverhulme Trust, the
David and Lucile Packard Foundation, the Research Corporation, and
the Alfred P. Sloan Foundation.

\clearpage

\appendix

\section{Credible Intervals}
\label{sec:credible-intervals}

In
Tables \ref{tab:BBH-HWINJ-credible}, \ref{tab:BNS-HWINJ-credible}, \ref{tab:BBH-SWINJ-credible}, \ref{tab:NSBH-SWINJ-credible}, \ref{tab:DOG-credible},
we report the median value and 90\% credible intervals for the
one-dimensional posteriors in $\mathcal{M}$, $\eta$, $m_1$, $m_2$, and
$d_L$ obtained in the context of the different waveform models for the
injections listed in Table \ref{tab:injections}.  Our reported
credible intervals are \emph{symmetric}, excluding an equal posterior
mass (in this case $0.05$) below and above the interval: a symmetric
$(100-\alpha)$\% credible interval
$\left[\theta_\mathrm{min}, \theta_\mathrm{max}\right]$ satisfies
\begin{equation}
\int_{-\infty}^{\theta_\mathrm{min}} d\theta \, p(\theta | {d}, H) = \int_{\theta_\mathrm{max}}^\infty d\theta \, p(\theta | {d}, H) = \frac{1}{2} \frac{\alpha}{100}.
\end{equation}

\begin{table}[h]
\begin{tabular}{|l||c|c|c|c|c|}
\hline & $\mathcal{M}$ & $\eta$ & $m_1$ & $m_2$ & $d$ \\ 
 \hline \hline 
ST & $3.85^{3.89}_{3.77}$  & $0.236^{0.25}_{0.195}$  & $5.67^{7.56}_{4.61}$  & $3.49^{4.27}_{2.71}$  & $41.2^{62.2}_{26.8}$ \\ 
ST\_SA & $3.88^{3.98}_{3.82}$  & $0.224^{0.25}_{0.121}$  & $6.27^{12.1}_{4.56}$  & $3.23^{4.31}_{1.98}$  & $36.2^{59.2}_{18.8}$ \\ 
ST\_NS & $3.85^{3.86}_{3.83}$  & $0.245^{0.25}_{0.221}$  & $5.13^{6.34}_{4.49}$  & $3.83^{4.37}_{3.13}$  & $35.7^{57.9}_{18.8}$ \\ 
TF2 & $3.85^{3.86}_{3.83}$  & $0.242^{0.25}_{0.224}$  & $5.29^{6.22}_{4.52}$  & $3.72^{4.35}_{3.18}$  & $35.4^{58.1}_{18.4}$ \\ 
TF2\_2 & $3.84^{3.86}_{3.82}$  & $0.161^{0.181}_{0.145}$  & $9.15^{10}_{8.21}$  & $2.32^{2.56}_{2.14}$  & $35.2^{57.1}_{19}$ \\ 
IMRPB & $3.88^{3.93}_{3.86}$  & $0.198^{0.249}_{0.0671}$  & $7.47^{18.4}_{4.73}$  & $2.78^{4.19}_{1.43}$  & $34.2^{61.5}_{16.3}$ \\ 
\hline 
\end{tabular}
\caption{\label{tab:BBH-HWINJ-credible} Median value and 90\% credible intervals on chirp mass ($\mathcal{M}$), symmetric mass ratio ($\eta$), masses
 ($m_1$ and $m_2$) and distance ($d_L$) obtained in the context of
 various waveform models (see Table \ref{tab:waveforms}) for the
 injection discussed in \S~\ref{hwinj:bbh}}
\end{table}

\begin{table}[h]
\begin{tabular}{|l||c|c|c|c|c|}
\hline & $\mathcal{M}$ & $\eta$ & $m_1$ & $m_2$ & $d$ \\ 
 \hline \hline 
ST & $1.5035^{1.503}_{1.504}$  & $0.248^{0.25}_{0.245}$  & $1.88^{2}_{1.76}$  & $1.59^{1.69}_{1.5}$  & $3.56^{4.01}_{3.23}$ \\ 
ST\_SA & $1.503^{1.504}_{1.502}$  & $0.248^{0.25}_{0.24}$  & $1.87^{2.13}_{1.74}$  & $1.59^{1.71}_{1.41}$  & $3.54^{4.01}_{2.38}$ \\ 
ST\_NS & $1.5017^{1.502}_{1.5016}$  & $0.25^{0.25}_{0.249}$  & $1.77^{1.85}_{1.73}$  & $1.68^{1.72}_{1.61}$  & $3.12^{3.56}_{2.41}$ \\ 
TF2 & $1.5018^{1.502}_{1.5017}$  & $0.25^{0.25}_{0.249}$  & $1.77^{1.84}_{1.73}$  & $1.68^{1.72}_{1.62}$  & $3.14^{3.53}_{2.47}$ \\ 
TF2\_2 & $1.5002^{1.5007}_{1.4998}$  & $0.185^{0.188}_{0.183}$  & $3.12^{3.16}_{3.06}$  & $1.01^{1.03}_{1}$  & $3.68^{4.18}_{3.31}$ \\ 
IMRPB & $1.509^{1.512}_{1.506}$  & $0.218^{0.228}_{0.187}$  & $2.56^{3.11}_{2.36}$  & $1.21^{1.29}_{1.03}$  & $3.66^{4.15}_{3.26}$ \\ 
\hline 
 \end{tabular}
\caption{\label{tab:BNS-HWINJ-credible} Median value and 90\% credible intervals on chirp mass ($\mathcal{M}$), symmetric mass ratio ($\eta$), masses
 ($m_1$ and $m_2$) and distance ($d_L$) obtained in the context of
 various waveform models (see Table \ref{tab:waveforms}) for the
 injection discussed in \S~\ref{hwinj:bns}}
\end{table}

\begin{table}[h]
\begin{tabular}{|l||c|c|c|c|c|}
\hline & $\mathcal{M}$ & $\eta$ & $m_1$ & $m_2$ & $d$ \\ 
 \hline \hline 
ST & $4.76^{4.82}_{4.7}$  & $0.244^{0.25}_{0.212}$  & $6.41^{8.44}_{5.53}$  & $4.68^{5.38}_{3.7}$  & $29.3^{36.7}_{16.8}$ \\ 
ST\_SA & $4.76^{4.81}_{4.72}$  & $0.244^{0.25}_{0.204}$  & $6.43^{8.92}_{5.54}$  & $4.67^{5.38}_{3.56}$  & $29.2^{36.7}_{16.4}$ \\ 
ST\_NS & $4.76^{4.77}_{4.73}$  & $0.247^{0.25}_{0.234}$  & $6.15^{7.11}_{5.54}$  & $4.86^{5.4}_{4.21}$  & $29.9^{37.3}_{17.1}$ \\ 
TF2 & $4.75^{4.76}_{4.72}$  & $0.246^{0.25}_{0.23}$  & $6.24^{7.31}_{5.52}$  & $4.78^{5.41}_{4.08}$  & $31.4^{39.2}_{18.1}$ \\ 
TF2\_2 & $4.76^{4.82}_{4.71}$  & $0.172^{0.205}_{0.15}$  & $10.7^{12}_{8.87}$  & $3.02^{3.59}_{2.7}$  & $31.6^{39.9}_{17.5}$ \\ 
IMRPB & $4.74^{4.78}_{4.71}$  & $0.241^{0.25}_{0.156}$  & $6.6^{11.6}_{5.54}$  & $4.52^{5.37}_{2.8}$  & $29.4^{37.2}_{16.2}$ \\ 
\hline 
 \end{tabular}
\caption{\label{tab:BBH-SWINJ-credible} Median value and 90\% credible intervals on chirp mass ($\mathcal{M}$), symmetric mass ratio ($\eta$), masses
 ($m_1$ and $m_2$) and distance ($d_L$) obtained in the context of
 various waveform models (see Table \ref{tab:waveforms}) for the
 injection discussed in \S~\ref{swinj:bbh}.}
\end{table}

\begin{table}[h]          
\begin{tabular}{|l||c|c|c|c|c|}
\hline & $\mathcal{M}$ & $\eta$ & $m_1$ & $m_2$ & $d$ \\ 
 \hline \hline 
ST & $3.16^{3.31}_{2.99}$  & $0.1^{0.176}_{0.0592}$  & $11.2^{15.7}_{7.13}$  & $1.44^{2.1}_{1.06}$  & $36.4^{56.7}_{22.5}$ \\ 
ST\_SA & $3.24^{3.51}_{3.14}$  & $0.187^{0.247}_{0.11}$  & $6.65^{11.1}_{4.24}$  & $2.21^{3.46}_{1.56}$  & $49.6^{82.9}_{26.7}$ \\ 
ST\_NS & $3.22^{3.27}_{3.03}$  & $0.133^{0.153}_{0.0733}$  & $9.1^{13.4}_{8.17}$  & $1.71^{1.9}_{1.16}$  & $41.8^{65.9}_{24.6}$ \\ 
TF2 & $3.25^{3.41}_{3.09}$  & $0.155^{0.228}_{0.101}$  & $8.02^{10.9}_{5.35}$  & $1.91^{2.92}_{1.39}$  & $47.7^{79}_{27.3}$ \\ 
TF2\_2 & $3.2^{3.42}_{3.1}$  & $0.0784^{0.163}_{0.0541}$  & $13.5^{16.8}_{8.09}$  & $1.27^{2.08}_{1.02}$  & $40.9^{73.8}_{22.1}$ \\ 
IMRPB & $3.25^{3.3}_{3.21}$  & $0.182^{0.248}_{0.122}$  & $6.88^{9.89}_{4.08}$  & $2.16^{3.4}_{1.64}$  & $41.6^{58.6}_{27.9}$ \\ 
\hline 
 \end{tabular}
\caption{\label{tab:NSBH-SWINJ-credible} Median value and
 90\% credible intervals on chirp mass ($\mathcal{M}$), symmetric mass
 ratio ($\eta$), masses ($m_1$ and $m_2$) and distance ($d_L$) obtained
 in the context of various waveform models (see
 Table \ref{tab:waveforms}) for the injection discussed
 in \S~\ref{swinj:nsbh}.}
\end{table}

\begin{table}[H]
\begin{tabular}{|l||c|c|c|c|c|c|}
\hline & $\mathcal{M}$ & $\eta$ & $m_1$ & $m_2$ & $d$ \\ 
 \hline \hline 
TF2\_25 & $4.63^{4.66}_{4.6}$  & $0.249^{0.25}_{0.242}$  & $5.67^{6.35}_{5.36}$  & $5^{5.3}_{4.45}$  & $45.7^{60}_{20.6}$  \\ 
ST\_25 & $4.94^{4.97}_{4.9}$  & $0.06^{0.0703}_{0.0534}$  & $25^{27.1}_{22.3}$  & $1.71^{1.84}_{1.62}$  & $36.4^{43.3}_{20.1}$  \\ 
\hline 
 \end{tabular}
\caption{\label{tab:DOG-credible} Median value and 90\%
 credible intervals on chirp mass ($\mathcal{M}$), symmetric mass
 ratio ($\eta$), masses ($m_1$ and $m_2$) and distance ($d_L$) obtained
 in the context of various waveform models (see
 Table \ref{tab:waveforms}) for the injection discussed
 in \S~\ref{bigdog}.}
\end{table}

\clearpage

\bibliography{../bibtex/iulpapers}

\begin{thebibliography}{69}
\expandafter\ifx\csname natexlab\endcsname\relax\def\natexlab#1{#1}\fi
\expandafter\ifx\csname bibnamefont\endcsname\relax
  \def\bibnamefont#1{#1}\fi
\expandafter\ifx\csname bibfnamefont\endcsname\relax
  \def\bibfnamefont#1{#1}\fi
\expandafter\ifx\csname citenamefont\endcsname\relax
  \def\citenamefont#1{#1}\fi
\expandafter\ifx\csname url\endcsname\relax
  \def\url#1{\texttt{#1}}\fi
\expandafter\ifx\csname urlprefix\endcsname\relax\def\urlprefix{URL }\fi
\providecommand{\bibinfo}[2]{#2}
\providecommand{\eprint}[2][]{\url{#2}}

\bibitem[{\citenamefont{Einstein}(1918)}]{Einstein:1918}
\bibinfo{author}{\bibfnamefont{A.}~\bibnamefont{Einstein}},
  \bibinfo{journal}{Preuss.\ Akad.\ Wiss.\ Berlin} pp.
  \bibinfo{pages}{154--167} (\bibinfo{year}{1918}).

\bibitem[{\citenamefont{Peters and Mathews}(1963)}]{PhysRev.131.435}
\bibinfo{author}{\bibfnamefont{P.~C.} \bibnamefont{Peters}} \bibnamefont{and}
  \bibinfo{author}{\bibfnamefont{J.}~\bibnamefont{Mathews}},
  \bibinfo{journal}{Phys. Rev.} \textbf{\bibinfo{volume}{131}},
  \bibinfo{pages}{435} (\bibinfo{year}{1963}).

\bibitem[{\citenamefont{{Weisberg} et~al.}(2010)\citenamefont{{Weisberg},
  {Nice}, and {Taylor}}}]{weisberg:2010}
\bibinfo{author}{\bibfnamefont{J.~M.} \bibnamefont{{Weisberg}}},
  \bibinfo{author}{\bibfnamefont{D.~J.} \bibnamefont{{Nice}}},
  \bibnamefont{and} \bibinfo{author}{\bibfnamefont{J.~H.}
  \bibnamefont{{Taylor}}}, \bibinfo{journal}{Astrophys. J.}
  \textbf{\bibinfo{volume}{722}}, \bibinfo{pages}{1030} (\bibinfo{year}{2010}),
  \eprint{arXiv:1011.0718}.

\bibitem[{\citenamefont{Abbott et~al.}(2009)}]{Abbott:2007kv}
\bibinfo{author}{\bibfnamefont{B.}~\bibnamefont{Abbott}} \bibnamefont{et~al.}
  (\bibinfo{collaboration}{LIGO Scientific Collaboration}),
  \bibinfo{journal}{Rep.\ Prog.\ Phys.} \textbf{\bibinfo{volume}{72}},
  \bibinfo{pages}{076901} (\bibinfo{year}{2009}).

\bibitem[{\citenamefont{Accadia et~al.}(2012)}]{Accadia:2012zz}
\bibinfo{author}{\bibfnamefont{T.}~\bibnamefont{Accadia}} \bibnamefont{et~al.}
  (\bibinfo{collaboration}{Virgo Collaboration}), \bibinfo{journal}{JINST}
  \textbf{\bibinfo{volume}{7}}, \bibinfo{pages}{P03012} (\bibinfo{year}{2012}).

\bibitem[{\citenamefont{{The LSC-Virgo Collaboration}}(In
  preparation)}]{Commissioning}
\bibinfo{author}{\bibnamefont{{The LSC-Virgo Collaboration}}}
  (\bibinfo{year}{In preparation}), \bibinfo{note}{{LIGO P1200087}}.

\bibitem[{\citenamefont{Abadie et~al.}(2012)}]{Collaboration:S6CBClowmass}
\bibinfo{author}{\bibfnamefont{J.}~\bibnamefont{Abadie}} \bibnamefont{et~al.}
  (\bibinfo{collaboration}{LIGO Scientific Collaboration and Virgo
  Collaboration}), \bibinfo{journal}{Phys.\ Rev.\ D}
  \textbf{\bibinfo{volume}{85}}, \bibinfo{pages}{082002}
  (\bibinfo{year}{2012}), \eprint{arXiv:1111.7314}.

\bibitem[{\citenamefont{{The LIGO Scientific Collaboration}
  et~al.}(2013)\citenamefont{{The LIGO Scientific Collaboration}, {the Virgo
  Collaboration}, {Aasi}, {Abadie}, {Abbott}, {Abbott}, {Abbott}, {Abernathy},
  {Accadia}, {Acernese} et~al.}}]{2013PhRvD..87b2002A}
\bibinfo{author}{\bibnamefont{{The LIGO Scientific Collaboration}}},
  \bibinfo{author}{\bibnamefont{{the Virgo Collaboration}}},
  \bibinfo{author}{\bibfnamefont{J.}~\bibnamefont{{Aasi}}},
  \bibinfo{author}{\bibfnamefont{J.}~\bibnamefont{{Abadie}}},
  \bibinfo{author}{\bibfnamefont{B.~P.} \bibnamefont{{Abbott}}},
  \bibinfo{author}{\bibfnamefont{R.}~\bibnamefont{{Abbott}}},
  \bibinfo{author}{\bibfnamefont{T.~D.} \bibnamefont{{Abbott}}},
  \bibinfo{author}{\bibfnamefont{M.}~\bibnamefont{{Abernathy}}},
  \bibinfo{author}{\bibfnamefont{T.}~\bibnamefont{{Accadia}}},
  \bibinfo{author}{\bibfnamefont{F.}~\bibnamefont{{Acernese}}},
  \bibnamefont{et~al.}, \bibinfo{journal}{\prd} \textbf{\bibinfo{volume}{87}},
  \bibinfo{eid}{022002} (\bibinfo{year}{2013}).

\bibitem[{\citenamefont{Harry and the LIGO
  Scientific~Collaboration}(2010)}]{Harry:2010zz}
\bibinfo{author}{\bibfnamefont{G.~M.} \bibnamefont{Harry}} \bibnamefont{and}
  \bibinfo{author}{\bibnamefont{the LIGO Scientific~Collaboration}},
  \bibinfo{journal}{Class. Quant. Grav.} \textbf{\bibinfo{volume}{27}},
  \bibinfo{pages}{084006} (\bibinfo{year}{2010}),
  \urlprefix\url{http://stacks.iop.org/0264-9381/27/i=8/a=084006}.

\bibitem[{PSD(2009)}]{PSD:AV}
\emph{\bibinfo{title}{{Advanced Virgo Baseline Design}}}
  (\bibinfo{year}{2009}),
  \bibinfo{note}{https://pub3.ego-gw.it/itf/tds/file.php?callFile=VIR-0027A-09.pdf}.

\bibitem[{\citenamefont{Abadie et~al.}(2010{\natexlab{a}})}]{ratesdoc}
\bibinfo{author}{\bibfnamefont{J.}~\bibnamefont{Abadie}} \bibnamefont{et~al.}
  (\bibinfo{collaboration}{LIGO Scientific Collaboration and Virgo
  Collaboration}), \bibinfo{journal}{Class. Quant. Grav.}
  \textbf{\bibinfo{volume}{27}}, \bibinfo{pages}{173001}
  (\bibinfo{year}{2010}{\natexlab{a}}).

\bibitem[{GW1(2011)}]{GW100916web}
 (\bibinfo{year}{2011}), \bibinfo{note}{lIGO Scientific Collaboration and Virgo
  Collaboration, The LIGO / Virgo Blind Injection GW100916 (2011),
  \url{http://www.ligo.org/science/GW100916/},
  \url{http://www.ligo.org/news/blind-injection.php}},
  \urlprefix\url{{http://www.ligo.org/science/GW100916/,
  http://www.ligo.org/news/blind-injection.php}}.

\bibitem[{\citenamefont{{Read} et~al.}(2009)\citenamefont{{Read}, {Markakis},
  {Shibata}, {Ury{\= u}}, {Creighton}, and {Friedman}}}]{Read:2009}
\bibinfo{author}{\bibfnamefont{J.~S.} \bibnamefont{{Read}}},
  \bibinfo{author}{\bibfnamefont{C.}~\bibnamefont{{Markakis}}},
  \bibinfo{author}{\bibfnamefont{M.}~\bibnamefont{{Shibata}}},
  \bibinfo{author}{\bibfnamefont{K.}~\bibnamefont{{Ury{\= u}}}},
  \bibinfo{author}{\bibfnamefont{J.~D.~E.} \bibnamefont{{Creighton}}},
  \bibnamefont{and} \bibinfo{author}{\bibfnamefont{J.~L.}
  \bibnamefont{{Friedman}}}, \bibinfo{journal}{Phys. Rev. D}
  \textbf{\bibinfo{volume}{79}}, \bibinfo{pages}{124033}
  (\bibinfo{year}{2009}), \eprint{arXiv:0901.3258}.

\bibitem[{\citenamefont{Ozel et~al.}(2010)\citenamefont{Ozel, Psaltis, Narayan,
  and McClintock}}]{Ozel:2010}
\bibinfo{author}{\bibfnamefont{F.}~\bibnamefont{Ozel}},
  \bibinfo{author}{\bibfnamefont{D.}~\bibnamefont{Psaltis}},
  \bibinfo{author}{\bibfnamefont{R.}~\bibnamefont{Narayan}}, \bibnamefont{and}
  \bibinfo{author}{\bibfnamefont{J.~E.} \bibnamefont{McClintock}},
  \bibinfo{journal}{Astrophys. J.} \textbf{\bibinfo{volume}{725}},
  \bibinfo{pages}{1918} (\bibinfo{year}{2010}), \eprint{1006.2834}.

\bibitem[{\citenamefont{{Farr} et~al.}(2011)\citenamefont{{Farr}, {Sravan},
  {Cantrell}, {Kreidberg}, {Bailyn}, {Mandel}, and {Kalogera}}}]{Farr:2010}
\bibinfo{author}{\bibfnamefont{W.~M.} \bibnamefont{{Farr}}},
  \bibinfo{author}{\bibfnamefont{N.}~\bibnamefont{{Sravan}}},
  \bibinfo{author}{\bibfnamefont{A.}~\bibnamefont{{Cantrell}}},
  \bibinfo{author}{\bibfnamefont{L.}~\bibnamefont{{Kreidberg}}},
  \bibinfo{author}{\bibfnamefont{C.~D.} \bibnamefont{{Bailyn}}},
  \bibinfo{author}{\bibfnamefont{I.}~\bibnamefont{{Mandel}}}, \bibnamefont{and}
  \bibinfo{author}{\bibfnamefont{V.}~\bibnamefont{{Kalogera}}},
  \bibinfo{journal}{Astrophys. J.} \textbf{\bibinfo{volume}{741}},
  \bibinfo{eid}{103} (\bibinfo{year}{2011}), \eprint{1011.1459}.

\bibitem[{\citenamefont{{Kreidberg} et~al.}(2012)\citenamefont{{Kreidberg},
  {Bailyn}, {Farr}, and {Kalogera}}}]{Kreidberg:2012}
\bibinfo{author}{\bibfnamefont{L.}~\bibnamefont{{Kreidberg}}},
  \bibinfo{author}{\bibfnamefont{C.~D.} \bibnamefont{{Bailyn}}},
  \bibinfo{author}{\bibfnamefont{W.~M.} \bibnamefont{{Farr}}},
  \bibnamefont{and}
  \bibinfo{author}{\bibfnamefont{V.}~\bibnamefont{{Kalogera}}},
  \bibinfo{journal}{Astrophys. J.} \textbf{\bibinfo{volume}{757}},
  \bibinfo{eid}{36} (\bibinfo{year}{2012}), \eprint{1205.1805}.

\bibitem[{\citenamefont{Fairhurst}(2009)}]{Fairhurst2009}
\bibinfo{author}{\bibfnamefont{S.}~\bibnamefont{Fairhurst}},
  \bibinfo{journal}{New Journal of Physics} \textbf{\bibinfo{volume}{11}},
  \bibinfo{pages}{123006} (\bibinfo{year}{2009}).

\bibitem[{\citenamefont{{Nissanke} et~al.}(2011)\citenamefont{{Nissanke},
  {Sievers}, {Dalal}, and {Holz}}}]{Nissanke:2011}
\bibinfo{author}{\bibfnamefont{S.}~\bibnamefont{{Nissanke}}},
  \bibinfo{author}{\bibfnamefont{J.}~\bibnamefont{{Sievers}}},
  \bibinfo{author}{\bibfnamefont{N.}~\bibnamefont{{Dalal}}}, \bibnamefont{and}
  \bibinfo{author}{\bibfnamefont{D.}~\bibnamefont{{Holz}}},
  \bibinfo{journal}{\apj} \textbf{\bibinfo{volume}{739}}, \bibinfo{pages}{99}
  (\bibinfo{year}{2011}), \eprint{1105.3184}.

\bibitem[{\citenamefont{{LIGO Scientific Collaboration}
  et~al.}(2012)\citenamefont{{LIGO Scientific Collaboration}, {Virgo
  Collaboration}, {Abadie}, {Abbott}, {Abbott}, {Abbott}, {Abernathy},
  {Accadia}, {Acernese}, {Adams} et~al.}}]{2011arXiv1109.3498T}
\bibinfo{author}{\bibnamefont{{LIGO Scientific Collaboration}}},
  \bibinfo{author}{\bibnamefont{{Virgo Collaboration}}},
  \bibinfo{author}{\bibfnamefont{J.}~\bibnamefont{{Abadie}}},
  \bibinfo{author}{\bibfnamefont{B.~P.} \bibnamefont{{Abbott}}},
  \bibinfo{author}{\bibfnamefont{R.}~\bibnamefont{{Abbott}}},
  \bibinfo{author}{\bibfnamefont{T.~D.} \bibnamefont{{Abbott}}},
  \bibinfo{author}{\bibfnamefont{M.}~\bibnamefont{{Abernathy}}},
  \bibinfo{author}{\bibfnamefont{T.}~\bibnamefont{{Accadia}}},
  \bibinfo{author}{\bibfnamefont{F.}~\bibnamefont{{Acernese}}},
  \bibinfo{author}{\bibfnamefont{C.}~\bibnamefont{{Adams}}},
  \bibnamefont{et~al.}, \bibinfo{journal}{\aap} \textbf{\bibinfo{volume}{539}},
  \bibinfo{eid}{A124} (\bibinfo{year}{2012}), \eprint{1109.3498}.

\bibitem[{\citenamefont{{Abadie} et~al.}(2012)\citenamefont{{Abadie}, {Abbott},
  {Abbott}, {Abbott}, {Abernathy}, {Accadia}, {Acernese}, {Adams}, {Adhikari},
  {Affeldt} et~al.}}]{Virgo:2011aa}
\bibinfo{author}{\bibfnamefont{J.}~\bibnamefont{{Abadie}}},
  \bibinfo{author}{\bibfnamefont{B.~P.} \bibnamefont{{Abbott}}},
  \bibinfo{author}{\bibfnamefont{R.}~\bibnamefont{{Abbott}}},
  \bibinfo{author}{\bibfnamefont{T.~D.} \bibnamefont{{Abbott}}},
  \bibinfo{author}{\bibfnamefont{M.}~\bibnamefont{{Abernathy}}},
  \bibinfo{author}{\bibfnamefont{T.}~\bibnamefont{{Accadia}}},
  \bibinfo{author}{\bibfnamefont{F.}~\bibnamefont{{Acernese}}},
  \bibinfo{author}{\bibfnamefont{C.}~\bibnamefont{{Adams}}},
  \bibinfo{author}{\bibfnamefont{R.}~\bibnamefont{{Adhikari}}},
  \bibinfo{author}{\bibfnamefont{C.}~\bibnamefont{{Affeldt}}},
  \bibnamefont{et~al.}, \bibinfo{journal}{Astron. \& Astrophys.}
  \textbf{\bibinfo{volume}{541}}, \bibinfo{eid}{A155} (\bibinfo{year}{2012}).

\bibitem[{\citenamefont{{Evans} et~al.}(2012)\citenamefont{{Evans},
  {Fridriksson}, {Gehrels}, {Homan}, {Osborne}, {Siegel}, {Beardmore},
  {Handbauer}, {Gelbord}, {Kennea} et~al.}}]{SwiftS6}
\bibinfo{author}{\bibfnamefont{P.~A.} \bibnamefont{{Evans}}},
  \bibinfo{author}{\bibfnamefont{J.~K.} \bibnamefont{{Fridriksson}}},
  \bibinfo{author}{\bibfnamefont{N.}~\bibnamefont{{Gehrels}}},
  \bibinfo{author}{\bibfnamefont{J.}~\bibnamefont{{Homan}}},
  \bibinfo{author}{\bibfnamefont{J.~P.} \bibnamefont{{Osborne}}},
  \bibinfo{author}{\bibfnamefont{M.}~\bibnamefont{{Siegel}}},
  \bibinfo{author}{\bibfnamefont{A.}~\bibnamefont{{Beardmore}}},
  \bibinfo{author}{\bibfnamefont{P.}~\bibnamefont{{Handbauer}}},
  \bibinfo{author}{\bibfnamefont{J.}~\bibnamefont{{Gelbord}}},
  \bibinfo{author}{\bibfnamefont{J.~A.} \bibnamefont{{Kennea}}},
  \bibnamefont{et~al.}, \bibinfo{journal}{\apjs}
  \textbf{\bibinfo{volume}{203}}, \bibinfo{eid}{28} (\bibinfo{year}{2012}),
  \eprint{1205.1124}.

\bibitem[{\citenamefont{{Nissanke} et~al.}(2013)\citenamefont{{Nissanke},
  {Kasliwal}, and {Georgieva}}}]{2013ApJ...767..124N}
\bibinfo{author}{\bibfnamefont{S.}~\bibnamefont{{Nissanke}}},
  \bibinfo{author}{\bibfnamefont{M.}~\bibnamefont{{Kasliwal}}},
  \bibnamefont{and}
  \bibinfo{author}{\bibfnamefont{A.}~\bibnamefont{{Georgieva}}},
  \bibinfo{journal}{\apj} \textbf{\bibinfo{volume}{767}}, \bibinfo{eid}{124}
  (\bibinfo{year}{2013}), \eprint{1210.6362}.

\bibitem[{\citenamefont{{Bloom} et~al.}(2009)}]{Bloom:2009vx}
\bibinfo{author}{\bibfnamefont{J.~S.} \bibnamefont{{Bloom}}}
  \bibnamefont{et~al.} (\bibinfo{year}{2009}), \eprint{0902.1527}.

\bibitem[{\citenamefont{{Mandel} et~al.}(2011)\citenamefont{{Mandel}, {Kelley},
  and {Ramirez-Ruiz}}}]{Mandel:2011}
\bibinfo{author}{\bibfnamefont{I.}~\bibnamefont{{Mandel}}},
  \bibinfo{author}{\bibfnamefont{L.~Z.} \bibnamefont{{Kelley}}},
  \bibnamefont{and}
  \bibinfo{author}{\bibfnamefont{E.}~\bibnamefont{{Ramirez-Ruiz}}},
  \bibinfo{journal}{ArXiv e-prints}  (\bibinfo{year}{2011}),
  \eprint{1111.0005}.

\bibitem[{\citenamefont{Metzger and Berger}(2012)}]{Metzger:2011bv}
\bibinfo{author}{\bibfnamefont{B.}~\bibnamefont{Metzger}} \bibnamefont{and}
  \bibinfo{author}{\bibfnamefont{E.}~\bibnamefont{Berger}},
  \bibinfo{journal}{Astrophys. J.} \textbf{\bibinfo{volume}{746}},
  \bibinfo{pages}{48} (\bibinfo{year}{2012}), \eprint{1108.6056}.

\bibitem[{\citenamefont{{Bulik} and
  {Belczy{\'n}ski}}(2003)}]{BulikBelczynski:2003}
\bibinfo{author}{\bibfnamefont{T.}~\bibnamefont{{Bulik}}} \bibnamefont{and}
  \bibinfo{author}{\bibfnamefont{K.}~\bibnamefont{{Belczy{\'n}ski}}},
  \bibinfo{journal}{Astrophys. J.} \textbf{\bibinfo{volume}{589}},
  \bibinfo{pages}{L37} (\bibinfo{year}{2003}), \eprint{astro-ph/0301470}.

\bibitem[{\citenamefont{{Mandel} and
  {O'Shaughnessy}}(2010)}]{MandelOshaughnessy:2010}
\bibinfo{author}{\bibfnamefont{I.}~\bibnamefont{{Mandel}}} \bibnamefont{and}
  \bibinfo{author}{\bibfnamefont{R.}~\bibnamefont{{O'Shaughnessy}}},
  \bibinfo{journal}{Class. Quant. Grav} \textbf{\bibinfo{volume}{27}},
  \bibinfo{pages}{114007} (\bibinfo{year}{2010}), \eprint{0912.1074}.

\bibitem[{\citenamefont{{Li} et~al.}(2012{\natexlab{a}})\citenamefont{{Li},
  {Del Pozzo}, {Vitale}, {Van Den Broeck}, {Agathos}, {Veitch}, {Grover},
  {Sidery}, {Sturani}, and {Vecchio}}}]{2012PhRvD..85h2003L}
\bibinfo{author}{\bibfnamefont{T.~G.~F.} \bibnamefont{{Li}}},
  \bibinfo{author}{\bibfnamefont{W.}~\bibnamefont{{Del Pozzo}}},
  \bibinfo{author}{\bibfnamefont{S.}~\bibnamefont{{Vitale}}},
  \bibinfo{author}{\bibfnamefont{C.}~\bibnamefont{{Van Den Broeck}}},
  \bibinfo{author}{\bibfnamefont{M.}~\bibnamefont{{Agathos}}},
  \bibinfo{author}{\bibfnamefont{J.}~\bibnamefont{{Veitch}}},
  \bibinfo{author}{\bibfnamefont{K.}~\bibnamefont{{Grover}}},
  \bibinfo{author}{\bibfnamefont{T.}~\bibnamefont{{Sidery}}},
  \bibinfo{author}{\bibfnamefont{R.}~\bibnamefont{{Sturani}}},
  \bibnamefont{and}
  \bibinfo{author}{\bibfnamefont{A.}~\bibnamefont{{Vecchio}}},
  \bibinfo{journal}{\prd} \textbf{\bibinfo{volume}{85}}, \bibinfo{eid}{082003}
  (\bibinfo{year}{2012}{\natexlab{a}}), \eprint{1110.0530}.

\bibitem[{\citenamefont{{Li} et~al.}(2012{\natexlab{b}})\citenamefont{{Li},
  {Del Pozzo}, {Vitale}, {Van Den Broeck}, {Agathos}, {Veitch}, {Grover},
  {Sidery}, {Sturani}, and {Vecchio}}}]{2012JPhCS.363a2028L}
\bibinfo{author}{\bibfnamefont{T.~G.~F.} \bibnamefont{{Li}}},
  \bibinfo{author}{\bibfnamefont{W.}~\bibnamefont{{Del Pozzo}}},
  \bibinfo{author}{\bibfnamefont{S.}~\bibnamefont{{Vitale}}},
  \bibinfo{author}{\bibfnamefont{C.}~\bibnamefont{{Van Den Broeck}}},
  \bibinfo{author}{\bibfnamefont{M.}~\bibnamefont{{Agathos}}},
  \bibinfo{author}{\bibfnamefont{J.}~\bibnamefont{{Veitch}}},
  \bibinfo{author}{\bibfnamefont{K.}~\bibnamefont{{Grover}}},
  \bibinfo{author}{\bibfnamefont{T.}~\bibnamefont{{Sidery}}},
  \bibinfo{author}{\bibfnamefont{R.}~\bibnamefont{{Sturani}}},
  \bibnamefont{and}
  \bibinfo{author}{\bibfnamefont{A.}~\bibnamefont{{Vecchio}}},
  \bibinfo{journal}{Journal of Physics Conference Series}
  \textbf{\bibinfo{volume}{363}}, \bibinfo{pages}{012028}
  (\bibinfo{year}{2012}{\natexlab{b}}), \eprint{1111.5274}.

\bibitem[{\citenamefont{Helmstrom}(1968)}]{helmstrom-1968}
\bibinfo{author}{\bibfnamefont{C.~W.} \bibnamefont{Helmstrom}},
  \emph{\bibinfo{title}{Statistical Theory of Signal Detection, 2nd edition}}
  (\bibinfo{publisher}{Pergamon Press, London}, \bibinfo{year}{1968}).

\bibitem[{\citenamefont{Abadie et~al.}(2010{\natexlab{b}})}]{LIGOS5}
\bibinfo{author}{\bibfnamefont{J.}~\bibnamefont{Abadie}} \bibnamefont{et~al.}
  (\bibinfo{collaboration}{LIGO Scientific Collaboration}),
  \bibinfo{journal}{Nucl.~Instrum.~Meth.~A} \textbf{\bibinfo{volume}{624}},
  \bibinfo{pages}{223} (\bibinfo{year}{2010}{\natexlab{b}}),
  \eprint{1007.3937}.

\bibitem[{\citenamefont{Accadia et~al.}(2011)}]{VirgoS2}
\bibinfo{author}{\bibfnamefont{T.}~\bibnamefont{Accadia}} \bibnamefont{et~al.}
  (\bibinfo{collaboration}{Virgo Collaboration}), \bibinfo{journal}{Class.\
  Quant.\ Grav.} \textbf{\bibinfo{volume}{28}}, \bibinfo{pages}{025005}
  (\bibinfo{year}{2011}), \eprint{arXiv:1009.5190}.

\bibitem[{\citenamefont{{Vitale} et~al.}(2011)\citenamefont{{Vitale}, {Del
  Pozzo}, {Li}, {Van Den Broeck}, {Mandel}, {Aylott}, and
  {Veitch}}}]{Vitale:2012}
\bibinfo{author}{\bibfnamefont{S.}~\bibnamefont{{Vitale}}},
  \bibinfo{author}{\bibfnamefont{W.}~\bibnamefont{{Del Pozzo}}},
  \bibinfo{author}{\bibfnamefont{T.~G.~F.} \bibnamefont{{Li}}},
  \bibinfo{author}{\bibfnamefont{C.}~\bibnamefont{{Van Den Broeck}}},
  \bibinfo{author}{\bibfnamefont{I.}~\bibnamefont{{Mandel}}},
  \bibinfo{author}{\bibfnamefont{B.}~\bibnamefont{{Aylott}}}, \bibnamefont{and}
  \bibinfo{author}{\bibfnamefont{J.}~\bibnamefont{{Veitch}}},
  \bibinfo{journal}{ArXiv e-prints}  (\bibinfo{year}{2011}),
  \eprint{1111.3044}.

\bibitem[{LAL()}]{LALPE}
\emph{\bibinfo{title}{{\normalfont LSC Algorithm Library,
  \url{http://www.lsc-group.phys.uwm.edu/lal}}}},
  \urlprefix\url{http://www.lsc-group.phys.uwm.edu/lal}.

\bibitem[{\citenamefont{{Raymond} et~al.}(2010)\citenamefont{{Raymond}, {van
  der Sluys}, {Mandel}, {Kalogera}, {R{\"o}ver}, and
  {Christensen}}}]{Raymond:2010}
\bibinfo{author}{\bibfnamefont{V.}~\bibnamefont{{Raymond}}},
  \bibinfo{author}{\bibfnamefont{M.~V.} \bibnamefont{{van der Sluys}}},
  \bibinfo{author}{\bibfnamefont{I.}~\bibnamefont{{Mandel}}},
  \bibinfo{author}{\bibfnamefont{V.}~\bibnamefont{{Kalogera}}},
  \bibinfo{author}{\bibfnamefont{C.}~\bibnamefont{{R{\"o}ver}}},
  \bibnamefont{and}
  \bibinfo{author}{\bibfnamefont{N.}~\bibnamefont{{Christensen}}},
  \bibinfo{journal}{Class. Quant. Grav} \textbf{\bibinfo{volume}{27}},
  \bibinfo{pages}{114009} (\bibinfo{year}{2010}), \eprint{0912.3746}.

\bibitem[{\citenamefont{{van der Sluys}
  et~al.}(2008{\natexlab{a}})}]{Sluys:2008a}
\bibinfo{author}{\bibfnamefont{M.}~\bibnamefont{{van der Sluys}}}
  \bibnamefont{et~al.}, \bibinfo{journal}{Class. Quant. Grav}
  \textbf{\bibinfo{volume}{25}}, \bibinfo{pages}{184011}
  (\bibinfo{year}{2008}{\natexlab{a}}), \eprint{0805.1689}.

\bibitem[{\citenamefont{{van der Sluys}
  et~al.}(2008{\natexlab{b}})}]{Sluys:2008b}
\bibinfo{author}{\bibfnamefont{M.~V.} \bibnamefont{{van der Sluys}}}
  \bibnamefont{et~al.}, \bibinfo{journal}{Astrophys. J.}
  \textbf{\bibinfo{volume}{688}}, \bibinfo{pages}{L61}
  (\bibinfo{year}{2008}{\natexlab{b}}), \eprint{0710.1897}.

\bibitem[{\citenamefont{{Veitch} and {Vecchio}}(2010)}]{Veitch:2010}
\bibinfo{author}{\bibfnamefont{J.}~\bibnamefont{{Veitch}}} \bibnamefont{and}
  \bibinfo{author}{\bibfnamefont{A.}~\bibnamefont{{Vecchio}}},
  \bibinfo{journal}{Phys. Rev. D} \textbf{\bibinfo{volume}{81}},
  \bibinfo{pages}{062003} (\bibinfo{year}{2010}), \eprint{0911.3820}.

\bibitem[{\citenamefont{{Feroz} et~al.}(2009)\citenamefont{{Feroz}, {Hobson},
  and {Bridges}}}]{Feroz:2009}
\bibinfo{author}{\bibfnamefont{F.}~\bibnamefont{{Feroz}}},
  \bibinfo{author}{\bibfnamefont{M.~P.} \bibnamefont{{Hobson}}},
  \bibnamefont{and}
  \bibinfo{author}{\bibfnamefont{M.}~\bibnamefont{{Bridges}}},
  \bibinfo{journal}{Monthly Notices of the Royal Astronomical Society}
  \textbf{\bibinfo{volume}{398}}, \bibinfo{pages}{1601} (\bibinfo{year}{2009}),
  \eprint{0809.3437}.

\bibitem[{\citenamefont{{Skilling}}(2006)}]{Skilling:2006}
\bibinfo{author}{\bibfnamefont{J.}~\bibnamefont{{Skilling}}},
  \bibinfo{journal}{Bayesian Analysis} \textbf{\bibinfo{volume}{1}},
  \bibinfo{pages}{833} (\bibinfo{year}{2006}).

\bibitem[{\citenamefont{{Weinberg}}(2009)}]{Weinberg:2009}
\bibinfo{author}{\bibfnamefont{M.~D.} \bibnamefont{{Weinberg}}},
  \bibinfo{journal}{ArXiv e-prints}  (\bibinfo{year}{2009}),
  \eprint{0911.1777}.

\bibitem[{\citenamefont{Sathyaprakash and Schutz}(2009)}]{lrr-2009-2}
\bibinfo{author}{\bibfnamefont{B.}~\bibnamefont{Sathyaprakash}}
  \bibnamefont{and} \bibinfo{author}{\bibfnamefont{B.~F.}
  \bibnamefont{Schutz}}, \bibinfo{journal}{Living Reviews in Relativity}
  \textbf{\bibinfo{volume}{12}} (\bibinfo{year}{2009}),
  \urlprefix\url{http://www.livingreviews.org/lrr-2009-2}.

\bibitem[{\citenamefont{Thorne}(1987)}]{thorne.k:1987}
\bibinfo{author}{\bibfnamefont{K.~S.} \bibnamefont{Thorne}}, in
  \emph{\bibinfo{booktitle}{Three hundred years of gravitation}}, edited by
  \bibinfo{editor}{\bibfnamefont{S.~W.} \bibnamefont{Hawking}}
  \bibnamefont{and} \bibinfo{editor}{\bibfnamefont{W.}~\bibnamefont{Israel}}
  (\bibinfo{publisher}{Cambridge University Press},
  \bibinfo{address}{Cambridge}, \bibinfo{year}{1987}),
  chap.~\bibinfo{chapter}{9}, pp. \bibinfo{pages}{330--458}.

\bibitem[{\citenamefont{Blanchet}(2006)}]{Blanchet:2006av}
\bibinfo{author}{\bibfnamefont{L.}~\bibnamefont{Blanchet}},
  \bibinfo{journal}{Living Rev. Rel.} \textbf{\bibinfo{volume}{9}},
  \bibinfo{pages}{3} (\bibinfo{year}{2006}), \eprint{gr-qc/0202016},
  \urlprefix\url{http://www.livingreviews.org/lrr-2006-4}.

\bibitem[{\citenamefont{Buonanno et~al.}(2009)\citenamefont{Buonanno, Iyer,
  Ochsner, Pan, and Sathyaprakash}}]{BuonannoIyerOchsnerYiSathya2009}
\bibinfo{author}{\bibfnamefont{A.}~\bibnamefont{Buonanno}},
  \bibinfo{author}{\bibfnamefont{B.~R.} \bibnamefont{Iyer}},
  \bibinfo{author}{\bibfnamefont{E.}~\bibnamefont{Ochsner}},
  \bibinfo{author}{\bibfnamefont{Y.}~\bibnamefont{Pan}}, \bibnamefont{and}
  \bibinfo{author}{\bibfnamefont{B.~S.} \bibnamefont{Sathyaprakash}},
  \bibinfo{journal}{Phys. Rev. D} \textbf{\bibinfo{volume}{80}},
  \bibinfo{pages}{084043} (\bibinfo{year}{2009}).

\bibitem[{\citenamefont{{Ajith} et~al.}(2012)\citenamefont{{Ajith},
  {Fotopoulos}, {Privitera}, {Neunzert}, and
  {Weinstein}}}]{2012arXiv1210.6666A}
\bibinfo{author}{\bibfnamefont{P.}~\bibnamefont{{Ajith}}},
  \bibinfo{author}{\bibfnamefont{N.}~\bibnamefont{{Fotopoulos}}},
  \bibinfo{author}{\bibfnamefont{S.}~\bibnamefont{{Privitera}}},
  \bibinfo{author}{\bibfnamefont{A.}~\bibnamefont{{Neunzert}}},
  \bibnamefont{and} \bibinfo{author}{\bibfnamefont{A.~J.}
  \bibnamefont{{Weinstein}}}, \bibinfo{journal}{ArXiv e-prints}
  (\bibinfo{year}{2012}), \eprint{1210.6666}.

\bibitem[{\citenamefont{Buonanno
  et~al.}(2003{\natexlab{a}})\citenamefont{Buonanno, Chen, and
  Vallisneri}}]{BuonannoChenVallisneri:2003b}
\bibinfo{author}{\bibfnamefont{A.}~\bibnamefont{Buonanno}},
  \bibinfo{author}{\bibfnamefont{Y.}~\bibnamefont{Chen}}, \bibnamefont{and}
  \bibinfo{author}{\bibfnamefont{M.}~\bibnamefont{Vallisneri}},
  \bibinfo{journal}{Phys.~Rev.~D} \textbf{\bibinfo{volume}{67}},
  \bibinfo{pages}{104025} (\bibinfo{year}{2003}{\natexlab{a}}),
  \bibinfo{note}{erratum-ibid. 74 (2006) 029904(E)}.

\bibitem[{\citenamefont{Ajith et~al.}(2011)}]{Ajith:2009bn}
\bibinfo{author}{\bibfnamefont{P.}~\bibnamefont{Ajith}} \bibnamefont{et~al.},
  \bibinfo{journal}{Phys. Rev. Lett.} \textbf{\bibinfo{volume}{106}},
  \bibinfo{pages}{241101} (\bibinfo{year}{2011}), \eprint{arXiv:0909.2867}.

\bibitem[{\citenamefont{{Kowalska} et~al.}(2011)\citenamefont{{Kowalska},
  {Bulik}, {Belczynski}, {Dominik}, and
  {Gondek-Rosinska}}}]{2011A&A...527A..70K}
\bibinfo{author}{\bibfnamefont{I.}~\bibnamefont{{Kowalska}}},
  \bibinfo{author}{\bibfnamefont{T.}~\bibnamefont{{Bulik}}},
  \bibinfo{author}{\bibfnamefont{K.}~\bibnamefont{{Belczynski}}},
  \bibinfo{author}{\bibfnamefont{M.}~\bibnamefont{{Dominik}}},
  \bibnamefont{and}
  \bibinfo{author}{\bibfnamefont{D.}~\bibnamefont{{Gondek-Rosinska}}},
  \bibinfo{journal}{\aap} \textbf{\bibinfo{volume}{527}}, \bibinfo{eid}{A70}
  (\bibinfo{year}{2011}), \eprint{1010.0511}.

\bibitem[{\citenamefont{Blanchet}(2002)}]{Blanchet:2002xy}
\bibinfo{author}{\bibfnamefont{L.}~\bibnamefont{Blanchet}}
  (\bibinfo{year}{2002}), \eprint{gr-qc/0207037}.

\bibitem[{\citenamefont{{Yunes} and {Berti}}(2008)}]{2008PhRvD..77l4006Y}
\bibinfo{author}{\bibfnamefont{N.}~\bibnamefont{{Yunes}}} \bibnamefont{and}
  \bibinfo{author}{\bibfnamefont{E.}~\bibnamefont{{Berti}}},
  \bibinfo{journal}{\prd} \textbf{\bibinfo{volume}{77}}, \bibinfo{eid}{124006}
  (\bibinfo{year}{2008}), \eprint{0803.1853}.

\bibitem[{\citenamefont{{Brown}}(2004)}]{Brown:2004vh}
\bibinfo{author}{\bibfnamefont{D.~A.} \bibnamefont{{Brown}}}, Ph.D. thesis,
  \bibinfo{school}{The University of Wisconsin - Milwaukee, Wisconsin, USA}
  (\bibinfo{year}{2004}), \eprint{0705.1514}.

\bibitem[{\citenamefont{Babak et~al.}(2013)}]{ihopePaper:2012}
\bibinfo{author}{\bibfnamefont{S.}~\bibnamefont{Babak}} \bibnamefont{et~al.},
  \bibinfo{journal}{Phys. Rev. D} \textbf{\bibinfo{volume}{87}},
  \bibinfo{pages}{024033} (\bibinfo{year}{2013}), \eprint{arXiv:1208.3491}.

\bibitem[{\citenamefont{Buonanno et~al.}(2007)}]{Buonanno:2007pf}
\bibinfo{author}{\bibfnamefont{A.}~\bibnamefont{Buonanno}}
  \bibnamefont{et~al.}, \bibinfo{journal}{Phys. Rev. D}
  \textbf{\bibinfo{volume}{76}}, \bibinfo{pages}{104049}
  (\bibinfo{year}{2007}), \eprint{arXiv:0706.3732}.

\bibitem[{\citenamefont{Cutler and Flanagan}(1994)}]{Cutler:1994}
\bibinfo{author}{\bibfnamefont{C.}~\bibnamefont{Cutler}} \bibnamefont{and}
  \bibinfo{author}{\bibfnamefont{E.}~\bibnamefont{Flanagan}},
  \bibinfo{journal}{Phys.~Rev.~D} \textbf{\bibinfo{volume}{49}},
  \bibinfo{pages}{2658} (\bibinfo{year}{1994}).

\bibitem[{\citenamefont{Poisson and Will}(1995)}]{Poisson:1995ef}
\bibinfo{author}{\bibfnamefont{E.}~\bibnamefont{Poisson}} \bibnamefont{and}
  \bibinfo{author}{\bibfnamefont{C.~M.} \bibnamefont{Will}},
  \bibinfo{journal}{Phys.~Rev.~D} \textbf{\bibinfo{volume}{52}},
  \bibinfo{pages}{848} (\bibinfo{year}{1995}), \eprint{arXiv:gr-qc/9502040}.

\bibitem[{\citenamefont{{Baird} et~al.}(2013)\citenamefont{{Baird},
  {Fairhurst}, {Hannam}, and {Murphy}}}]{2013PhRvD..87b4035B}
\bibinfo{author}{\bibfnamefont{E.}~\bibnamefont{{Baird}}},
  \bibinfo{author}{\bibfnamefont{S.}~\bibnamefont{{Fairhurst}}},
  \bibinfo{author}{\bibfnamefont{M.}~\bibnamefont{{Hannam}}}, \bibnamefont{and}
  \bibinfo{author}{\bibfnamefont{P.}~\bibnamefont{{Murphy}}},
  \bibinfo{journal}{\prd} \textbf{\bibinfo{volume}{87}}, \bibinfo{eid}{024035}
  (\bibinfo{year}{2013}), \eprint{1211.0546}.

\bibitem[{\citenamefont{Buonanno
  et~al.}(2003{\natexlab{b}})\citenamefont{Buonanno, Chen, and
  Vallisneri}}]{BuonannoChenVallisneri:2003a}
\bibinfo{author}{\bibfnamefont{A.}~\bibnamefont{Buonanno}},
  \bibinfo{author}{\bibfnamefont{Y.}~\bibnamefont{Chen}}, \bibnamefont{and}
  \bibinfo{author}{\bibfnamefont{M.}~\bibnamefont{Vallisneri}},
  \bibinfo{journal}{Phys.~Rev.~D} \textbf{\bibinfo{volume}{67}},
  \bibinfo{pages}{024016} (\bibinfo{year}{2003}{\natexlab{b}}),
  \bibinfo{note}{erratum-ibid. 74 (2006) 029903(E)}.

\bibitem[{\citenamefont{{Raymond} et~al.}(2009)\citenamefont{{Raymond}, {van
  der Sluys}, {Mandel}, {Kalogera}, {R{\"o}ver}, and
  {Christensen}}}]{Raymond:2009}
\bibinfo{author}{\bibfnamefont{V.}~\bibnamefont{{Raymond}}},
  \bibinfo{author}{\bibfnamefont{M.~V.} \bibnamefont{{van der Sluys}}},
  \bibinfo{author}{\bibfnamefont{I.}~\bibnamefont{{Mandel}}},
  \bibinfo{author}{\bibfnamefont{V.}~\bibnamefont{{Kalogera}}},
  \bibinfo{author}{\bibfnamefont{C.}~\bibnamefont{{R{\"o}ver}}},
  \bibnamefont{and}
  \bibinfo{author}{\bibfnamefont{N.}~\bibnamefont{{Christensen}}},
  \bibinfo{journal}{Class. Quant. Grav} \textbf{\bibinfo{volume}{26}},
  \bibinfo{pages}{114007} (\bibinfo{year}{2009}), \eprint{0812.4302}.

\bibitem[{\citenamefont{Littenberg and Cornish}(2010)}]{PhysRevD.82.103007}
\bibinfo{author}{\bibfnamefont{T.~B.} \bibnamefont{Littenberg}}
  \bibnamefont{and} \bibinfo{author}{\bibfnamefont{N.~J.}
  \bibnamefont{Cornish}}, \bibinfo{journal}{Phys. Rev. D}
  \textbf{\bibinfo{volume}{82}}, \bibinfo{pages}{103007}
  (\bibinfo{year}{2010}),
  \urlprefix\url{http://link.aps.org/doi/10.1103/PhysRevD.82.103007}.

\bibitem[{\citenamefont{{Smith} et~al.}(2012)\citenamefont{{Smith}, {Cannon},
  {Hanna}, {Keppel}, and {Mandel}}}]{SmithSVD:2012}
\bibinfo{author}{\bibfnamefont{R.~J.~E.} \bibnamefont{{Smith}}},
  \bibinfo{author}{\bibfnamefont{K.}~\bibnamefont{{Cannon}}},
  \bibinfo{author}{\bibfnamefont{C.}~\bibnamefont{{Hanna}}},
  \bibinfo{author}{\bibfnamefont{D.}~\bibnamefont{{Keppel}}}, \bibnamefont{and}
  \bibinfo{author}{\bibfnamefont{I.}~\bibnamefont{{Mandel}}},
  \bibinfo{journal}{ArXiv e-prints}  (\bibinfo{year}{2012}),
  \eprint{1211.1254}.

\bibitem[{\citenamefont{Rao}(2009)}]{rao2009linear}
\bibinfo{author}{\bibfnamefont{C.}~\bibnamefont{Rao}},
  \emph{\bibinfo{title}{Linear Statistical Inference and its Applications}},
  Wiley Series in Probability and Statistics (\bibinfo{publisher}{Wiley},
  \bibinfo{year}{2009}), ISBN \bibinfo{isbn}{9780470317143},
  \urlprefix\url{http://books.google.com/books?id=gKXYxdw-24YC}.

\bibitem[{\citenamefont{Arun et~al.}(2005)\citenamefont{Arun, Iyer,
  Sathyaprakash, and Sundararajan}}]{Arun:2006yw}
\bibinfo{author}{\bibfnamefont{K.~G.} \bibnamefont{Arun}},
  \bibinfo{author}{\bibfnamefont{B.~R.} \bibnamefont{Iyer}},
  \bibinfo{author}{\bibfnamefont{B.~S.} \bibnamefont{Sathyaprakash}},
  \bibnamefont{and} \bibinfo{author}{\bibfnamefont{P.~A.}
  \bibnamefont{Sundararajan}}, \bibinfo{journal}{Phys.~Rev.~D}
  \textbf{\bibinfo{volume}{71}}, \bibinfo{eid}{084008}
  (pages~\bibinfo{numpages}{16}) (\bibinfo{year}{2005}),
  \urlprefix\url{http://link.aps.org/abstract/PRD/v71/e084008}.

\bibitem[{\citenamefont{{Vitale} and {Zanolin}}(2011)}]{2011PhRvD..84j4020V}
\bibinfo{author}{\bibfnamefont{S.}~\bibnamefont{{Vitale}}} \bibnamefont{and}
  \bibinfo{author}{\bibfnamefont{M.}~\bibnamefont{{Zanolin}}},
  \bibinfo{journal}{\prd} \textbf{\bibinfo{volume}{84}}, \bibinfo{eid}{104020}
  (\bibinfo{year}{2011}), \eprint{1108.2410}.

\bibitem[{\citenamefont{{Veitch} et~al.}(2012)\citenamefont{{Veitch}, {Mandel},
  {Aylott}, {Farr}, {Raymond}, {Rodriguez}, {van der Sluys}, {Kalogera}, and
  {Vecchio}}}]{2012PhRvD..85j4045V}
\bibinfo{author}{\bibfnamefont{J.}~\bibnamefont{{Veitch}}},
  \bibinfo{author}{\bibfnamefont{I.}~\bibnamefont{{Mandel}}},
  \bibinfo{author}{\bibfnamefont{B.}~\bibnamefont{{Aylott}}},
  \bibinfo{author}{\bibfnamefont{B.}~\bibnamefont{{Farr}}},
  \bibinfo{author}{\bibfnamefont{V.}~\bibnamefont{{Raymond}}},
  \bibinfo{author}{\bibfnamefont{C.}~\bibnamefont{{Rodriguez}}},
  \bibinfo{author}{\bibfnamefont{M.}~\bibnamefont{{van der Sluys}}},
  \bibinfo{author}{\bibfnamefont{V.}~\bibnamefont{{Kalogera}}},
  \bibnamefont{and}
  \bibinfo{author}{\bibfnamefont{A.}~\bibnamefont{{Vecchio}}},
  \bibinfo{journal}{\prd} \textbf{\bibinfo{volume}{85}}, \bibinfo{eid}{104045}
  (\bibinfo{year}{2012}), \eprint{1201.1195}.

\bibitem[{\citenamefont{{Vallisneri}}(2008)}]{2008PhRvD..77d2001V}
\bibinfo{author}{\bibfnamefont{M.}~\bibnamefont{{Vallisneri}}},
  \bibinfo{journal}{\prd} \textbf{\bibinfo{volume}{77}}, \bibinfo{eid}{042001}
  (\bibinfo{year}{2008}), \eprint{arXiv:gr-qc/0703086}.

\bibitem[{\citenamefont{{Zanolin} et~al.}(2010)\citenamefont{{Zanolin},
  {Vitale}, and {Makris}}}]{2010PhRvD..81l4048Z}
\bibinfo{author}{\bibfnamefont{M.}~\bibnamefont{{Zanolin}}},
  \bibinfo{author}{\bibfnamefont{S.}~\bibnamefont{{Vitale}}}, \bibnamefont{and}
  \bibinfo{author}{\bibfnamefont{N.}~\bibnamefont{{Makris}}},
  \bibinfo{journal}{\prd} \textbf{\bibinfo{volume}{81}}, \bibinfo{eid}{124048}
  (\bibinfo{year}{2010}), \eprint{0912.0065}.

\bibitem[{\citenamefont{{R{\"o}ver} et~al.}(2011)\citenamefont{{R{\"o}ver},
  {Meyer}, and {Christensen}}}]{RoverMeyerChristensen:2011}
\bibinfo{author}{\bibfnamefont{C.}~\bibnamefont{{R{\"o}ver}}},
  \bibinfo{author}{\bibfnamefont{R.}~\bibnamefont{{Meyer}}}, \bibnamefont{and}
  \bibinfo{author}{\bibfnamefont{N.}~\bibnamefont{{Christensen}}},
  \bibinfo{journal}{Class. Quant. Grav} \textbf{\bibinfo{volume}{28}},
  \bibinfo{pages}{015010} (\bibinfo{year}{2011}).

\bibitem[{\citenamefont{Littenberg et~al.}(2013)\citenamefont{Littenberg,
  Coughlin, Farr, and Farr}}]{Littenberg:2013gja}
\bibinfo{author}{\bibfnamefont{T.~B.} \bibnamefont{Littenberg}},
  \bibinfo{author}{\bibfnamefont{M.}~\bibnamefont{Coughlin}},
  \bibinfo{author}{\bibfnamefont{B.}~\bibnamefont{Farr}}, \bibnamefont{and}
  \bibinfo{author}{\bibfnamefont{W.~M.} \bibnamefont{Farr}}
  (\bibinfo{year}{2013}), \eprint{1307.8195}.

\end{thebibliography}

\end{document}